\documentclass[aps,preprint,floatfix,nofootinbib,showpacs]{revtex4-1}
\pdfoutput=1
\usepackage{graphicx,color}
\usepackage{hyperref}
\usepackage{amsmath}
\usepackage{amsfonts}
\usepackage{amssymb}
\usepackage{setspace}
\newcommand{\real}{\Re {\rm e}}

\begin{document}

{\small
\begin{flushright}
CNU-HEP-15-03
\end{flushright} }

\title{An Exploratory study of Higgs-boson pair production}

\def\slash#1{#1\!\!/}

\renewcommand{\thefootnote}{\arabic{footnote}}

\author{
Chih-Ting Lu$^1$, Jung Chang$^1$, Kingman Cheung$^{1,2,3}$, and 
Jae Sik Lee$^{3,4}$
}
\affiliation{
$^1$ Department of Physics, National Tsing Hua University,
Hsinchu 300, Taiwan \\
$^2$ Division of Quantum Phases and Devices, School of Physics, 
Konkuk University, Seoul 143-701, Republic of Korea \\
$^3$ Physics Division, National Center for Theoretical Sciences,
Hsinchu, Taiwan \\
$^4$ Department of Physics, Chonnam National University, \\
300 Yongbong-dong, Buk-gu, Gwangju, 500-757, Republic of Korea
}
\date{\today}

\begin{abstract}
Higgs-boson pair production is well known being capable to probe the
trilinear self-coupling of the Higgs boson, which is one of the
important ingredients of the Higgs sector itself.  Pair production then
depends on the top-quark Yukawa coupling $g_t^{S,P}$, 
Higgs trilinear coupling $\lambda_{3H}$,
and a possible dim-5 contact-type $ttHH$ coupling $g_{tt}^{S,P}$,
which may appear in some
higher representations of the Higgs sector. We take into account the
possibility that the top-Yukawa and the $ttHH$ couplings 
involved can be CP violating.  We calculate the
cross sections and the interference terms as coefficients of the
square or the 4th power of each coupling $(g_t^{S,P},\,
\lambda_{3H},\,g_{tt}^{S,P})$ at various stages of cuts,
such that the desired cross section under various cuts can be obtained
by simply inputing the couplings.
We employ the $H H \to \gamma\gamma b \bar b$ decay mode of the 
Higgs-boson pair to investigate the possibility of disentangle the
triangle diagram from the box digram so as to have a clean probe of
the trilinear coupling at the LHC. We found that the angular separation 
between the
$b$ and $\bar b$ and that between the two photons is useful.
We obtain the sensitivity reach of each pair of couplings at the 14 TeV
LHC and the future 100 TeV $pp$ machine. 
Finally, we also comment on using the $b\bar b \tau^+ \tau^-$ 
decay mode in Appendix.
\end{abstract}

\maketitle

\section{Introduction}

A  boson was discovered at the Large
Hadron Collider (LHC) \cite{atlas,cms}. After almost all the Run I data were
analyzed, the measured properties of the  new particle
are best described 
by the standard-model (SM) Higgs boson \cite{higgcision,update2014}, which was 
proposed in 1960s \cite{higgs}.
The most constrained is the gauge-Higgs coupling
$C_v \equiv g_{HWW} = 0.94\,^{+0.11}_{-0.12}$, 
which is very close to the 
SM value \cite{latest}.  
On the other hand, the relevant top- and bottom-Yukawa couplings
are not determined as precisely as $C_v$ by the current data. Nevertheless,
they are within $30-40\%$ of the SM values \cite{latest}.

Until now there is no information at all about the self-couplings
of the Higgs boson, which emerges from the inner dynamics of the Higgs
sector. For example, the trilinear couplings from the SM, two-Higgs doublet
models (2HDM), and MSSM are very different from one another. 
Thus, investigations of the
trilinear coupling will shed lights on the dynamics of the Higgs sector. 
One of the best probes is Higgs-boson-pair production at the LHC.
There have been a large number of works of Higgs-pair production
in the SM \cite{plehn,baglio,hh-sm,loop}, in model-independent 
fashion~\cite{hh-mi,1301.3492,contino,low}, 
and in special models beyond the SM \cite{hh-bey} and in SUSY \cite{hh-susy}.
In the SM, Higgs-pair production receives contributions from two
entangled sources, the triangle and box diagrams.  The
triangle diagram involves the Higgs self-trilinear coupling
and the top-Yukawa coupling 
while the box diagram involves only the top-Yukawa coupling.
In order to probe the effects of the Higgs trilinear coupling,
we have to disentangle the triangle diagram from the box diagram.
We anticipate that the triangle diagram, which contains an $s$-channel
Higgs propagator, does not increase as much as the box diagram as
the center-of-mass energy $\sqrt{\hat s} \equiv M_{HH}$ increases.
Therefore, the box diagram tends to give more energetic Higgs-boson pairs
than the triangle diagram. Thus, the opening angle in the decay products
of each Higgs boson can be used to isolate the triangle-diagram
contribution.  Indeed, we found that the angular separation
$\Delta R_{\gamma\gamma}$ and $\Delta R_{bb}$ between the decay
products of the Higgs-boson pair are very useful variables to disentangle
the two sources.

Here we also entertain the possibility of a dimension-5 operator 
$ttHH$, which can arise from a number of extended Higgs models,
including composite Higgs models or some general 2HDM's. For example,
in a general 2HDM we can have a diagram with a 
$(\bar t_L t_R \varphi)$ vertex and a $(\varphi H H)$
vertex connected by the heavy $\varphi$. When the heavy $\varphi$ is integrated
out, we are left with the contact diagram 
$\bar t_L t_R H H$.
The anomalous $ttHH$ coupling can contribute to 
Higgs-pair production via a triangle diagram.
This triangle diagram is
similar to the triangle diagram with the trilinear Higgs self coupling, except
that it does not have the $s$-channel Higgs propagator. 
We shall show that the new contact diagram will give terms that can be 
combined with the terms of the triangle diagram, as in Eq.~(\ref{eq4}).
We note that the kinematic behavior of the triangle diagram induced
by the dim-5 $ttHH$ contact interaction is different from that induced by the
trilinear Higgs self coupling, because of the absence of the
Higgs propagator in the contact diagram.

In this work, we adopt the effective Lagrangian approach, taking the liberty
that the involved Higgs-boson couplings can be varied freely within reasonable
ranges.  The relevant couplings considered in this work are (i) 
the top-quark Yukawa coupling,
(ii) the trilinear Higgs self-coupling, and (iii) 
the contact-type $ttHH$ coupling.
In the top-quark Yukawa and contact-type $ttHH$ interactions, 
we take into account the possibility of the simultaneous presence of
the scalar and pseudoscalar couplings which can signal CP violation.
The rationale behind the CP-odd part is that the current data,
other than the EDM constraints, cannot restrict the CP-odd part. The
EDM constraints, however, depend on a number of assumptions and may
therefore be weaken because of cancellation among various CP-violating
sources \cite{edm}.  
On the other hand, $ttHH$ is a dimension-5
operator, which may originate from a genuine dim-6 operator, e.g., 
$(\overline{Q_L} \Phi t_R)(\Phi^\dagger \Phi)$
after electroweak symmetry breaking, 
$\Phi =(0,\, (v+H)/\sqrt{2} )^T$. 
This operator is thus suppressed only by two powers of higher scale, such
that it can give a significant contribution at
the LHC energies. 

Our strategy is first to find a useful expression for
Higgs-boson pair production cross sections in terms of 
these couplings, see Eq.~(\ref{cdef}). The coefficient of each
term depends on the collider energy $\sqrt{s}$, Higgs decay channels,
experimental cuts, etc. Thus, such an expression enables us to 
easily obtain the cross 
section under various experimental conditions for arbitrary
values of the couplings. Our aim is to extract the 
information on the Higgs couplings,
especially on the Higgs self coupling by exploiting the expression.
It is helpful to consider some experimental cuts which can 
isolate the contribution from the Higgs self coupling and lead to
various cross sections with different dependence on 
the Higgs couplings. 
In this work, specifically, we employ the $b\bar b
\gamma\gamma$ decay mode of the Higgs-boson pair
and look into the angular separation between the $b$ and $\bar b$ and that
between the photons. It is shown that one can map out the
possible regions of Higgs couplings assuming
certain values of measured cross sections,
though it is channel dependent. 
Thanks to the largest branching ratio of the Higgs boson into $b\bar b$,
the angular separation between 
the bottom-quark pair is an useful tool in for most of the proposed channel
at the LHC.

In summary, the current work marks a number of improvements over 
previous published works as listed in the following:
\begin{enumerate}
\item We have included the CP-odd part in the top-Yukawa coupling. The 
CP-even and CP-odd parts are constrained by an elliptical-like equation
by the current Higgs-boson data, as shown in Fig.~4 of Ref.~\cite{update2014}.
Note that the effects of the CP-odd part of the top-Yukawa coupling
at the LHC, ILC, and photon colliders were studied in Ref.~\cite{cpodd},
though we study the effects in depth for the LHC here.

\item We have included the dim-5 anomalous $ttHH$ contact coupling, as done in
Ref.~\cite{contino,low}. Furthermore, we also include the CP-odd part of this
contact coupling.

\item We have calculated an easy-to-use expression to obtain the
cross sections as a function of the involved couplings at each 
center-of-mass energy. We also obtain similar expressions in 
various kinematic regions such that one can easily obtain the cross
sections under 
the proposed experimental cuts for arbitrary values of the Higgs
couplings.

\item 
With assumed uncertainties in the measurements of cross sections,
we can map out the sensitivity regions of parameter space that 
can be probed at the LHC.

\end{enumerate}

The organization of the work is as follows. 
In the next section, we describe the formalism for our exploratory
approach and present an expression for
the Higgs-boson pair production cross section in terms of 
various combinations of the Higgs couplings
under consideration. In Section III, we examine the behavior of each term of the
cross section versus energies. In Section IV, 
employing the $HH\to(\gamma\gamma)(b\bar{b})$ decay mode,
we illustrate how to extract the information on the Higgs couplings by 
exploiting
the angular separations between the Higgs decay products.
There we also
discuss the prospect for the 100 TeV $pp$ machine.
We conclude in Section V and offer a few comments with regard to our findings.
In Appendix, we 
compare the SM cross sections at 14 TeV with those at 100 TeV
for the process $pp\to H\to \gamma\gamma b \bar b$
and give some comments on the $ \tau^+ \tau^- b\bar b$ decay mode.

\section{Formalism}

Higgs-boson pair production via gluon fusion goes through a triangle
diagram with a Higgs-boson propagator and also through a box diagram 
with colored particles running in it. The relevant couplings involved 
are top-Yukawa and the Higgs trilinear self coupling. We further explore
the possibility of a dim-5 anomalous $ttHH$ contact coupling \cite{low}. They
are given in this Lagrangian:
\begin{equation}
\label{lag}
-{\cal L}=\frac{1}{3!}\left(\frac{3M_H^2}{v}\right)\,\lambda_{3H}\,H^3
\ + \
\frac{m_t}{v}\,\bar{t}\left(g^S_t+i\gamma_5 g^P_t\right)t\,H \ + \ 
\frac{1}{2}\frac{m_t}{v^2}\bar{t}\left(g^S_{tt}+i\gamma_5
g^P_{tt}\right)t\,H^2\,.
\end{equation}
In the SM, $\lambda_{3H}=g^S_t=1$ and $g^P_t=0$ and $g^{S,P}_{tt}=0$. 
The differential cross section for
the process $g(p_1)g(p_2) \to H(p_3)H(p_4)$ 
in the SM was obtained in Ref.~\cite{plehn}
as
\begin{equation}
\frac{d\hat\sigma(gg\to HH)}{d\hat{t}}=\frac{G_F^2 \alpha_s^2}{512(2\pi)^3}
\left[ \Big|
\lambda_{3H}g^S_tD(\hat{s})F_\triangle^S
+(g^S_t)^2F_\Box^{SS} \Big|^2 
+\Big| (g^S_t)^2G_\Box^{SS} \Big|^2 \right] 
\end{equation}
where 
\begin{equation}
D(\hat{s})= \frac{3M_H^2}{\hat{s}-M_H^2+iM_H\Gamma_H}
\end{equation}
and
$\hat{s}=(p_1+p_2)^2$,
$\hat{t}=(p_1-p_3)^2$, and
$\hat{u}=(p_2-p_3)^2$ with $p_1+p_2=p_3+p_4$.%
The loop functions $F_\triangle^S=F_\triangle$, 
$F_\Box^{SS}=F_\Box$, and $G_\Box^{SS}=G_\Box$ 
with $F_{\triangle\,,\Box}$ and $G_\Box$
given in Appendix A.1 of Ref.~\cite{plehn}.

Here we extend the result to including the CP-odd top-Yukawa and the 
anomalous $ttHH$ couplings:
\begin{eqnarray}
\frac{d\hat\sigma(gg\to HH)}{d\hat{t}}
& =  &
\frac{G_F^2 \alpha_s^2}{512(2\pi)^3} \biggr\{
\Big| \left(\lambda_{3H}g^S_tD(\hat{s})+g^S_{tt}\right)F_\triangle^S
+(g^S_t)^2F_\Box^{SS}+(g^P_t)^2F_\Box^{PP} \Big|^2 
\nonumber \\[2mm]
&+&
\Big| (g^S_t)^2G_\Box^{SS}+(g^P_t)^2G_\Box^{PP} \Big|^2
\nonumber \\[2mm]
&+&
\Big| \left(\lambda_{3H}g^P_tD(\hat{s})+g^P_{tt}\right)F_\triangle^P
+g^S_tg^P_tF_\Box^{SP} \Big|^2  +
\Big| g^S_tg^P_tG_\Box^{SP} \Big|^2  \biggr \}\,. \label{eq4}
\end{eqnarray}
More explicitly in terms of each combination of couplings and 
ignoring the proportionality constant at the beginning of the equation,
the above equation becomes
\begin{eqnarray}
\frac{d\hat\sigma(gg\to HH)}{d\hat{t}}
&\propto&\lambda_{3H}^2|D(\hat{s})|^2\left[
|F_\triangle^S|^2(g_t^S)^2+|F_\triangle^P|^2(g_t^P)^2
\right]
\nonumber \\[2mm]
&+& 2 \lambda_{3H} g_t^S \real\left\{D(\hat{s})\left[
F_\triangle^S F_\Box^{SS*} (g_t^S)^2+
(F_\triangle^S F_\Box^{PP*}+F_\triangle^P F_\Box^{SP*}) (g_t^P)^2
\right]\right\}
\nonumber \\[2mm]
&+& 
\left[ |F_\Box^{SS}|^2+|G_\Box^{SS}|^2 \right] (g_t^S)^4
\nonumber \\[2mm]
&+& 
\left\{ 2\real\left[F_\Box^{SS}F_\Box^{PP*} +G_\Box^{SS}G_\Box^{PP*}\right]
+|F_\Box^{SP}|^2 + |G_\Box^{SP}|^2\right\} (g_t^S)^2 (g_t^P)^2
\nonumber \\[2mm]
&+& 
\left[|F_\Box^{PP}|^2+|G_\Box^{PP}|^2 \right] (g_t^P)^4
\nonumber \\[2mm] &+& 
|F_\Delta^S|^2 (g_{tt}^S)^2+ |F_\Delta^P|^2 (g_{tt}^P)^2
\nonumber \\[2mm] &+& 
2g_{tt}^S\left[
\lambda_{3H}g^S_t\real(D)|F_\Delta^S|^2+
(g_t^S)^2F_\Delta^S F_\Box^{SS*}+
(g_t^P)^2F_\Delta^S F_\Box^{PP*} \right]
\nonumber \\[2mm] &+& 
2g_{tt}^P\left[
\lambda_{3H}g^P_t\real(D)|F_\Delta^P|^2+
g_s^Sg_t^PF_\Delta^P F_\Box^{SP*} \right] \;,
\end{eqnarray}
where $F_\triangle^P=F^A_\triangle$, 
$F_\Box^{SP}=F_\Box$, and $G_\Box^{SP}=G_\Box$ 
with $F_\triangle^A$, $F_{\Box}$ and $G_\Box$
given in Appendix A.2  of Ref.~\cite{plehn} while
$F_\Box^{PP}=F_\Box$, and $G_\Box^{PP}=G_\Box$ 
with $F_{\Box}$ and $G_\Box$
in Appendix A.3 of Ref.~\cite{plehn}.
In the heavy quark limit, one may have~\cite{plehn}
\begin{eqnarray}
F_\triangle^S &=&+\frac{2}{3}+{\cal O}(\hat{s}/m_Q^2)\,, \ \ \
F_\Box^{SS}  = -\frac{2}{3}+{\cal O}(\hat{s}/m_Q^2)\,, \ \ \
F_\Box^{PP}  = +\frac{2}{3}+{\cal O}(\hat{s}/m_Q^2)\,, \nonumber \\[2mm]
F_\triangle^P &=&+1+{\cal O}(\hat{s}/m_Q^2)\,, \ \ \
F_\Box^{SP}  = -1+{\cal O}(\hat{s}/m_Q^2)\,, \nonumber \\[2mm]
G_\Box^{SS}  &=& G_\Box^{PP}  = G_\Box^{SP}  =  {\cal O}(\hat{s}/m_Q^2)
\end{eqnarray}
leading to
large cancellation between the triangle and box diagrams.

The production cross section normalized to the corresponding SM cross
section, with or without cuts, can be parameterized as follows:
\begin{eqnarray}
\frac{\sigma(gg\to HH)}{\sigma_{\rm SM}(gg\to HH)}&=&
\lambda_{3H}^2\left[c_1(s)(g^S_t)^2+d_1(s)(g^P_t)^2\right]+
\lambda_{3H}g^S_t\left[c_2(s)(g^S_t)^2+d_2(s)(g^P_t)^2\right]
\nonumber \\[2mm] && +
\left[c_3(s)(g^S_t)^4+d_3(s)(g^S_t)^2(g^P_t)^2+d_4(s)(g^P_t)^4\right]
\nonumber \\ &&
+\lambda_{3H}\left[e_1(s)g^S_t g^S_{tt}+f_1(s)g^P_t g^P_{tt}\right]
+g^S_{tt}\left[e_2(s)(g^S_t)^2+f_2(s)(g^P_t)^2\right]
\nonumber \\ &&
+\left[
 e_3(s)(g^S_{tt})^2
+f_3(s)g^S_tg^P_tg^P_{tt}
+f_4(s)(g^P_{tt})^2
\right] \label{cdef}
\end{eqnarray}
where the numerical coefficients $c_{1,2,3}(s)$, $d_{1,2,3,4}(s)$,
 $e_{1,2,3}(s)$, and $f_{1,2,3,4}(s)$
depend on $s$ and 
experimental selection cuts.  Upon our normalization,
the ratio
should be equal to $1$ when $g_t^S=\lambda_{3H}=1$ and
$g_t^P=g_{tt}^{S,P}=0$
or $c_1(s)+c_2(s)+c_3(s)=1$. The coefficients $c_1(s)$ and $c_3(s)$ are for
the SM contributions from the triangle and box diagrams, respectively, and
the coefficient $c_2(s)$ for the interference between them.

Once we have the coefficients $c_i, d_i, e_i$, and $f_i$'s,
the cross sections can be easily obtained for 
any combinations of couplings. Our first task is to obtain
the dependence of the coefficients 
on the collider energy $\sqrt{s}$, Higgs decay channels,
experimental cuts, etc. 

\section{Behavior of the cross sections}
\begin{figure}[ht!]
\centering
\includegraphics[width=4in]{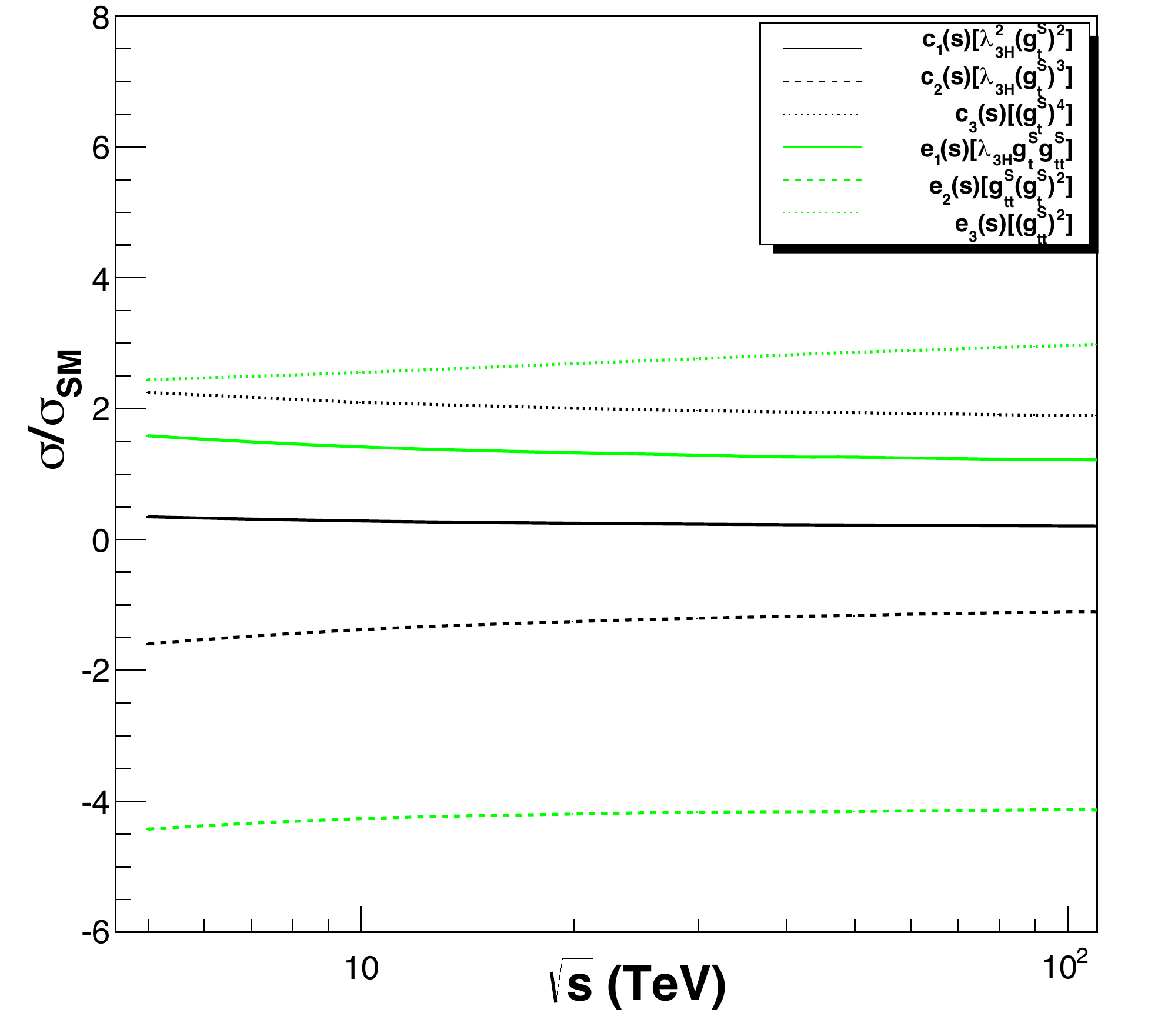}
\includegraphics[width=4in]{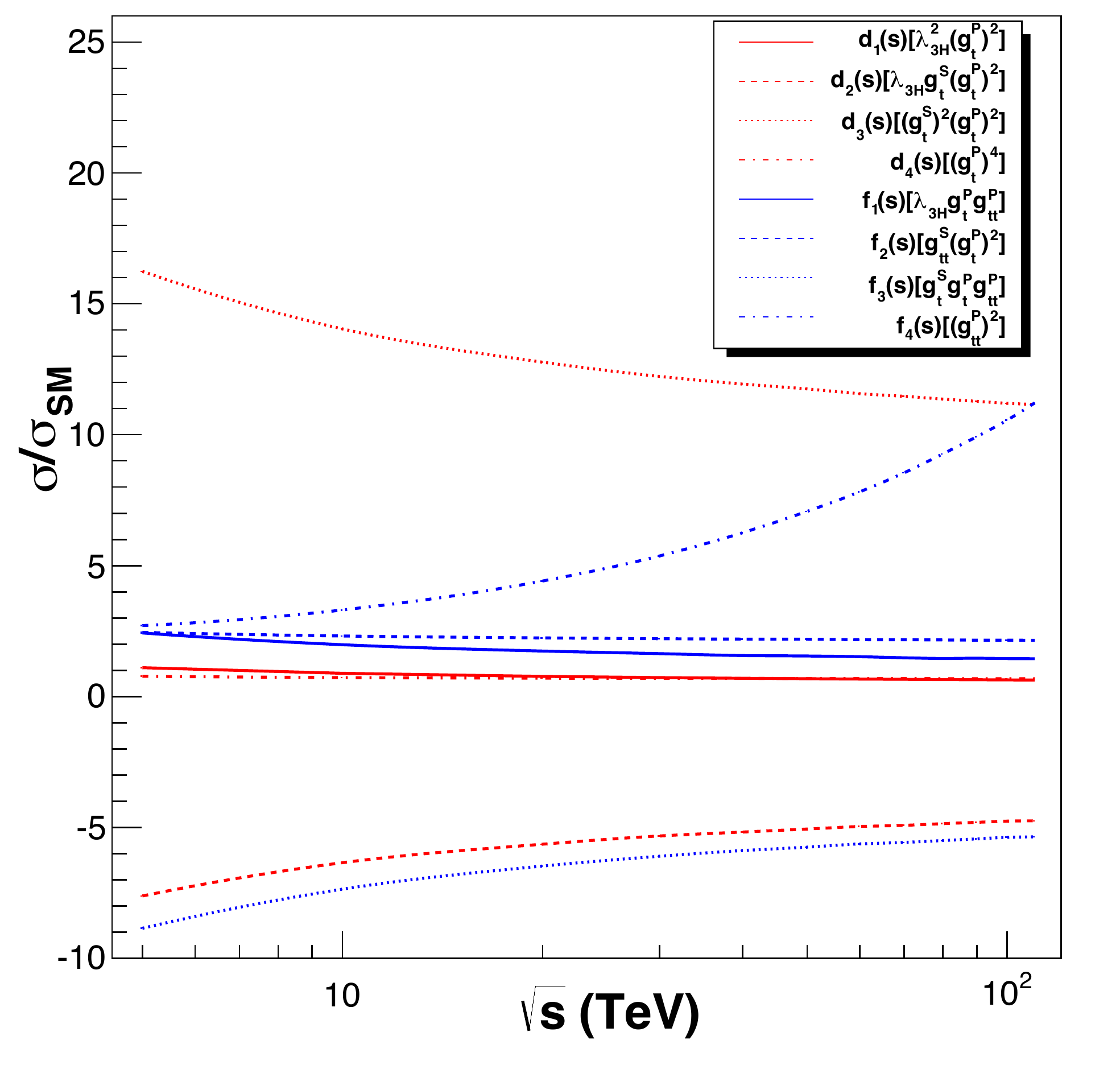}
\caption{\small \label{fig:coef} 
The coeffcients $c_i(s)$ and $e_i(s)$ (upper) with $i=1,2,3$ and
$d_j(s)$ and $f_j(s)$ (lower) with $j=1,2,3,4$ as functions of $s$.
}
\end{figure}

We first examine the behavior of each piece of cross sections versus 
energies. We show the coefficients $c_i, d_i, e_i, f_i$'s 
at $\sqrt{s} = 8,14,33,100$ TeV in Table~\ref{tab:coef}
and also in Fig.~\ref{fig:coef}
\footnote{We note our results are in well accord with those in 
literature when
the comparison is possible. The values of $c_{1,2,3}$ and $e_{1,2,3}$
at $\sqrt{s}=100$ TeV, for example, are in good agreement with those presented
in Ref.~\cite{low}.}.
The triangle diagram involves an $s$-channel Higgs-boson propagator.
At center-of-mass energy $\sqrt{\hat s}$ considerably higher than $M_H$,
it behaves like $1/\hat s$. Thus the triangle diagram decreases more 
rapidly than the box and contact  diagrams, as reflected from the
coefficients $c_1$ and $d_1$ when $\sqrt{s}$ goes from 8 TeV to 100 TeV.
The coefficients $c_3$, $d_3$ and $d_4$ associated with the box diagram
more or less remain the same as the energy goes up.
On the other hand, the contact diagram involving the coefficients 
$e_3$ and $f_4$ increases with energy, in particular tremendous increases
in $f_4$. It is easy to see that the contact diagram is dim-5 and 
obviously grows with energy.  At certain high enough energy, 
it may upset unitarity.

We can examine the validity of the anomalous $ttHH$ contact coupling
by projecting out the leading partial-wave coefficient for the scattering
$t \bar t \to H H$. At high energy, the amplitude 
\[
i {\cal M}(t \bar t \to HH) \sim g^S_{tt} \frac{m_t \sqrt{\hat s}}{v^2} \;.
\]
The leading partial-wave coefficient is given by 
\[
 a_0 = \frac{1}{64\pi} \int_{-1}^1 d(\cos\theta) P_0 (\cos\theta) (i {\cal M})
     = g^S_{tt}  \frac{m_t \sqrt{\hat s}}{32 \pi v^2}\;.
\]
Requiring $|a_0|< 1/2$ for unitarity we obtain 
\[
   \sqrt{\hat s} \le \frac{17.6}{ g_{tt}^S } \; {\rm TeV}  \;.
\]
Therefore, the anomalous $ttHH$ contact term can be safely applied
at the LHC
for $g_{tt}^S \alt 3-5$ as most of the
collisions occur at $\sqrt{\hat s} \alt $ a few TeV.

\begin{table}[t!]
\caption{\small
\label{tab:coef}
The behavior of the coefficients $c_i(s)$, $d_i(s)$,
$e_i(s)$, and $f_i(s)$ versus collider energy $\sqrt{s}$, 
see Eq.~(\ref{cdef}).
}
\vspace{2.0mm}
\begin{tabular}{|c|c|c|c||c|c|c|c|}
\hline
$\sqrt{s}$ (TeV) & $c_1(s)$ & $c_2(s)$ & $c_3(s)$ &
$d_1(s)$ & $d_2(s)$ & $d_3(s)$ & $d_4(s)$\\
$$  & $\left[\lambda_{3H}^2(g_t^S)^2\right]$ &
$\left[\lambda_{3H}(g_t^S)^3\right]$ &
$\left[(g_t^S)^4\right]$ &
$\left[\lambda_{3H}^2(g_t^P)^2\right]$ &
$\left[\lambda_{3H}g_t^S(g_t^P)^2\right]$ &
$\left[(g_t^S)^2(g_t^P)^2\right]$ &
$\left[(g_t^P)^4\right]$\\ \hline
$8$ & $0.300$ & $-1.439$ & $2.139$ &
$0.942$ & $-6.699$ & $14.644$ & $0.733$\\ \hline
$14$ & $0.263$ & $-1.310$ & $2.047$ &
$0.820$ & $-5.961$ & $13.348$ & $0.707$\\ \hline
$33$ & $0.232$ & $-1.193$ & $1.961$ &
$0.713$ & $-5.274$ & $12.126$ & $0.690$\\ \hline
$100$ & $0.208$ & $-1.108$ & $1.900$ &
$0.635$ & $-4.789$ & $11.225$ & $0.683$\\ \hline
\end{tabular}
\\
\vspace{2.0mm}
\begin{tabular}{|c|c|c|c||c|c|c|r|}
\hline
$\sqrt{s}$ (TeV) & $e_1(s)$ & $e_2(s)$ & $e_3(s)$ &
$f_1(s)$ & $f_2(s)$ & $f_3(s)$ & $f_4(s)$\\
$$  & $\left[\lambda_{3H}g_t^S g_{tt}^S\right]$ &
$\left[g_{tt}^S (g_t^S)^2\right]$ &
$\left[(g_{tt}^S)^2\right]$ &
$\left[\lambda_{3H}g_t^P g_{tt}^P\right]$ &
$\left[g_{tt}^S (g_t^P)^2\right]$ &
$\left[g_t^S g_t^P g_{tt}^P\right]$ &
$\left[(g_{tt}^P)^2\right]$\\ \hline
$8$ & $1.460$ & $-4.313$ & $2.519$ &
$2.104$ & $2.350$ & $-7.761$ & $3.065$\\ \hline
$14$ & $1.364$ & $-4.224$ & $2.617$ &
$1.848$ & $2.269$ & $-6.886$ & $3.769$\\ \hline
$33$ & $1.281$ & $-4.165$ & $2.783$ &
$1.622$ & $2.207$ & $-6.033$ & $5.635$\\ \hline
$100$ & $1.214$ & $-4.137$ & $2.974$ &
$1.474$ & $2.154$ & $-5.342$ & $10.568$\\ \hline
\end{tabular}
\end{table}

To some extent we have understood the behavior of the triangle, box,
and contact diagrams with the center-of-mass energy, which is 
kinematically equal to the invariant mass  $M_{HH}$ of the Higgs-boson pair.
One can then uses $M_{HH}$ to enhance or reduce the relative contributions
 of triangle or box diagrams. The higher the $M_{HH}$ the relatively larger
proportion comes from the box and contact diagrams.  
Since $M_{HH}$ correlates with
the boost energy of each Higgs boson, a more energetic Higgs boson
will decay into a pair of particles, which have a smaller angular separation
between them than a less energetic Higgs boson. Therefore, the angular
separation $\Delta R_{ij}$ between the decay products $i,j$ is another
useful kinematic variable to separate the contributions among the triangle,
box, and contact diagrams. 

\section{Numerical Analysis}

The Lagrangian in Eq.~(\ref{lag}) consists of five parameters:
the scalar and pseudoscalar parts of the top-Yukawa coupling $g_t^{S,P}$,
the scalar and pseudoscalar parts of the anomalous contact coupling 
$g_{tt}^{S,P}$, and the Higgs trilinear coupling $\lambda_{3H}$. 
In order to facilitate the presentation and understanding of the physics,
we study a few scenarios:
\begin{enumerate}
\item {\bf CPC1}--the top-Yukawa 
coupling involves only the scalar part and the scale in the
anomalous contact coupling is very large -- only $g_t^S$ and $\lambda_{3H}$
are relevant.
The relevant coefficients are $c_1$, $c_2$, and $c_3$.

\item {\bf CPC2}--the top-Yukawa and the anomalous contact couplings involve 
only the scalar part -- $g_t^S$, $g_{tt}^S$, and $\lambda_{3H}$ are relevant.
The relevant coefficients are $c_1$, $c_2$, $c_3$,
$e_1$, $e_2$, and $e_3$.

\item {\bf CPV1}--the top-Yukawa coupling involves both the scalar and
pseudoscalar parts -- $g_t^S$, $g_t^P$, and $\lambda_{3H}$ are relevant.
The relevant coefficients are $c_1$, $c_2$, $c_3$,
$d_1$, $d_2$, $d_3$, and $d_4$.

\item {\bf CPV2}--the contact $ttHH$ coupling involves both the
scalar and pseudoscalar parts -- $g_{tt}^S$, $g_{tt}^P$, and $\lambda_{3H}$ 
are relevant while the top-Yukawa coupling is kept at fixed values.
In this case, all the coefficients become relevant.
In one of the simplest cases with $g^S_t=1$ and $g^P_t=0$, for example,
the relevant coefficients are 
$c_1$, $c_2$, $c_3$, $e_1$, $e_2$, $e_3$, and $f_4$.

\end{enumerate}

Note that the above scenarios have been studied using 
different approaches and separately in literature:
{\bf CPC1} in Ref.~\cite{1301.3492}, {\bf CPC2} in Ref.~\cite{contino},
and {\bf CPV1} in Ref.~\cite{cpodd}.
The scenario {\bf CPV2} is new with the CP-odd component of the $ttHH$,
in which the CP-odd component $g_{tt}^P$ of the $ttHH$ coupling
appears either with $g_t^P$ or in square.

  We used the CTEQ6L1 \cite{cteq} with both the renormalization 
and factorization scales $\mu= M_H$ for the parton distribution 
functions. Since we focus on the ratios relative to the SM predictions, 
we anticipate the uncertainties due to scale dependence, choice of 
parton distribution functions, and experimental acceptance are reduced 
to a minimal level
\footnote{
We have observed that the absolute vaues of the production cross sections
decrease by about 20 \% if we take $\mu=M_{HH}$, well within the 
theoretical uncertainty estimated in Ref.~\protect\cite{baglio}.}. 
 For the branching ratios of the Higgs boson we employ the values for
the SM Higgs boson listed in the LHC Higgs Cross Section Working Group
\cite{group}. 
Here we ignore the slight variation in the diphoton
branching ratio due to the change in the top-Yukawa coupling. 
\footnote
{
According to our most recent analysis using all the Run I data 
\cite{update2014}, the allowed range of the top-Yukawa coupling is, 
at 95\% C.L., $[-1.1, -0.8] \cup [0.7,1.4]$ 
in the scenario where gauge-Higgs coupling and top-, bottom-, and 
tau-Yukawa couplings are varied; and additionally if we 
included non-SM contributions to $Hgg$ and $H\gamma\gamma$ vertices,
the range becomes $[-2,2]$.
The values of top-Yukawa coupling used in this
study are in the allowed ranges such that the diphoton signal
strength is kept within	the uncertainty of the 
experimental data, for example, $\mu_{ggH} = 1.32 \pm 0.38$ (ATLAS)
\cite{atlas-pp}
and $\mu_{ggH} = 1.12 \,^{+0.37}_{-0.32}$ (CMS) \cite{cms-pp}
for $gg \to H\to \gamma\gamma$.
}
This is because we do not want to interfere with the more important goal of
the work -- interference effects between the triangle and box diagrams
and the sensitivity to the trilinear Higgs coupling.
For Higgs-pair production we used a modified MADGRAPH implementation
\cite{loop,website}, which allows us to vary the top-Yukawa, Higgs
trilinear, and the contact $ttHH$  couplings.

  In this work, we make a working assumption that the $b \bar b\gamma\gamma$
background can be estimated with a reasonable accuracy and can be
extracted from the experimental data. Once the background is subtracted
from the data, we are left with the signal events.  In order to do so,
we impose a set of basic cuts for the acceptance of the $b$ quarks
and photons, and use $B$-tagging according to the ATLAS template in
Delphes v.3 \cite{delphes3}, 
as well as an invariant mass window on the $b\bar b$ pair and
the photon pair around the Higgs boson mass. 
The basic cuts are
\begin{eqnarray}
&&  p_{T_\gamma} > 10 \, {\rm GeV}, \;\;\; 
 p_{T_b} > 10 \, {\rm GeV}, \;\;\; 
|\eta_\gamma| < 2.5 ,\;\;\;
|\eta_b| < 2.5 ,\;\;\;
\Delta R_{\gamma\gamma} > 0.4,\;\;\;
\Delta R_{bb} > 0.4, \nonumber \\
&& 
 |M_H - M_{\gamma\gamma} |< 15 \;{\rm GeV} ,\;\;\;
 |M_H - M_{bb}| < 25 \;{\rm GeV} ,\;\;\;
 |M_H - M_{\tau\tau} |< 25 \;{\rm GeV} .\;\;\; \label{basic}
\end{eqnarray}
Note that in the appendix we shall discuss the feasibility of using
the $\tau^+\tau^- b\bar b$ mode.  
With this set of basic cuts we continue to study the signal events
in various kinematical regions separated by $\Delta R_{\gamma\gamma}$ and
$\Delta R_{bb}$.

\subsection{CPC1: $g_t^S$ and $\lambda_{3H}$}

This is the simplest scenario to investigate the variation of the
triangle and box diagrams with respect to changes in the Yukawa
and trilinear couplings. The corresponding coefficients in Eq.~(\ref{cdef})
are $c_1, c_2, c_3$. We let the Higgs boson pair decay into
\[
 H H \to (\gamma\gamma) \;(b\bar b) \;.
\]
Note that if the process is studied without detector simulation, the
distributions for other decay channels, like 
$b\bar b \tau\tau$ or
$\gamma\gamma \tau\tau$ would be the same.  Nevertheless, the 
resolutions for $b$, $\tau$, and $\gamma$ are quite different 
in detectors, and so are 
the backgrounds considered for each decay channel. In the following,
we focus on the $\gamma\gamma b \bar b$ channel, which
has been considered in a number of works.

We use MADGRAPH v.5 \cite{madgraph} with parton showering by Pythia v.6
\cite{pythia}, detector simulations using Delphes v.3\cite{delphes3}, and 
the analysis tools by MadAnalysis 5 \cite{madana}. 
We have verified that the coefficients $c_{1,2,3}$ using the default,
ATLAS, and CMS templates inside Delphes v.3 are within 10\% of one
another. From now on we employ the ATLAS template in detector simulations.
We show the coefficients in Table~\ref{tab:all}. We will 
come back to this table a bit later.

\begin{figure}[t!]
\centering
\includegraphics[width=5.2in]{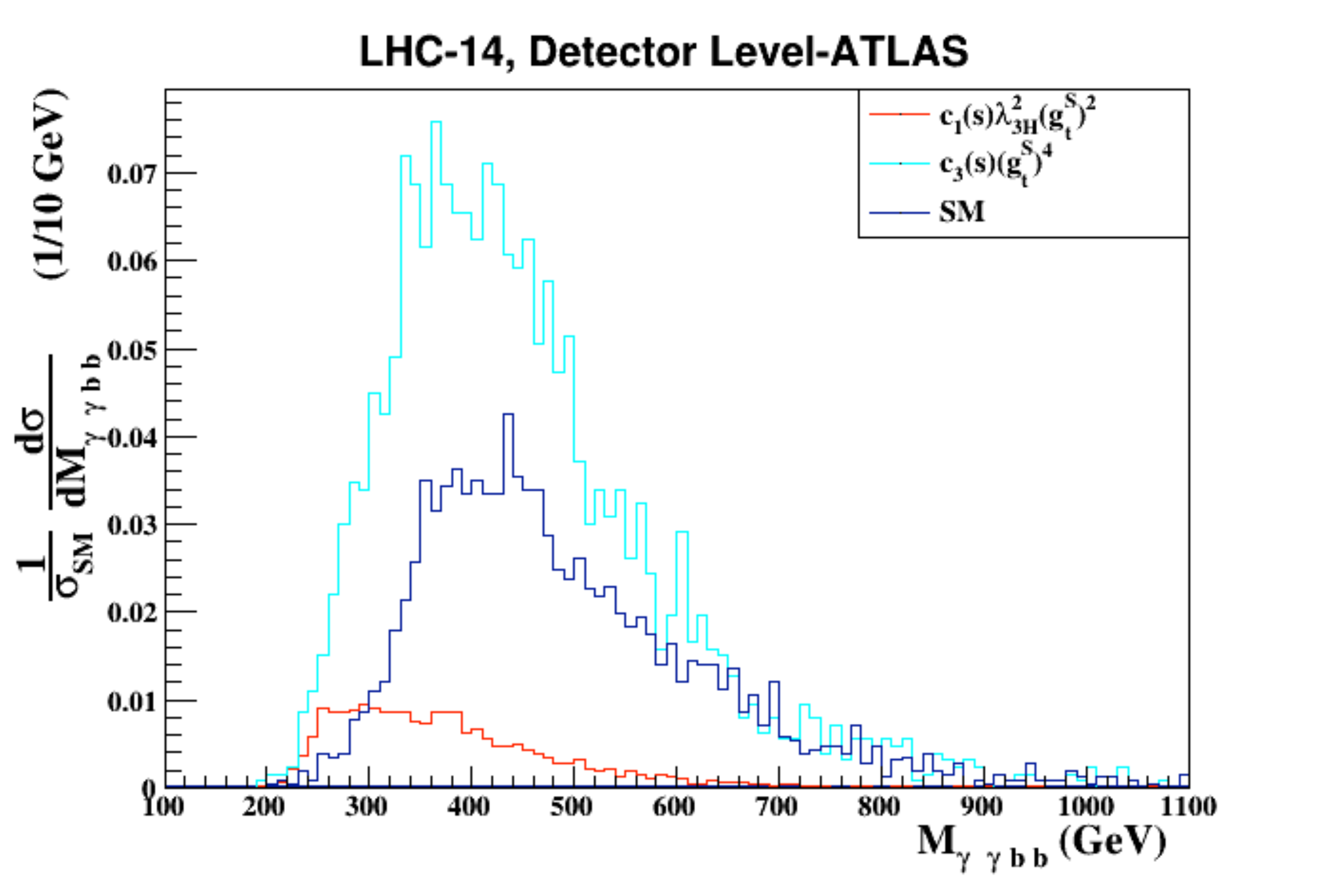}
\includegraphics[width=5.2in]{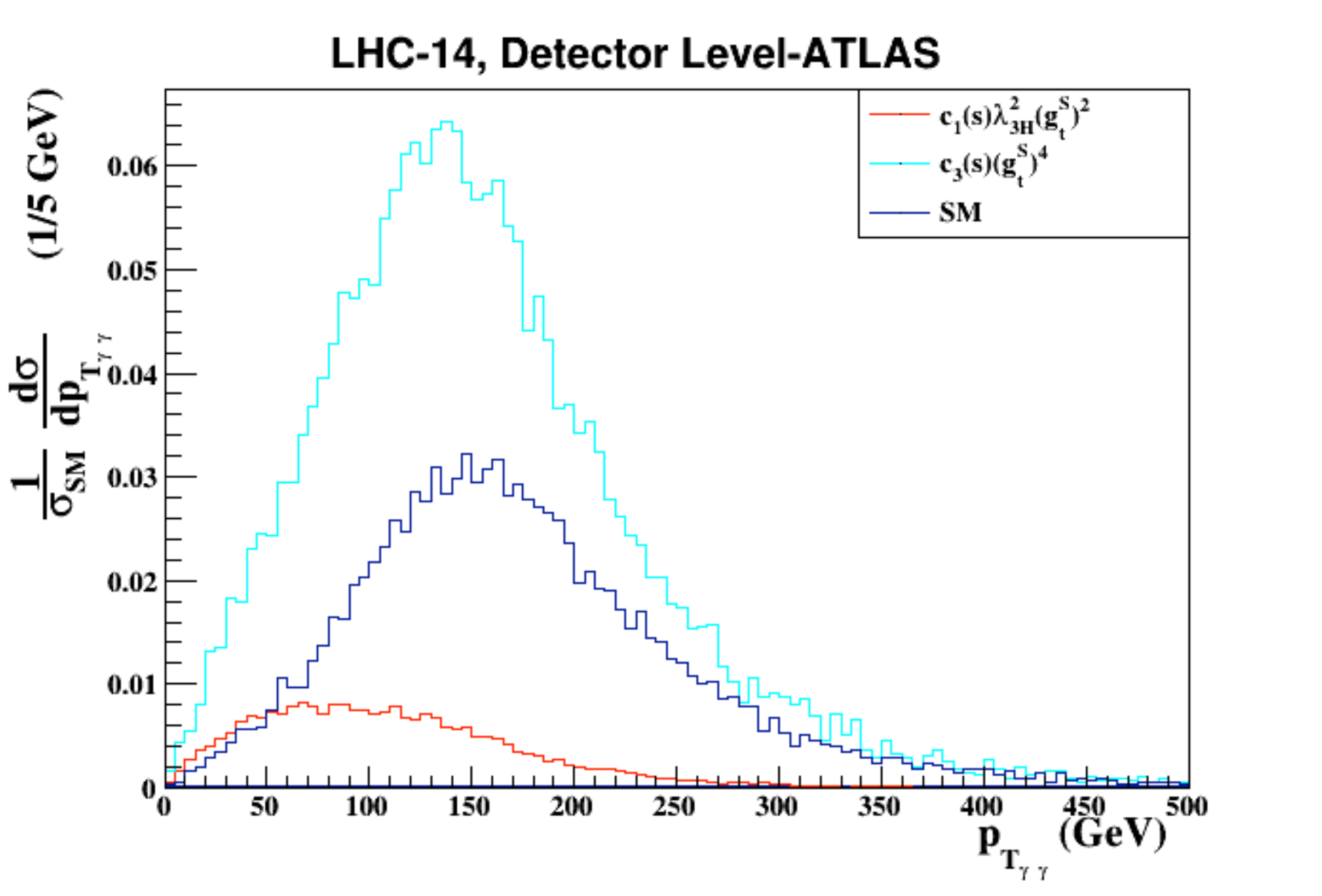}
\caption{\small \label{cpc1-dist-m}
{\bf CPC1:} Distributions in $M_{\gamma\gamma b b}$ (upper) and
$p_{T_{\gamma\gamma}}$ (lower) in the decay products of the Higgs boson pair
with detector simulation.}
\end{figure}

We first look at the distribution of events versus the invariant
mass of the Higgs-boson pair via $H H \to \gamma\gamma b \bar b$
in Fig.~\ref{cpc1-dist-m} (upper panel) and versus the transverse
momentum of the photon pair $p_{T_{\gamma\gamma}}$ (lower panel). 
In principle, up to detector simulations and higher-order emissions,
the transverse momentum distribution of the photon pair is the same as 
that of the $b\bar b$ pair.
The triangle diagram (red line) peaks at the lower invariant 
mass and decreases with $M_{\gamma\gamma b \bar b}$, 
because of the $s$-channel Higgs propagator.  
The box diagram (skyblue line), 
on the other hand, is larger than the triangle
diagram at high invariant mass. 
The (darkblue) SM line represents the whole contributions including
the destructive interference between the triangle and box diagrams.
Similar behavior can be seen in the $p_T$ 
distribution of the photon pair.

\begin{figure}[t!]
\centering
\includegraphics[width=5.2in]{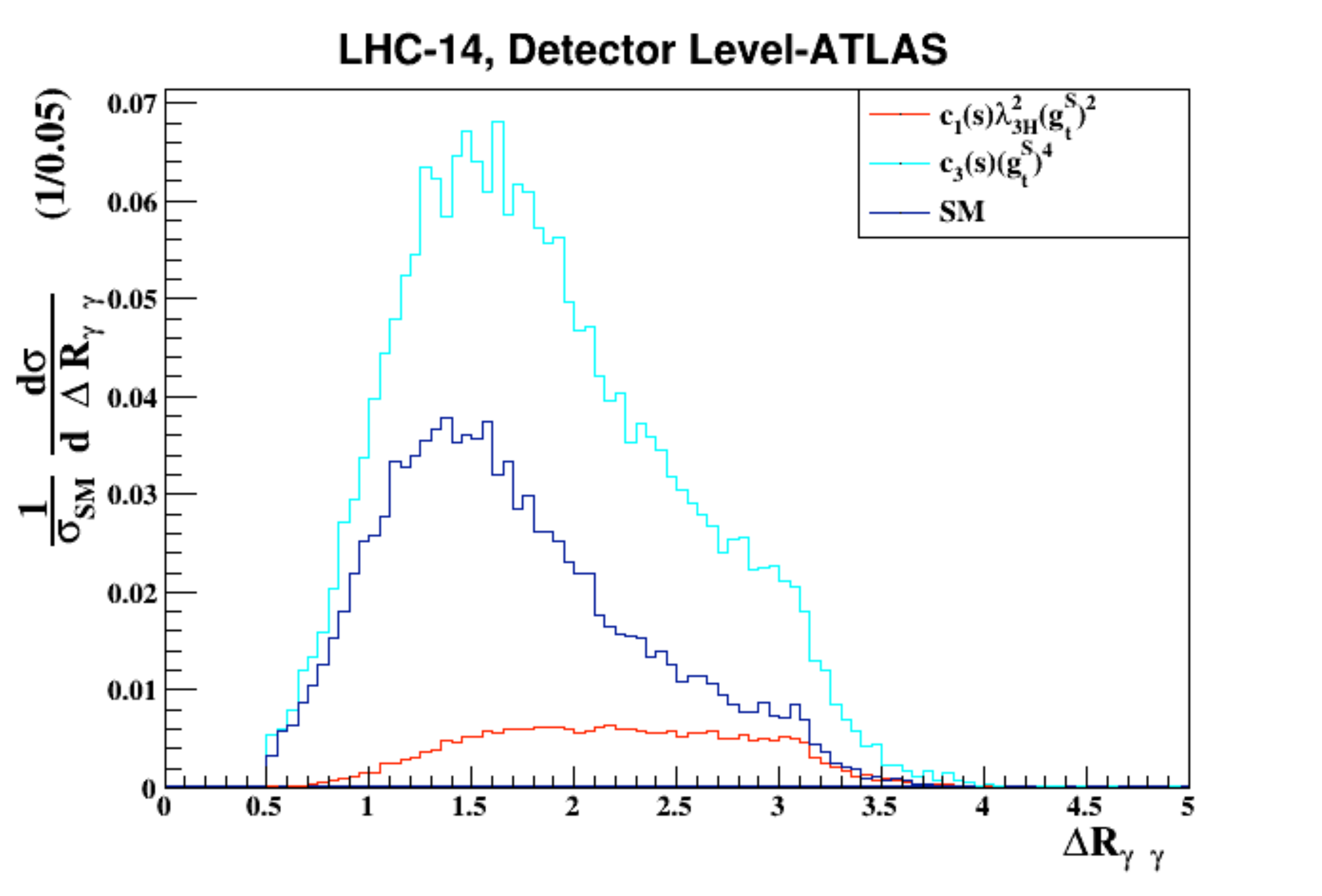}
\includegraphics[width=5.2in]{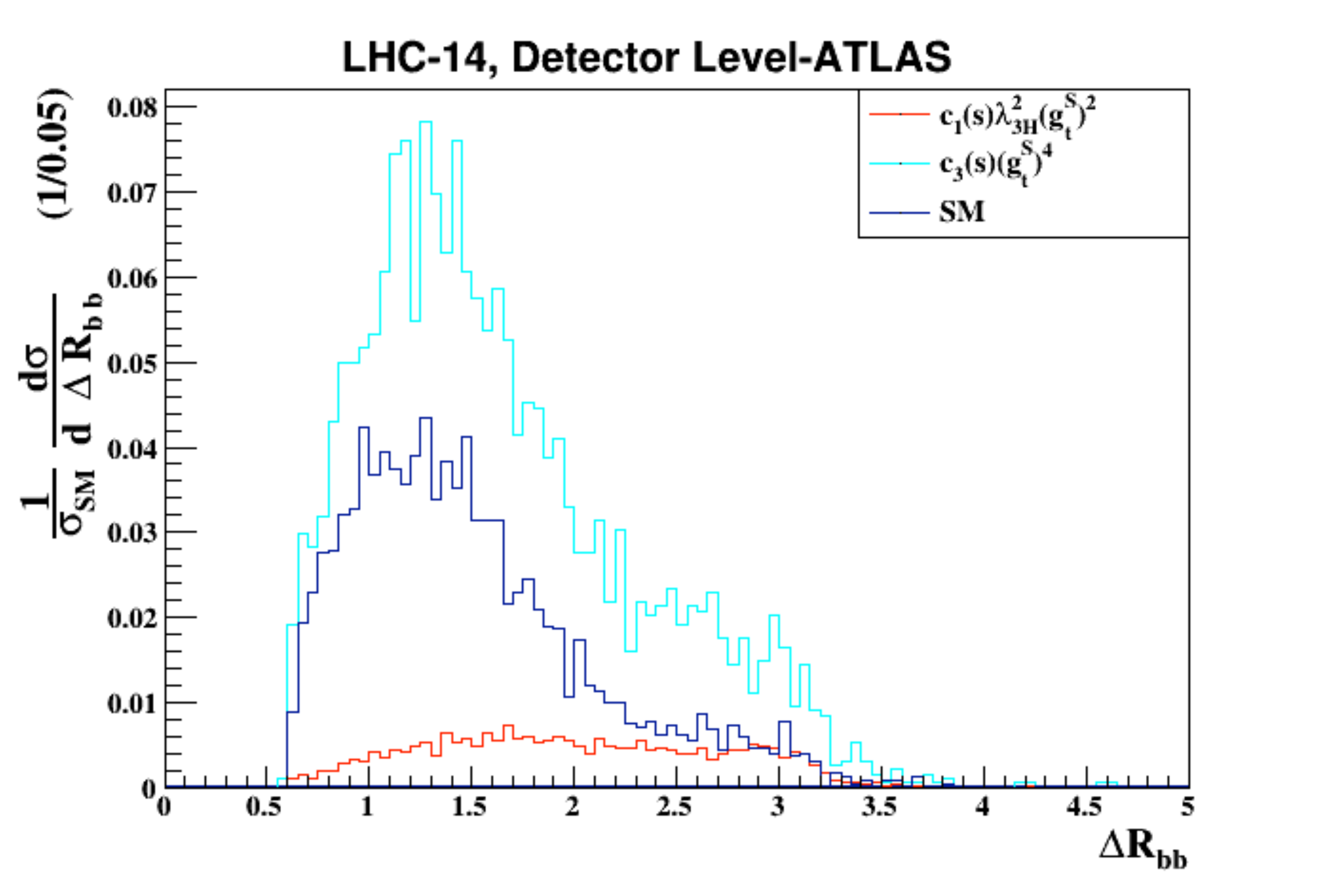}
\caption{\small \label{cpc1-dist}
{\bf CPC1:} Distributions in $\Delta R_{\gamma\gamma}$ (upper) and
$\Delta R_{b b}$ (lower) in the decay products of the Higgs boson pair
with detector simulation.}
\end{figure}

Next, we look at the distributions versus 
$\Delta R_{\gamma\gamma}$ and $\Delta R_{b b}$ in Fig.~\ref{cpc1-dist}.
The lines are the same as in Fig.~\ref{cpc1-dist-m}.
As we have explained in the previous section, $\Delta R_{\gamma\gamma}$
and $\Delta R_{b b}$ between the decay products of each Higgs boson are
useful variables to separate the triangle and box contributions. 
The angular distribution $\Delta R$ between the two decay products of
each Higgs boson correlates with the energy of the Higgs boson, which
in turns correlates with the invariant of the Higgs-boson pair. The
higher the invariant mass, the more energetic the Higgs boson will be,
and the smaller the angular separation between the decay products will
be.  Therefore, the triangle diagram has wider separation than 
the box diagram.

It is clear that 
the distributions of $\Delta R_{\gamma\gamma}$ and $\Delta R_{b b}$
have similar behavior within uncertainties. The box diagram and also the
SM, which is dominated by the box contribution, have a peak at 
$\Delta R_{\gamma\gamma}$ or $\Delta R_{b b}$ less than $2.0$, 
while the triangle diagram prefers to have the majority at larger
$\Delta R_{\gamma\gamma}$ or $\Delta R_{b b}$, say between $2$ and $3$.
We therefore come up with (i) $\Delta R_{\gamma\gamma} > 2$ ($<2$),
(ii) $\Delta R_{b b} > 2$ ($<2$), and
(iii) $\Delta R_{\gamma\gamma} > 2$ and $\Delta R_{b b} > 2$ (both $<2$)
to enrich the sample of triangle (box) contribution.
In the following, we use $\Delta R$ to denote either $\Delta R_{\gamma\gamma}$ 
or $\Delta R_{b b}$, unless stated distinctively.

We can now look at Table~\ref{tab:all}, where the 
coefficients for the ratio 
of the cross sections $\sigma(gg \to HH)/\sigma_{\rm SM}(gg \to
HH)$ as in Eq.~(\ref{cdef}) are shown. In the {\bf CPC1} case,
the relevant coefficients are $c_1$, $c_2$, and $c_3$
in which $c_1$ is induced by the triangle diagram, $c_3$ by the
box diagram, and $c_2$ by the interference between them.
The rows labeled ``Basic Cuts'' are the ratio of cross sections
under the set of cuts in Eq.~(\ref{basic}).
In the same Table, we also show the coefficients obtained
after applying the angular-separation cuts of
$\Delta R > 2$ or $<2$ 
and both $\Delta R_{\gamma\gamma, b b} > 2$ or $<2$.
It is clear that $\Delta R_{\gamma\gamma} > 2$ ($<2$) enriches the
triangle-diagram (box-diagram) contribution. Similar is true for
$\Delta R_{b b } > 2$ ($<2$). Further enhancement of triangle diagram
can be obtained with both $\Delta R_{\gamma\gamma} > 2$ and
$\Delta R_{b b } > 2$, and vice versa for box diagram.

In the following, we investigate the sensitivity in the parameter space
$(g_t^S,\; \lambda_{3H})$ that one can reach at the 14 TeV with 
3000 fb$^{-1}$
luminosity by using the measurements of cross sections in various
kinematical regions.  Since we have found that the triangle and box 
contributions can be distinguished using the $\Delta R$ cuts, we make use of the measured 
cross sections in the kinematical regions separated by these cuts.

There are two issues that we have to considered when we take the measured
cross sections in the kinematical regions. First, the SM backgrounds
for the decay channel that we consider, and second, the 
Next-to-Leading-Order (NLO) corrections \cite{hh-nlo}. 
It was shown in Ref.~\cite{hh-nlo} that the
NLO and NNLO corrections can be as large as 100\% with uncertainty of order
10--20\%. The SM backgrounds, on the other hand, can be estimated with
uncertainties less than the NNLO corrections.  
We therefore adopt an 
approach that the signal cross sections (after background subtraction) 
are measured with uncertainties of order 25--50\%.

About the signal cross sections,
the first and the second columns of Table~\ref{tab:cx} show
the SM cross sections for the process
$pp \to HH \to \gamma\gamma b\bar b$ with 
detector simulations under various cuts at the 14 TeV LHC
\footnote{Before applying the basic cuts, we find
the cross sections are $8.92\times 10^{-2}$ and $2.41$ in fb 
for SM-14($\gamma\gamma b \bar b$) and
SM-14($\tau^+\tau^- b \bar b$), respectively, which agree very well with
those in Ref.~\cite{baglio}. While our cross sections after applying
the basic cuts are smaller than those presented in Ref.~\cite{baglio}
by a factor of $\sim 3$ for $\gamma\gamma b \bar b$ and
a factor of $\sim 30$ for $\tau^+\tau^- b \bar b$. 
This is basically because we have implemented full
detector simulation to reconstruct $b$ quarks, photons, and $\tau$ leptons
from Higgs 
and partly due to different experimental cuts applied and 
different $b$- and $\tau$-tagging efficiencies taken.
One may need to optimize the cuts to increase the signal to background ratio
but it is beyond the scope of this paper and we will pursue this issue 
later in our future publication.
Incidentally the SM-100($\gamma\gamma b \bar b$) cross section
is $3.73$ fb before applying the basic cuts.
}.  
We have taken account of the SM NLO cross section
$\sigma_{\rm SM}(pp\to HH) \simeq 34$ fb, the Higgs branching fractions,
and both the photon and $b$-quark reconstruction efficiencies
with angular-separation cuts of $\Delta R_{bb\,,\gamma\gamma}$.
In various kinematical regions depending on 
the angular cuts, the cross sections range from
$\sim 0.001$ fb  to $\sim 0.01$ fb.  
With an integrated luminosity of 3000 fb$^{-1}$,
we expect of order $30$ signal events 
when the cross section is 0.01 fb.
An estimate
of the statistical error is given by the square root of the number
of events $\sqrt{N}$, which is then roughly $20\%$
of the total number.
Taking into account the uncertainty of order $10-20\%$ from
NLO and NNLO corrections, in this work,
we use a total uncertainty of $25-50\%$ in
the signal cross section in the estimation of sensitivity of the 
couplings.
 Our approach is more or less valid except for the case in 
which both the $\Delta R_{b b}>2$ and $\Delta R_{\gamma\gamma}>2$ cuts
are imposed simultaneously. It would be challenging to measure this 
size  of cross section only in the $HH\to \gamma\gamma b \bar b$ mode
and one may need to combine the measurements in 
different Higgs-decay channels.
Or, one may rely on the future colliders 
such as a 100 TeV $pp$ machine with larger cross sections
and/or higher luminosities.

\begin{figure}[th!]
\centering
\includegraphics[width=3.2in]{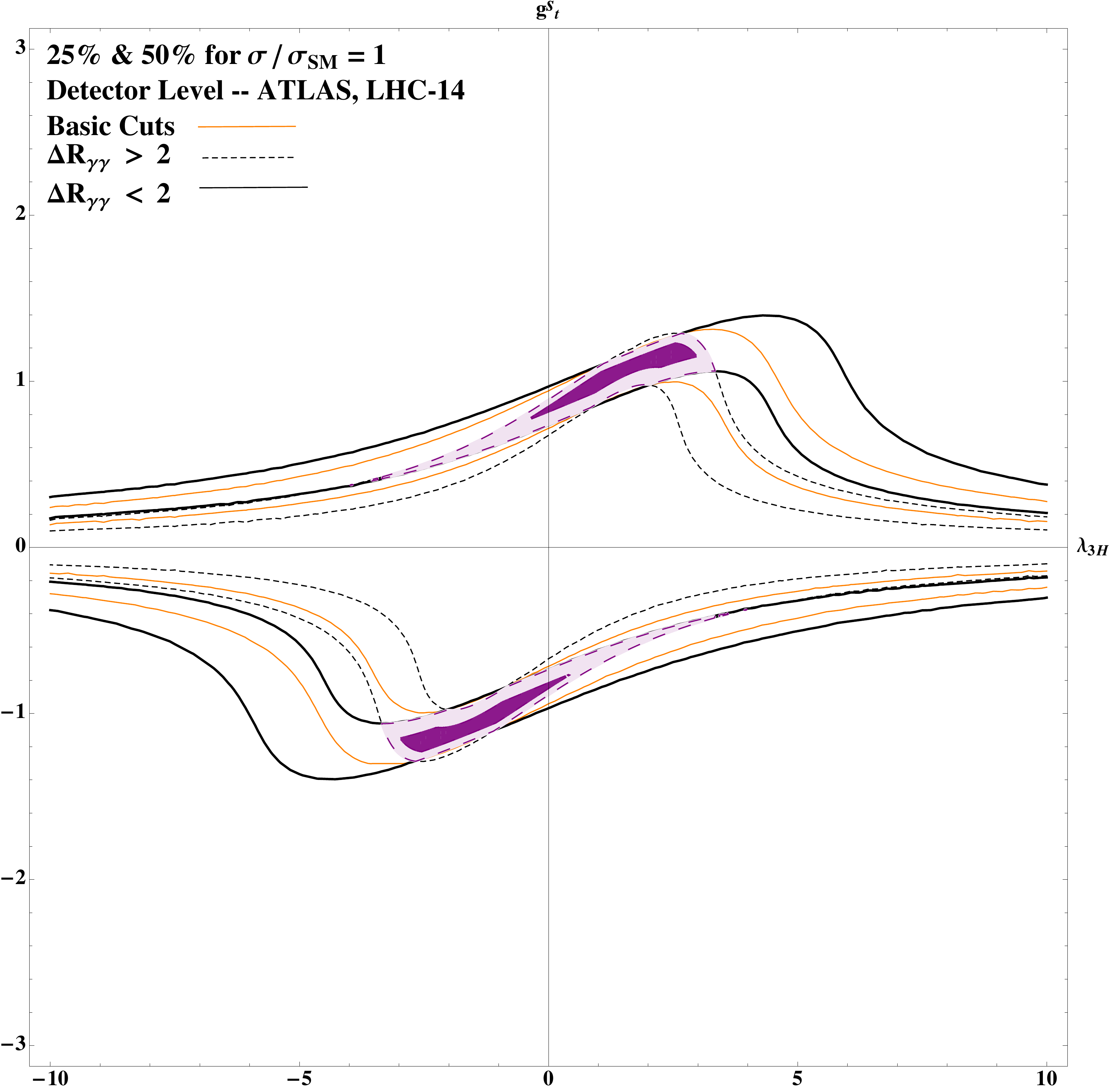}
\includegraphics[width=3.2in]{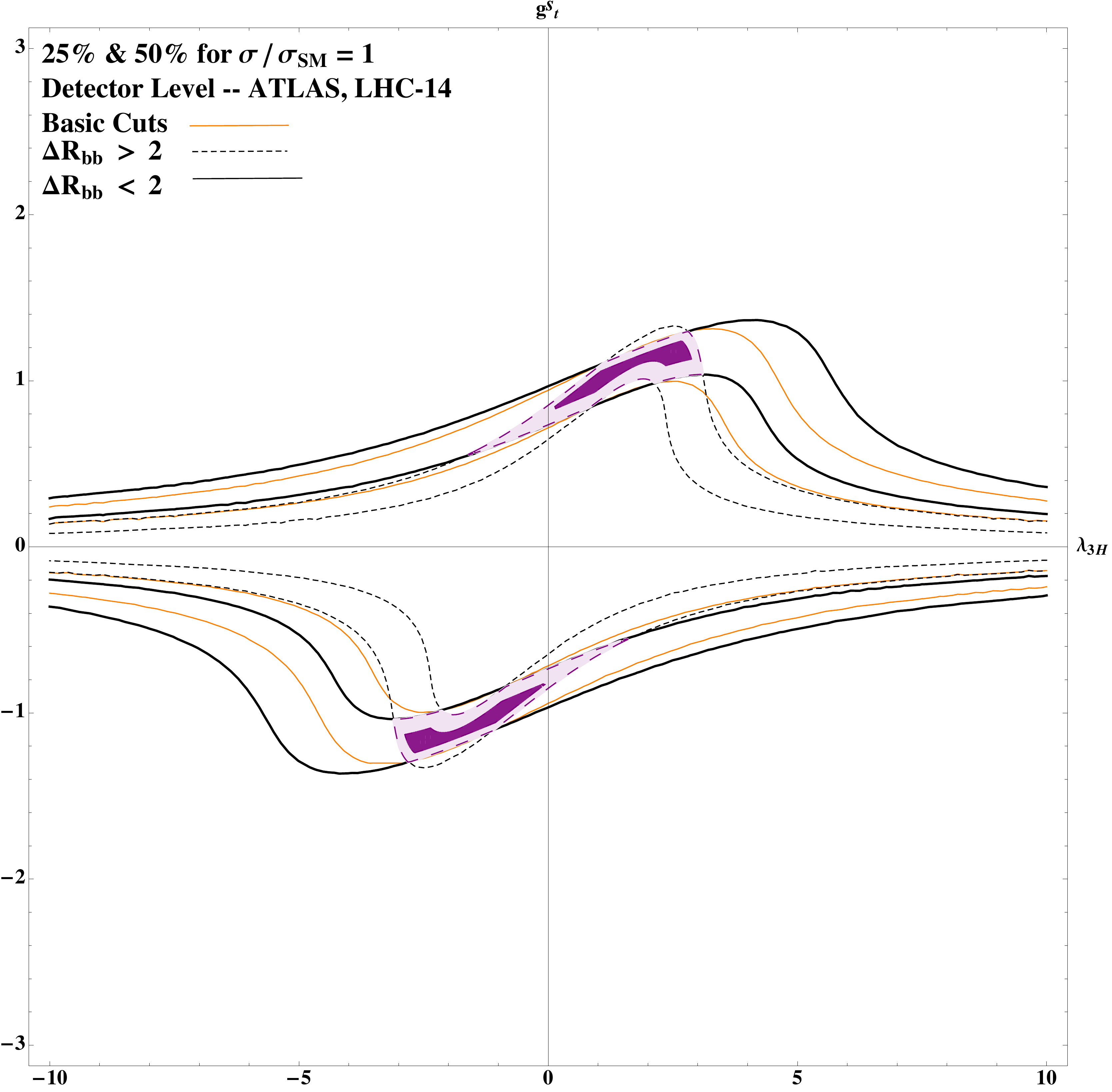}
\includegraphics[width=3.2in]{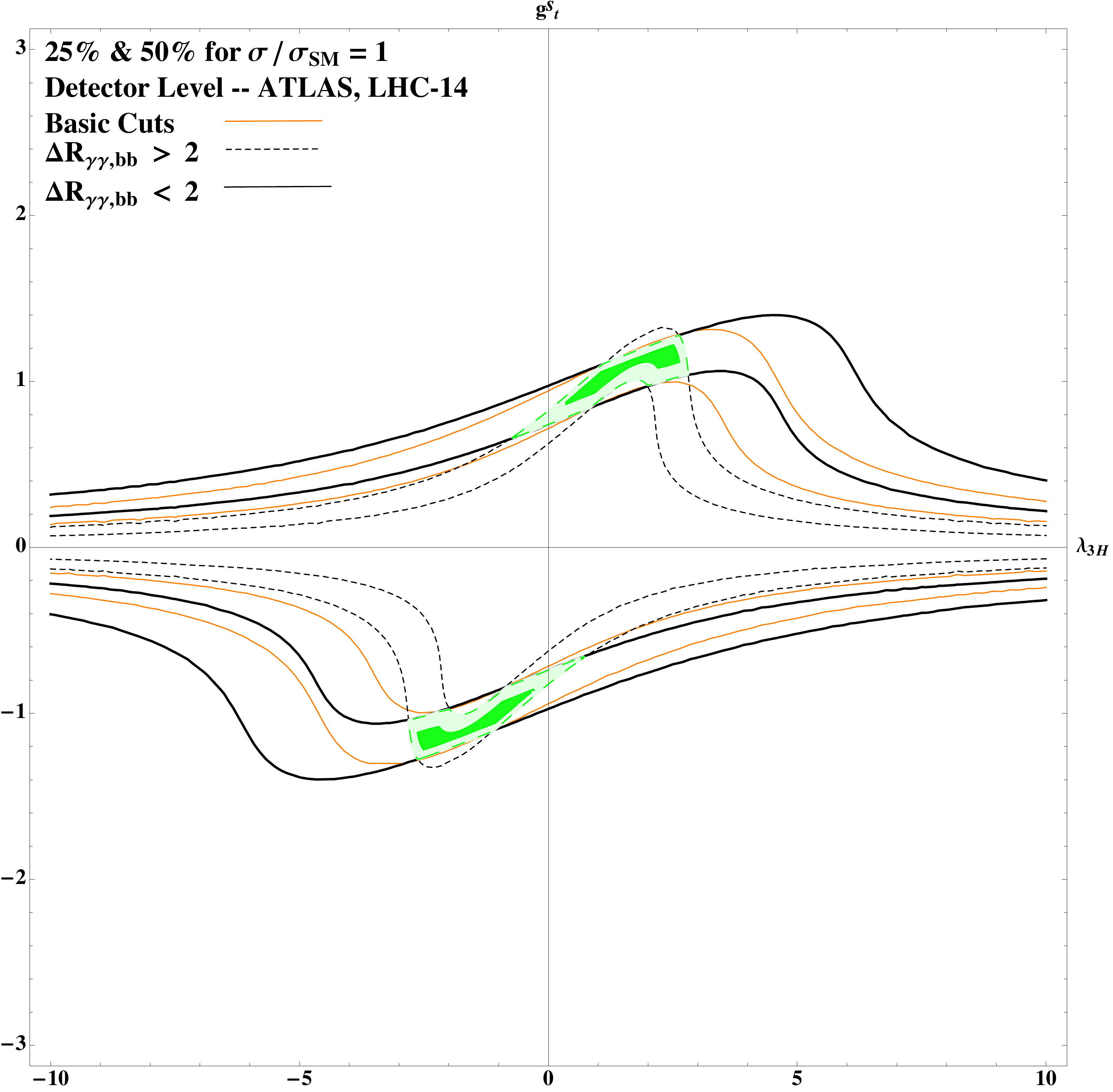}
\caption{\small \label{fig:cpc1}
{\bf CPC1:} The 25\% and 50\% sensitivity regions bounded by three
measurements of cross sections with basic cuts, 
$\Delta R_{\gamma\gamma} > 2$, and 
$\Delta R_{\gamma\gamma} < 2$ (the upper-left panel);
with basic cuts, 
$\Delta R_{b b} > 2$, and 
$\Delta R_{b b } < 2$ (the upper-right panel); and
basic cuts,
$\Delta R_{\gamma\gamma},\,\Delta R_{b b} > 2$, and 
$\Delta R_{\gamma\gamma},\,\Delta R_{b b} < 2$ (the lower panel).
We assume that the measurements agree with the SM values with 
uncertainties of 25\% and 50\%, respectively.
}
\end{figure}

In Fig.~\ref{fig:cpc1}, we show the contour lines of 
$\sigma(gg\to HH)/\sigma_{\rm SM} (gg \to HH) = 0.5$ and $1.5$ with
the Higgs boson pair decaying into $(\gamma\gamma) (b\bar b)$.
In each panel, we assume three measurements of the ratios
corresponding to basic cuts (orange lines), 
$\Delta R > 2$ (dashed black lines), 
and $\Delta R  < 2$ (solid black lines):
here $\Delta R$ represents $\Delta R_{\gamma\gamma}$ (upper left),
$\Delta R_{b b}$ (upper right), or $\Delta R_{\gamma\gamma,bb}$
(lower). Therefore, for example,
if the bacis-cuts cross section ratio is measured
to be consistent with the SM prediction within $50\%$ error,
any points in the two bands 
bounded by the two pairs of orange lines are allowed.
In each band, a rather wide range of $g_t^S$ and $\lambda_{3H}$ is allowed
although they are correlated. Suppose we only make one measurement
of the cross section without or with a cut on $\Delta R$, 
we would not be able to 
pin down useful values for $g_t^S$ and $\lambda_{3H}$.
However, since the shape of the three bands are not exactly the same,
we can make use of three simultaneous measurements in order to obtain
more useful information for the couplings
$g_t^S$ and $\lambda_{3H}$.

In the upper-left panel of Fig.~\ref{fig:cpc1}, 
we suppose that one can make three measurements 
of cross sections: with basic cuts,
 $\Delta R_{\gamma\gamma} > 2$, and 
$\Delta R_{\gamma\gamma} < 2$.  We assume that the measurements
agree with the SM predictions within 25\% or 50\% uncertainty. 
The region of parameter space in $(g_t^S,\lambda_{3H})$ bounded 
by the three measurements
is shown by the lighter purple region for 50\% uncertainty and
darker purple region for 25\% uncertainty.
Similarly the upper-right panel is for the regions with the
$\Delta R_{b b}$ cut. 
In the lower panel, we show the regions with the combined
cuts of $\Delta R_{\gamma\gamma}$ and $\Delta R_{b b}$: both larger
than or smaller than 2. 
The implications from the measurements are 
very significant. First, all panels show that $g_t^S$ is
significantly away from  zero if one can simultaneously measure the cross 
sections (no matter with 25\% or 50\% uncertainties) 
with basic cuts, $\Delta R_{\gamma\gamma} > 2$, and 
$\Delta R_{\gamma\gamma} < 2$; and similarly for $\Delta R_{b b}$ and
using both distributions. 
Second, as shown in the lower panel, the
value for $\lambda_{3H}$ is statistically distinct from  zero if one
measures the cross sections with a 25\% uncertainty. This is achieved
by using both $\Delta R_{\gamma\gamma}$ and $\Delta R_{b b} > 2$ or $<2$. 
Furthermore, from the lower panel in Fig.~\ref{fig:cpc1} we can see that
with 25\% level uncertainty the values of $\lambda_{3H}$ sensitivity 
regions are $0.3 \alt |\lambda_{3H}| \alt 2.6$.

\begin{figure}[th!]
\centering
\includegraphics[width=3.2in]{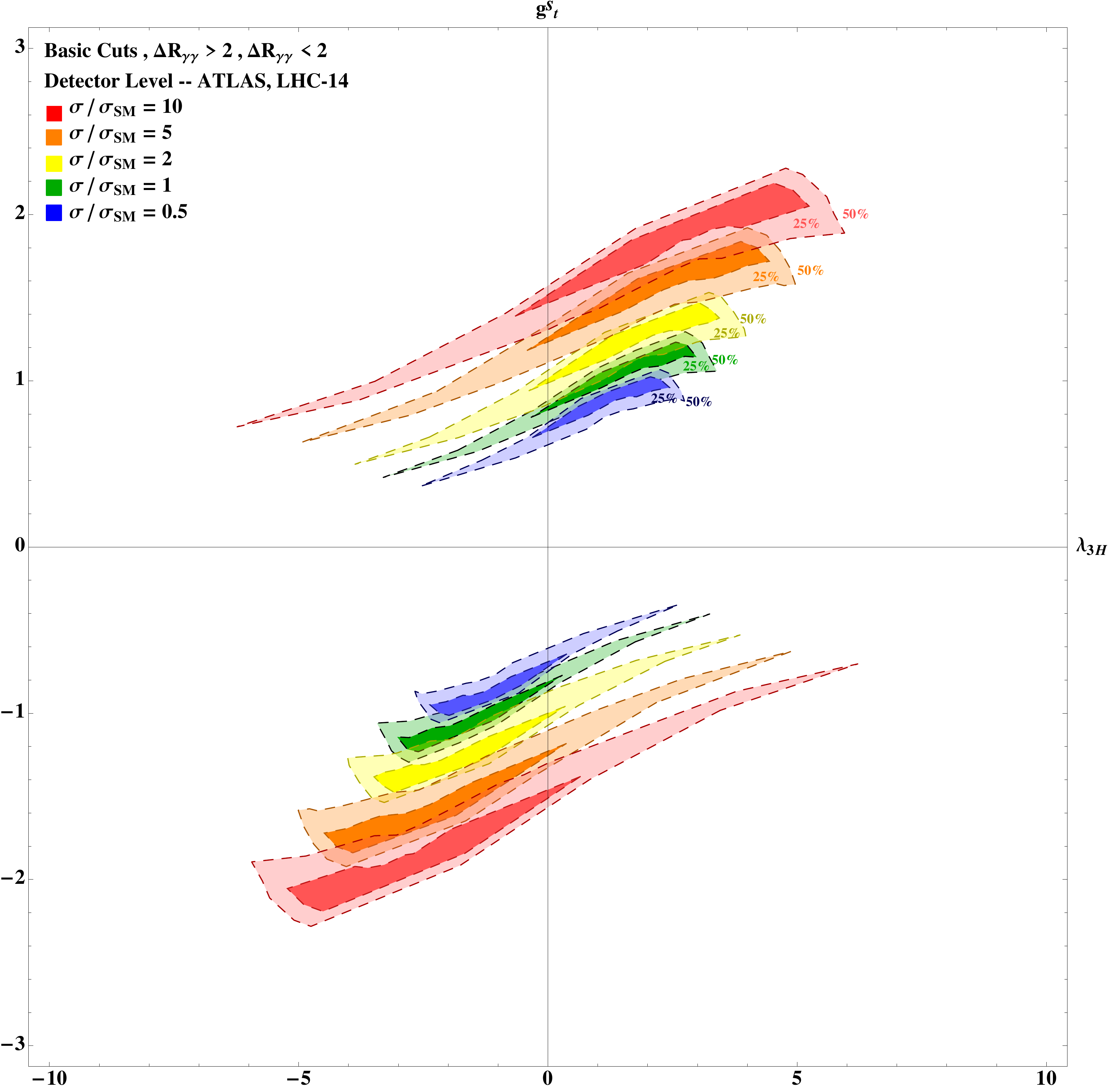}
\includegraphics[width=3.2in]{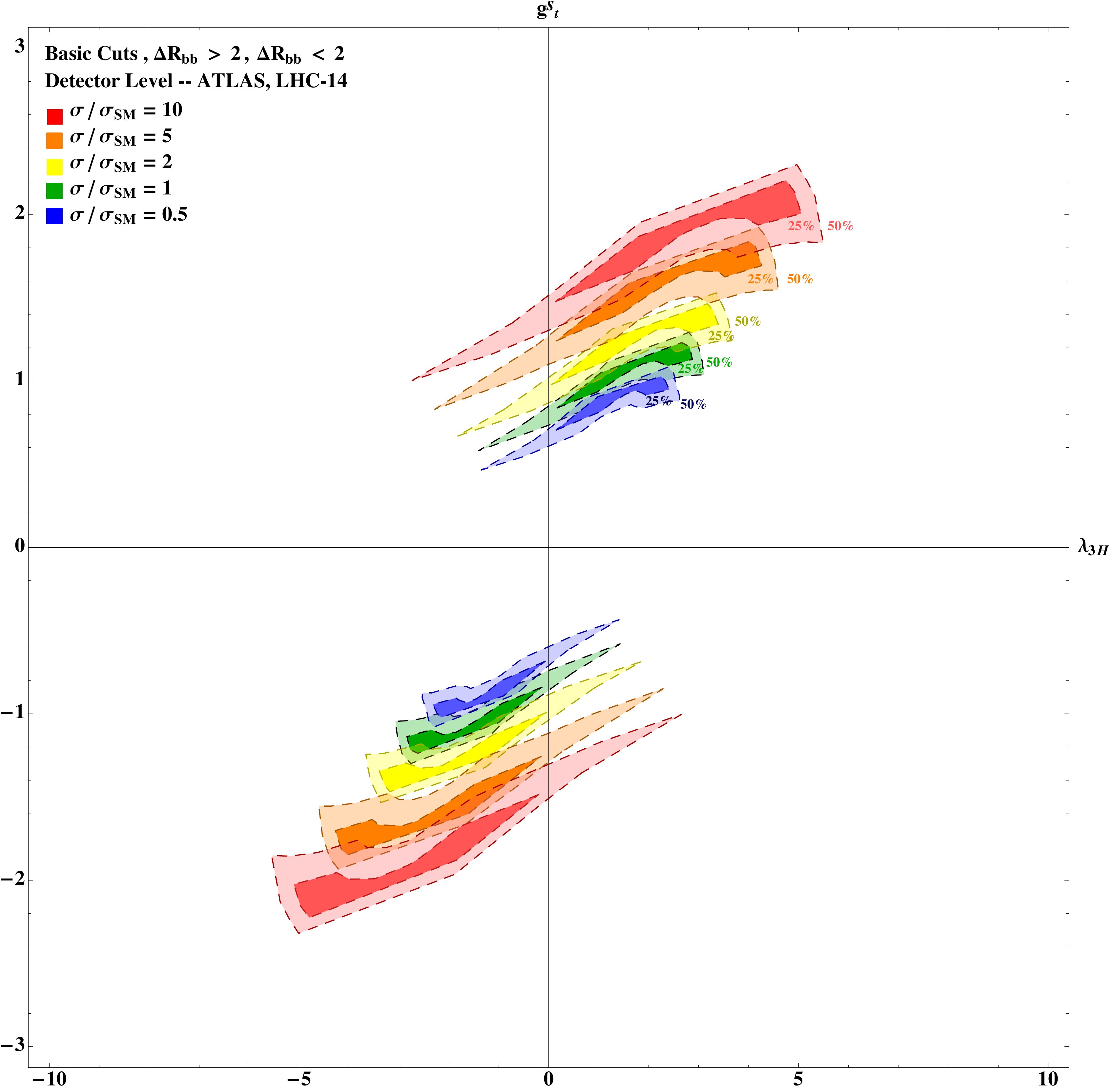}
\includegraphics[width=3.2in]{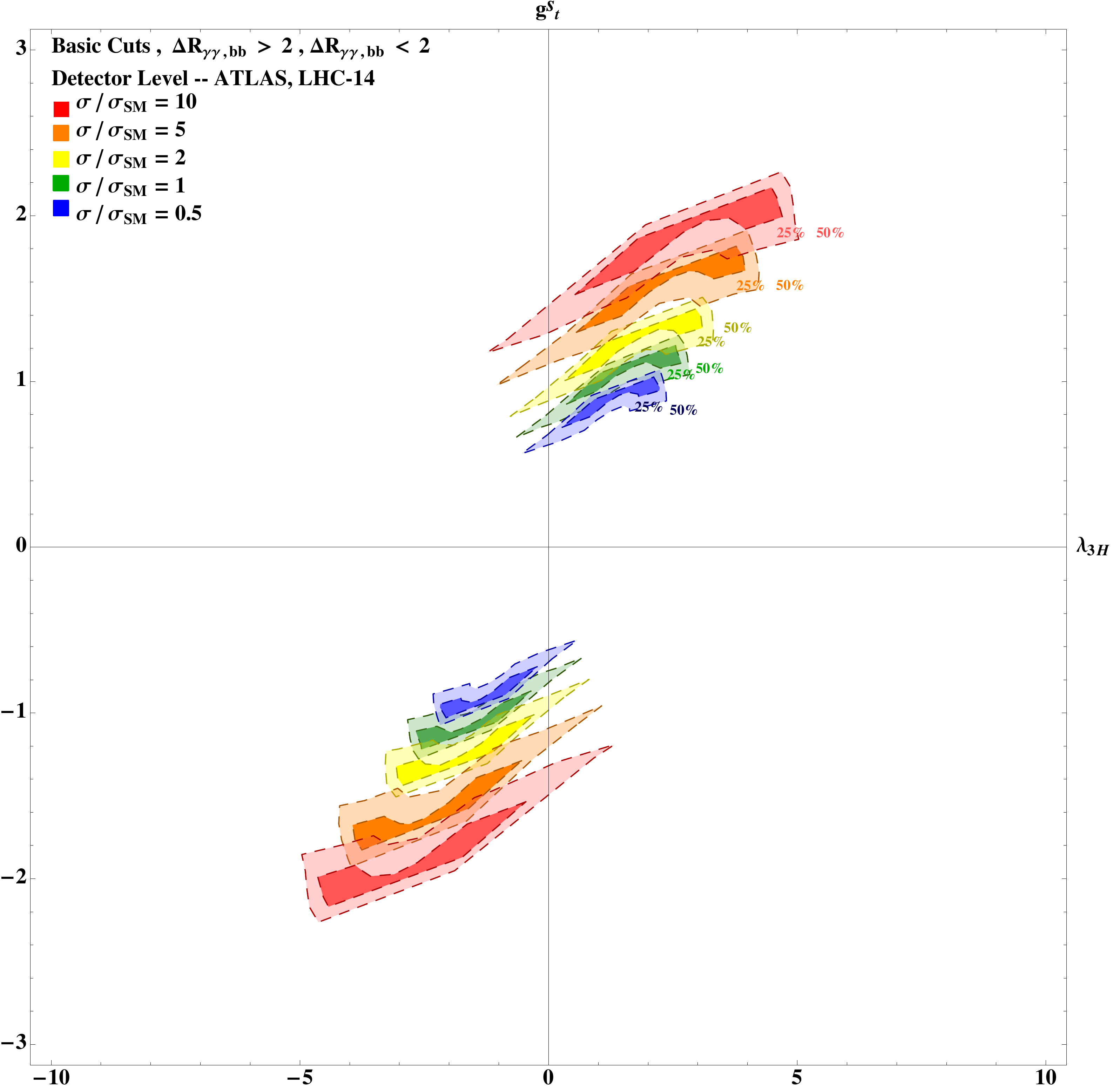}
\caption{\small \label{fig:cpc1-2}
{\bf CPC1:} The same as Fig.~\ref{fig:cpc1} but we assume 
$\sigma / \sigma_{\rm SM} = 0.5 - 10$.
}
\end{figure}

We can repeat the exercise with the measured cross sections being
multiples of the SM predictions. We show the corresponding 25\% and 
50\% regions in Fig.~\ref{fig:cpc1-2} for 
$\sigma/\sigma_{\rm SM} = 0.5,\, 1,\, 2,\,5,\, 10$.
Only with both $\Delta R_{\gamma\gamma}$ and $\Delta R_{b b}$, one can
really tell if $\lambda_{3H}$ is significantly distinct from zero.
The sensitivity regions for $\lambda_{3H}$ for various
$\sigma/\sigma_{\rm SM}$ are indicated by the darker color areas.

If the top-Yukawa coupling can be constrained more effectively 
by Higgs production, by $t\bar t H$ production, or by single top 
with Higgs production in the future measurements, say 
$g^S_t=1\pm 0.1$ (10\% uncertainty), it can help pinning down the 
acceptable range of $\lambda_{3H}$.
However, even in this case,
we emphasize the importance of simultaneous independent measurements, 
as illustrated in the following argument.
In the limit of $g^S_t=1$, the ratio of the cross sections is given by
\begin{equation}
\frac{\sigma(gg\to HH)}{\sigma_{\rm SM}(gg\to HH)}=
c_1(s)\,\lambda_{3H}^2+ c_2(s)\,\lambda_{3H}+ c_3(s)\,.
\end{equation}
Suppose $\sigma(gg\to HH)$ is measured to be the same as 
$\sigma_{\rm SM}(gg\to HH)$ and then, using the relation
$c_1(s)+c_2(s)+c_3(s)=1$, 
one may find the two
solutions for $\lambda_{3H}$: $1$ or $-c_2(s)/c_1(s)-1$. 
For example, one may have $\lambda_{3H}=1$ or $4$ at most if
only the basic-cuts ratio is measured, see Table~\ref{tab:all}.
Therefore,
one cannot determine $\lambda_{3H}$ uniquely
with only one measurement
even when the measurement is very precise and the exact value of
$g^S_t$ is known. 
It is unlikely to resolve this two-fold ambiguity at the LHC even 
we assume the three measurements of the ratios, as shown in Fig.~\ref{fig:cpc1}.
Also, the situation remains the same at the 100 TeV $pp$  machine in which
we have $\lambda_{3H}=1$ or $5$ in the bacis-cuts case
when $\sigma(gg\to HH)=\sigma_{\rm SM}(gg\to HH)$,
see Table~\ref{tab:100} and Fig.~\ref{fig:cpc1_100}. 
If a future $e^+e^-$ linear collider and/or the 100 TeV $pp$ machine
are operating in the era of the high-luminosity LHC, combined efforts are
desirable to determine the value of $\lambda_{3H}$ uniquely~\cite{future}.

\subsection{CPC2: $g_t^S$, $\lambda_{3H}$, and $g_{tt}^S$ }
This is the scenario that involves all scalar-type couplings in 
the triangle, box, and contact diagrams.
The corresponding coefficients in Eq.~(\ref{cdef})
are $c_1, c_2, c_3, e_1, e_2, e_3$. 
Results at the detector level using the ATLAS template in Delphes v.3
are shown in Table~\ref{tab:all}.

\begin{figure}[th!]
\centering
\includegraphics[width=3.2in]{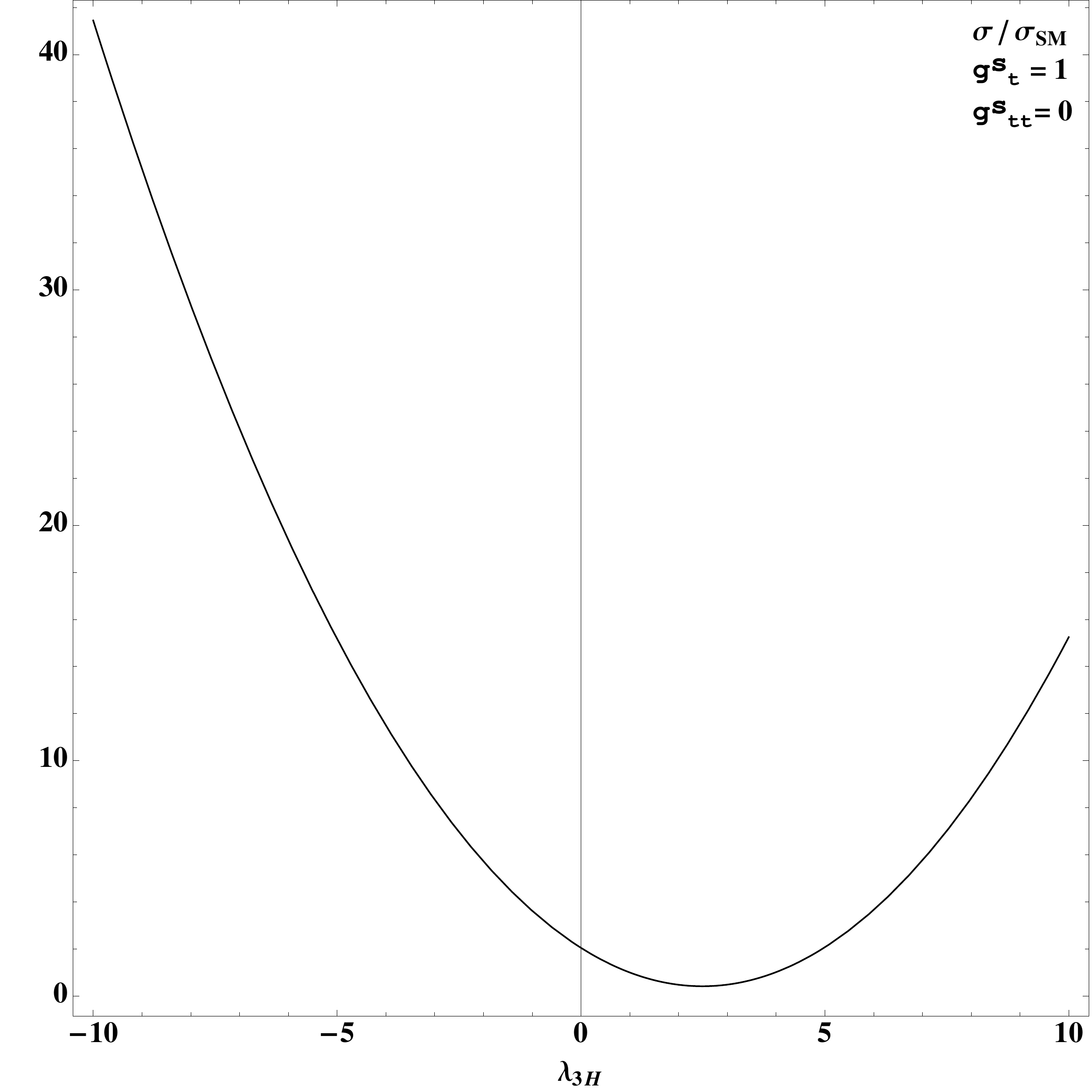}
\includegraphics[width=3.2in]{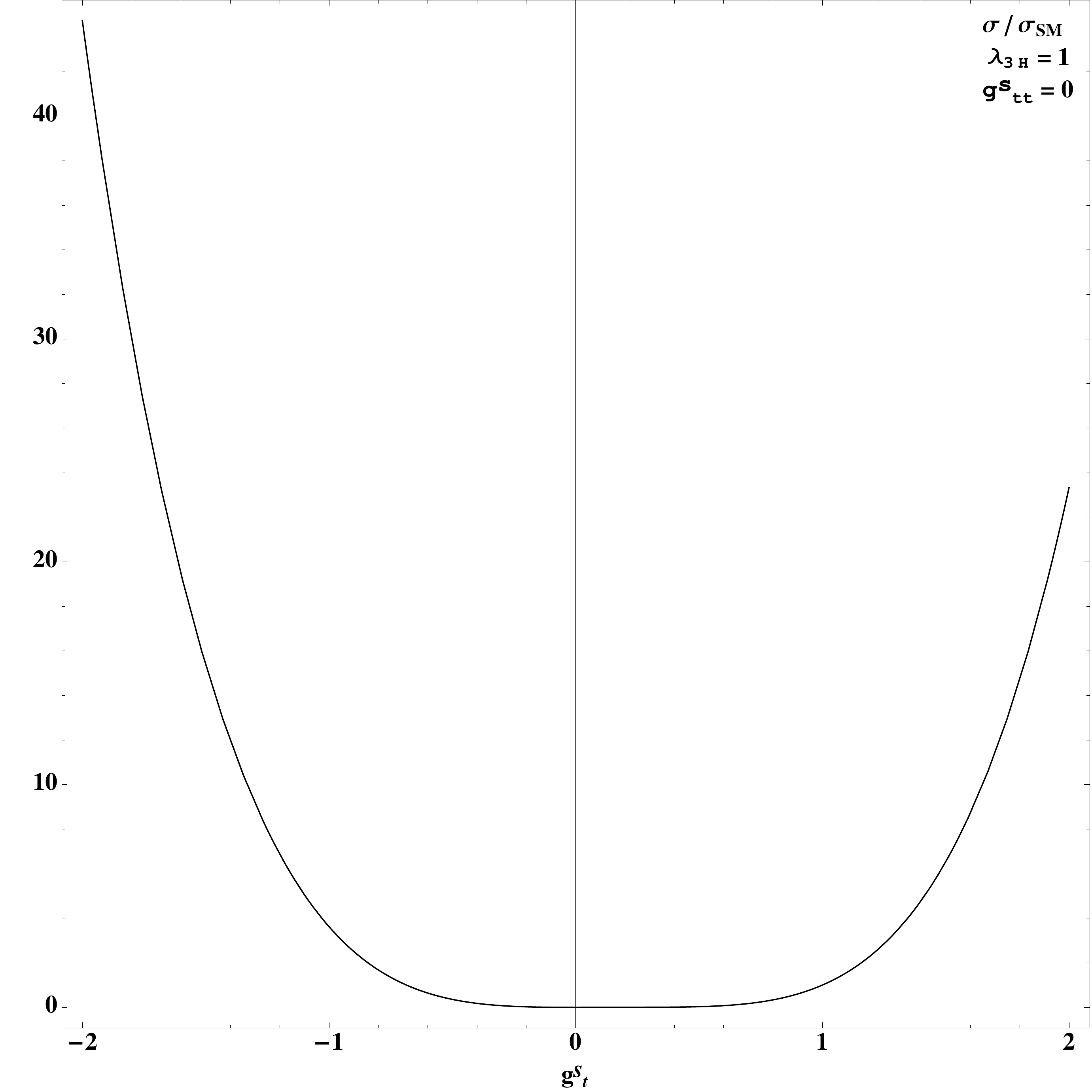}
\includegraphics[width=3.2in]{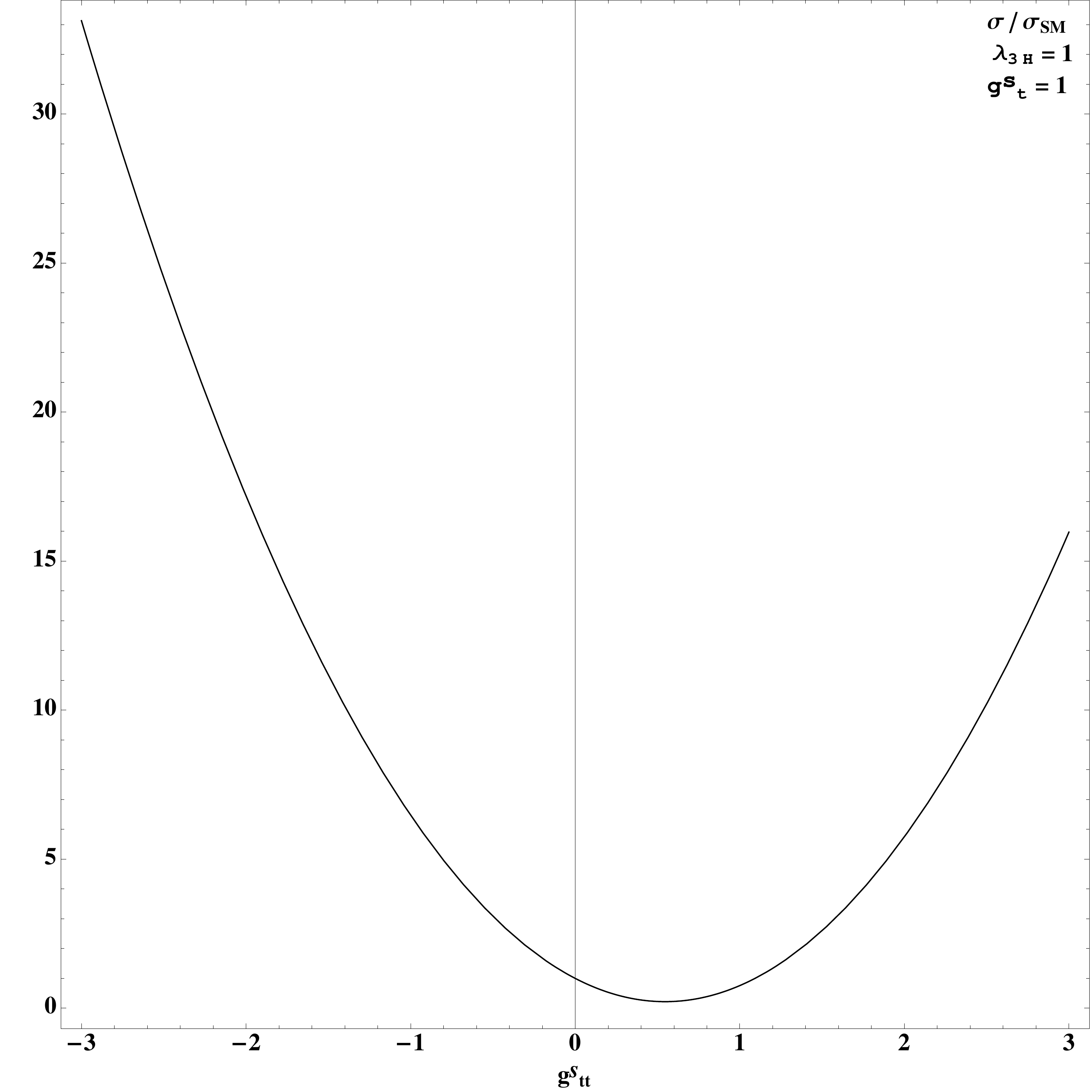}
\caption{\small \label{fig:cpc2-3}
{\bf CPC2:} Variation of the 
ratio $\sigma / \sigma_{\rm SM}$ using Eq.~(\ref{cdef})
versus $\lambda_{3H}$ (upper-left), $g_t^S$ (upper-right), 
and $g_{tt}^S$ (lower) 
while keeping the other two parameters at their corresponding SM values.
}
\end{figure}

We first examine the cross section versus one input parameter at a
time, shown in Fig.~\ref{fig:cpc2-3}, while keeping the two parameters
at their corresponding SM values. 
In the upper-left panel for $\sigma/\sigma_{\rm SM}$ 
versus $\lambda_{3H}$,
the lowest point occurs at
$\lambda_{3H} \approx 2.5$ when the interference term strongly cancels
the triangle and box diagrams. Then the ratio increases from the
lowest point on either side of $\lambda_{3H} \approx 2.5$.  Negative
$\lambda_{3H}$s give constructive interference while positive
$\lambda_{3H}$s give destructive interference.  
One may observe similar behavior when $g^S_{tt}$ is
varied as shown in the lower panel. Taking $\lambda_{3H}=g^S_t=1$,
\begin{equation}
\frac{\sigma(gg\to HH)}{\sigma_{\rm SM}(gg\to HH)}=
e_3(s)\,(g^S_{tt})^2+ [e_2(s)+e_1(s)]\,g^S_{tt}+ 1\,.
\end{equation}
Since $e_1(s)>0$ and $e_2(s)<0$, we see that the contact diagram interferes
constructively with the triangle diagram but
destructively with the box diagram. The dominance of the box diagram leads to
the totally destructive interference when $g^S_{tt}>0$, 
resulting in the minimum at $g^S_{tt}\approx 0.5$.

\begin{figure}[th!]
\centering
\includegraphics[width=3.2in]{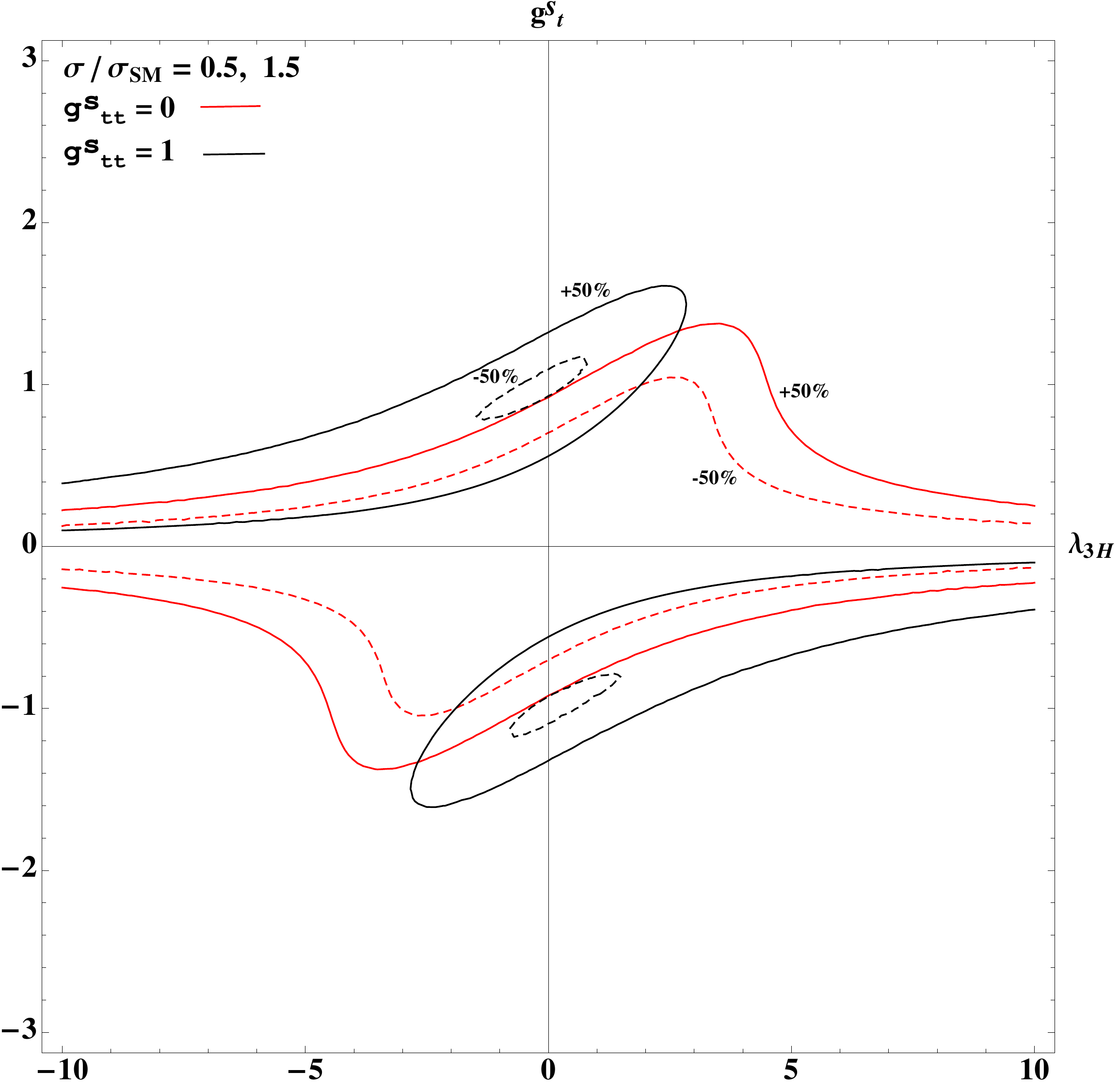}
\includegraphics[width=3.2in]{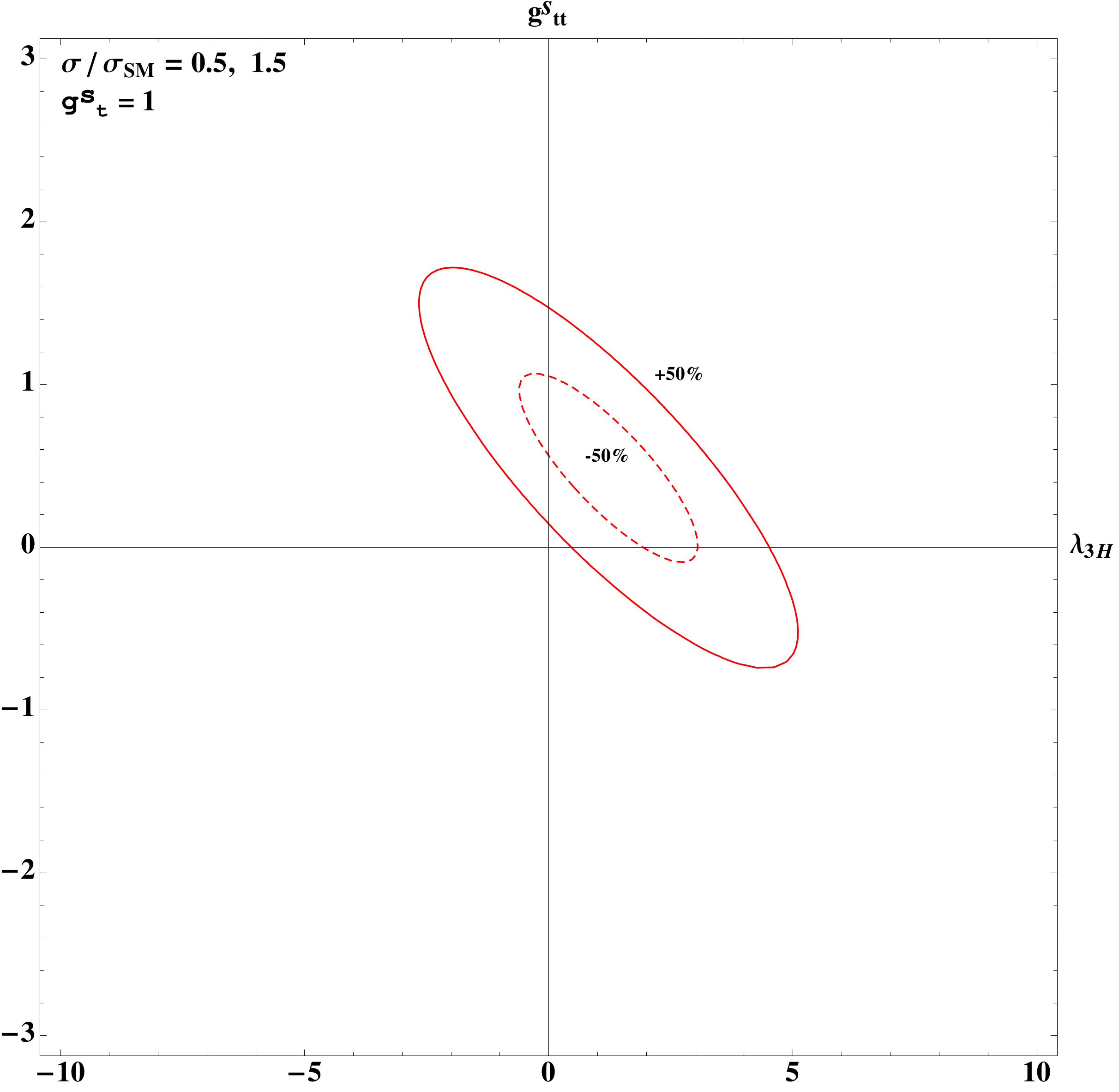}
\includegraphics[width=3.2in]{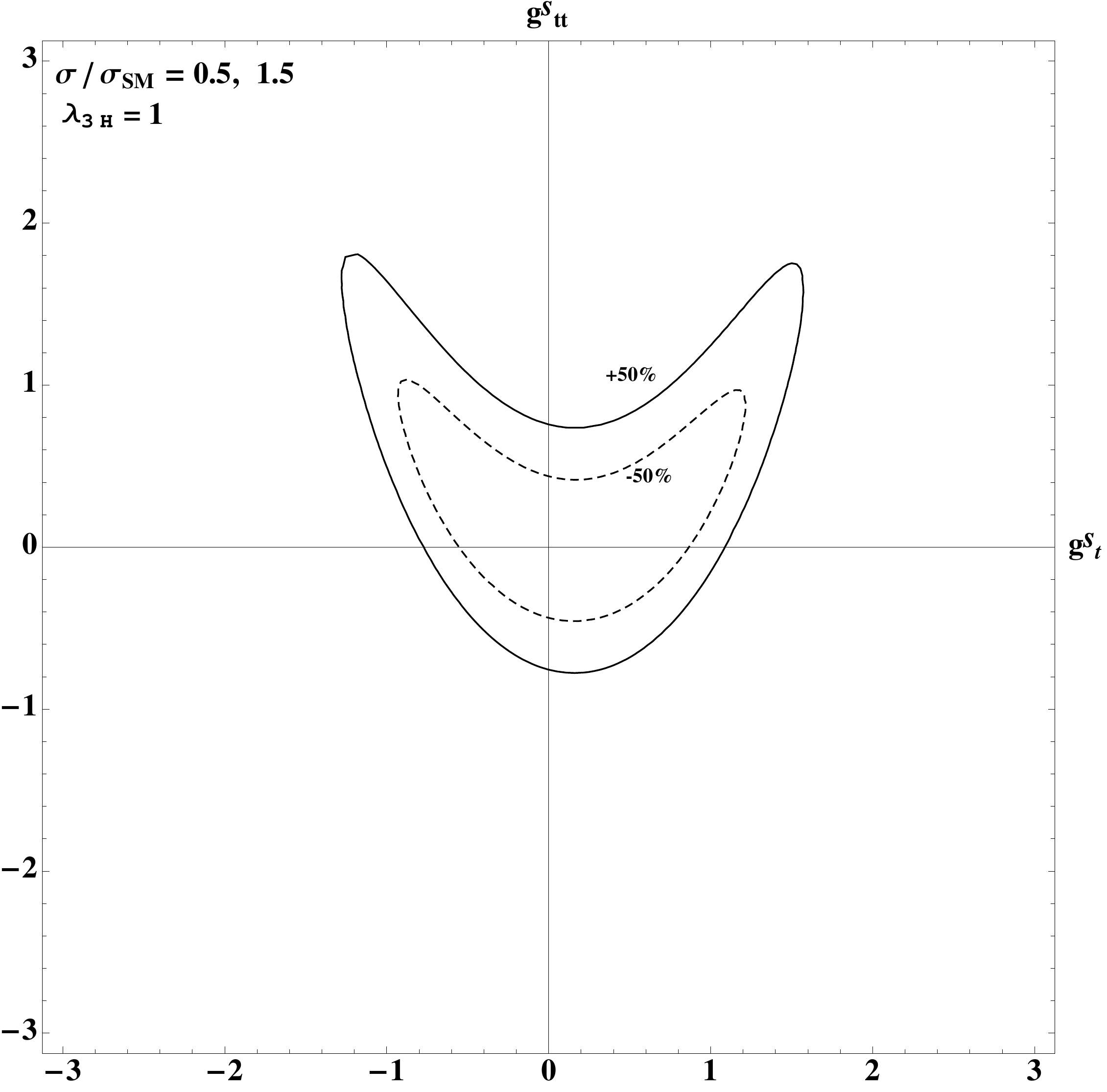}
\caption{\small \label{fig:cpc2-4}
{\bf CPC2:} Contours marked by $\sigma / \sigma_{\rm SM} = 0.5, 1.5$
for $(\lambda_{3H},\, g_t^S)$ (upper-left), $(\lambda_{3H},\, g_{tt}^S)$ (upper-right), 
and $( g_t^S,\, g_{tt}^S)$ (lower).
}
\end{figure}

We show contours for the ratio $\sigma/\sigma_{\rm SM} = 0.5, 1.5$ in
the plane of $(\lambda_{3H} ,\, g_t^S)$ (upper-left),
$(\lambda_{3H},\, g_{tt}^S)$ (upper-right), and $( g_t^S,\, g_{tt}^S)$
(lower) in Fig.~\ref{fig:cpc2-4}. 
The dashed lines denoted by $-50\%$ is 
for $\sigma/\sigma_{\rm SM} = 0.5$ and those by $+50\%$ 
for $\sigma/\sigma_{\rm SM} = 1.5$.
In the upper-left panel in the plane
of $(\lambda_{3H} ,\, g_t^S)$, we show contours for $g_{tt}^S = 0,1$.
The $g_{tt}^S= 0$ is the same as the SM so that the contours are
exactly the same as in Fig.~\ref{fig:cpc1}, while
with $g_{tt}^S=1$, the contact diagram contributes significantly to
the cross section, so that the contours shift more to negative
(positive) $\lambda_{3H}$ for positive (negative) $g_t^S$. In the
upper-right panel, where we show the contours in the plane of
$(\lambda_{3H},\, g_{tt}^S)$, the $g_{tt}^S$ negatively correlates
with $\lambda_{3H}$ because $e_1(s) > 0$.  In the lower panel, where
we fix $\lambda_{3H}=1$, somewhat nontrivial correlation between
$g_t^S$ and $g_{tt}^S$ exists.

\begin{figure}[th!]
\centering
\includegraphics[width=5.2in]{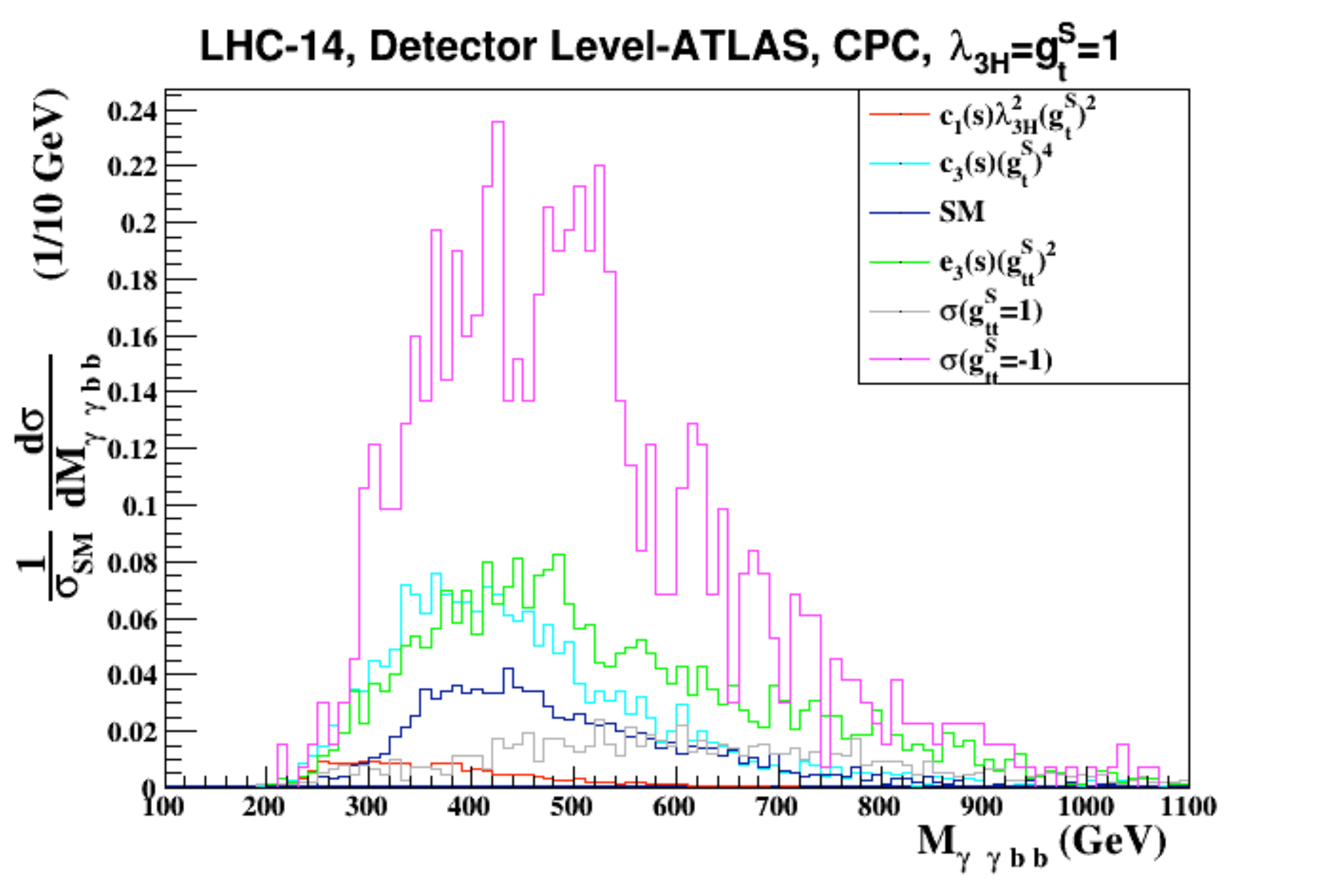}
\includegraphics[width=5.2in]{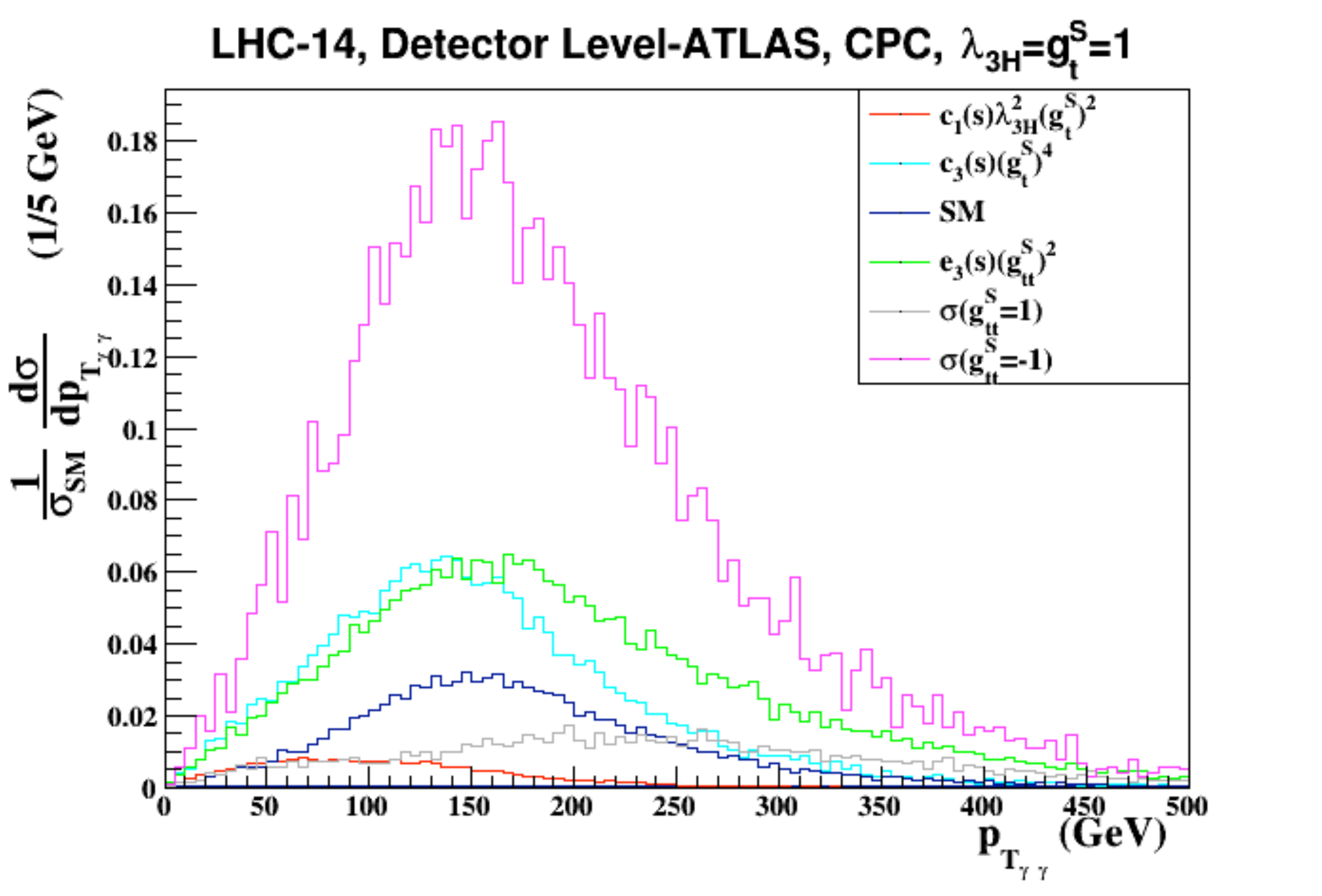}
\caption{\small \label{fig:cpc2-m}
{\bf CPC2:} 
Distributions in $M_{\gamma\gamma bb}$ (upper) and 
$p_{T_{\gamma\gamma}}$ (lower)
in the decay products of the Higgs boson pair with detector simulation.
}
\end{figure}

We use the same tools as in the {\bf CPC1} case to investigate the decay channel
$HH \to \gamma\gamma b \bar b$ with parton showering and detector simulations.
We first show the invariant mass ($M_{\gamma\gamma b \bar b}$) distribution of 
the Higgs-boson pair and the transverse momentum distribution of the 
photon pair 
in Fig.~\ref{fig:cpc2-m}.
The figure clearly illustrates the behavior of each
contributing diagram. The triangle diagram (red lines)
peaks at the lower invariant mass and
decreases with $M_{\gamma\gamma b \bar b}$, because of the $s$-channel Higgs 
propagator.  Then followed by the box diagram (skyblue lines)
which is larger than the triangle diagram at high invariant mass.  
The (darkblue) SM line represents the contributions from the triangle
and box diagram including the destructive interference between them.
These three distributions are the same as in the {\bf CPC1} case.
The contact diagram (green lines) shows similar behavior as
the box diagram (skyblue lines) at low $M_{\gamma\gamma b \bar b}$
and $P_{T_{\gamma\gamma}}$ but with higher and larger tails.
The grey and magenta lines represent the full contributions including
the destructive and constructive interferences 
among the triangle, box, and contact diagrams when $g^S_{tt}=+1$ and
$g^S_{tt}=-1$, respectively. 
The contact diagram is the largest 
at the high invariant mass. It demonstrates what we describe earlier that the
contact diagram grows with energy. 

\begin{figure}[th!]
\centering
\includegraphics[width=5.2in]{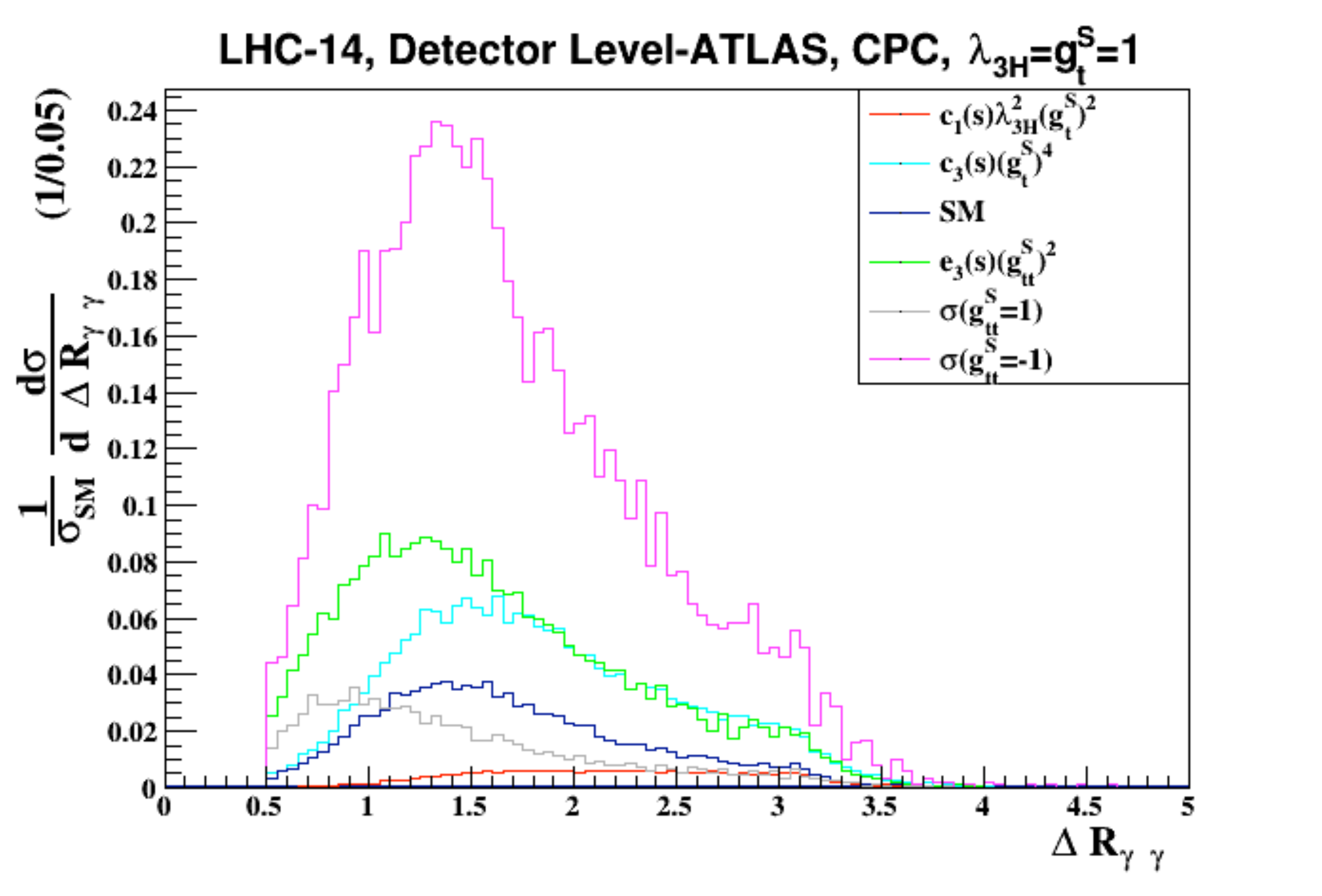}
\includegraphics[width=5.2in]{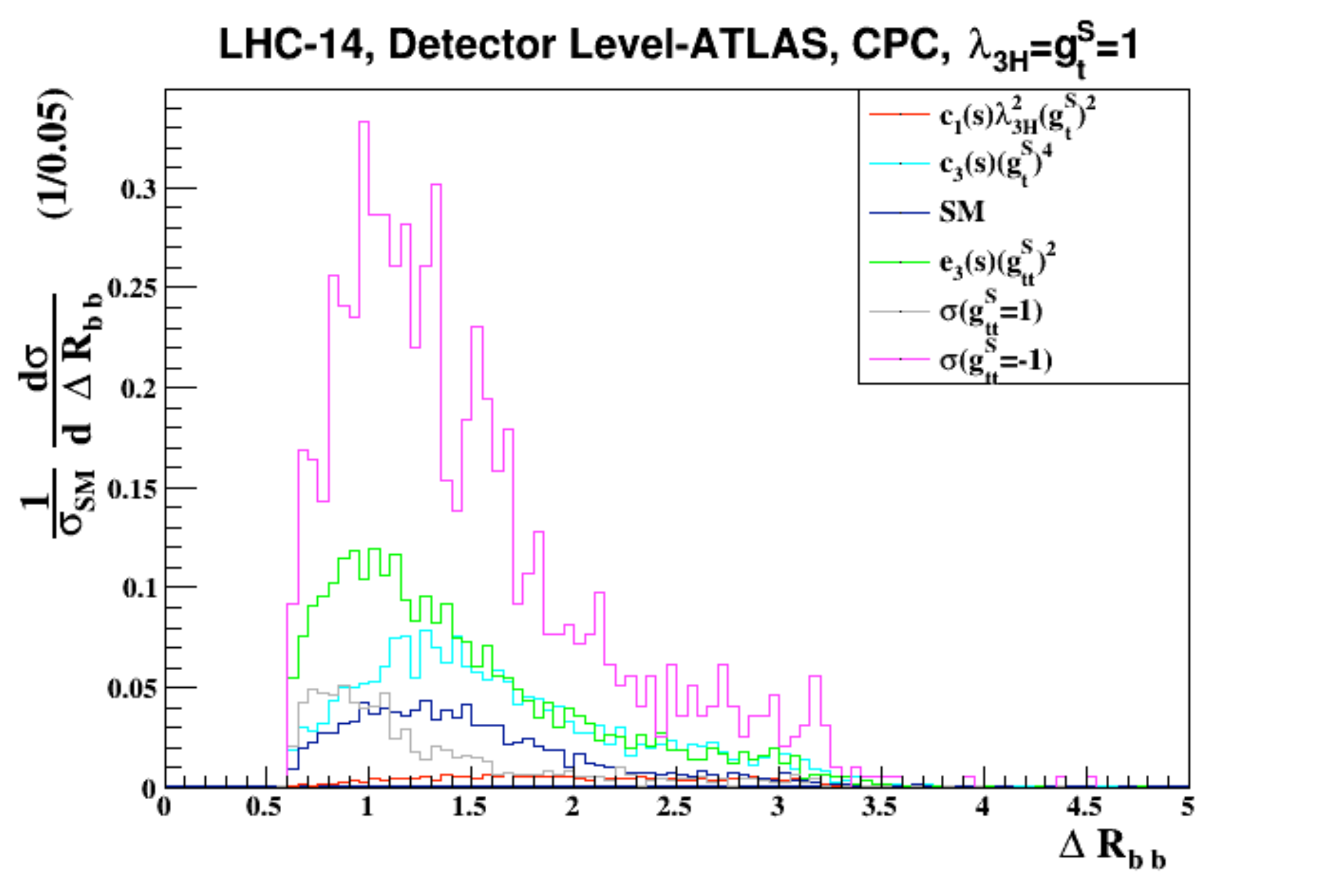}
\caption{\small \label{fig:cpc2-2}
{\bf CPC2:} 
Angular distributions of $\Delta R_{\gamma\gamma}$ and $\Delta R_{b b}$ between
the two photons and between the two $b$ quarks.
}
\end{figure}

We show the angular distributions $\Delta R_{\gamma\gamma}$ 
and $\Delta R_{b b}$ between the two decay products of 
each Higgs boson in Fig.~\ref{fig:cpc2-2}.
The lines are the same as in Fig.~\ref{fig:cpc2-m}.
Similar to the {\bf CPC1} case, the higher the invariant mass, the more
energetic the Higgs boson, and the smaller the angular separation
between the decay products will be.  Therefore, triangle diagram (red lines) has
the widest separation, then followed by the box diagram (skyblue lines), 
and finally
the contact diagram (green lines) has the smallest angular separation. We come up
with the similar cuts as in the {\bf CPC1} case: $\Delta R$ larger or smaller
than $2$ to discriminate the triangle, box, and contact diagrams. We
show in Table~\ref{tab:all} the coefficients $c_1,\, c_2,\, c_3,\,
e_1,\, e_2,\, e_3$ such that the ratio of cross sections to the SM
predictions can be given by Eq.~(\ref{cdef}).

\begin{figure}[th!]
\centering
\includegraphics[width=3in,height=2.7in]{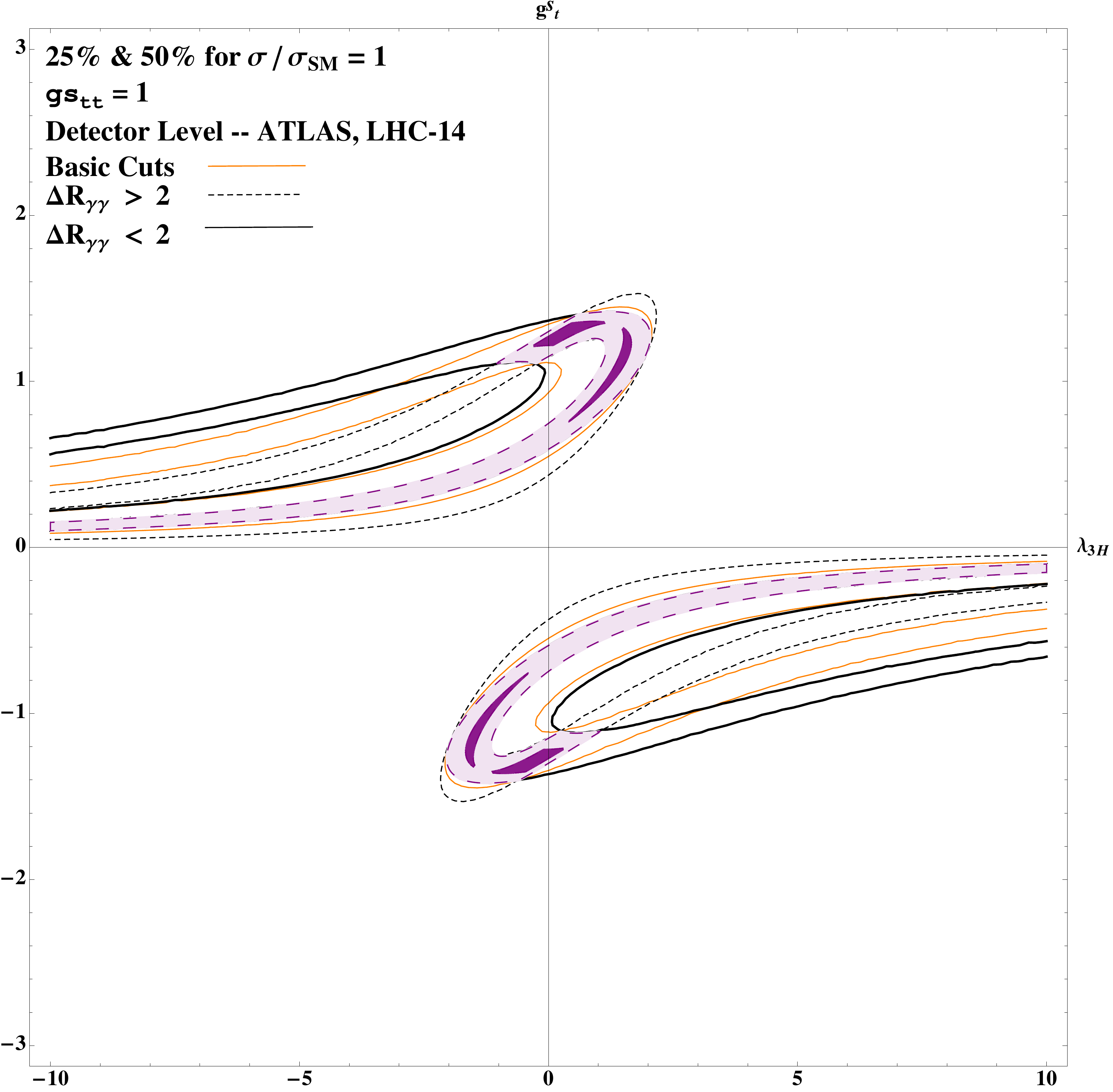}
\includegraphics[width=3in,height=2.7in]{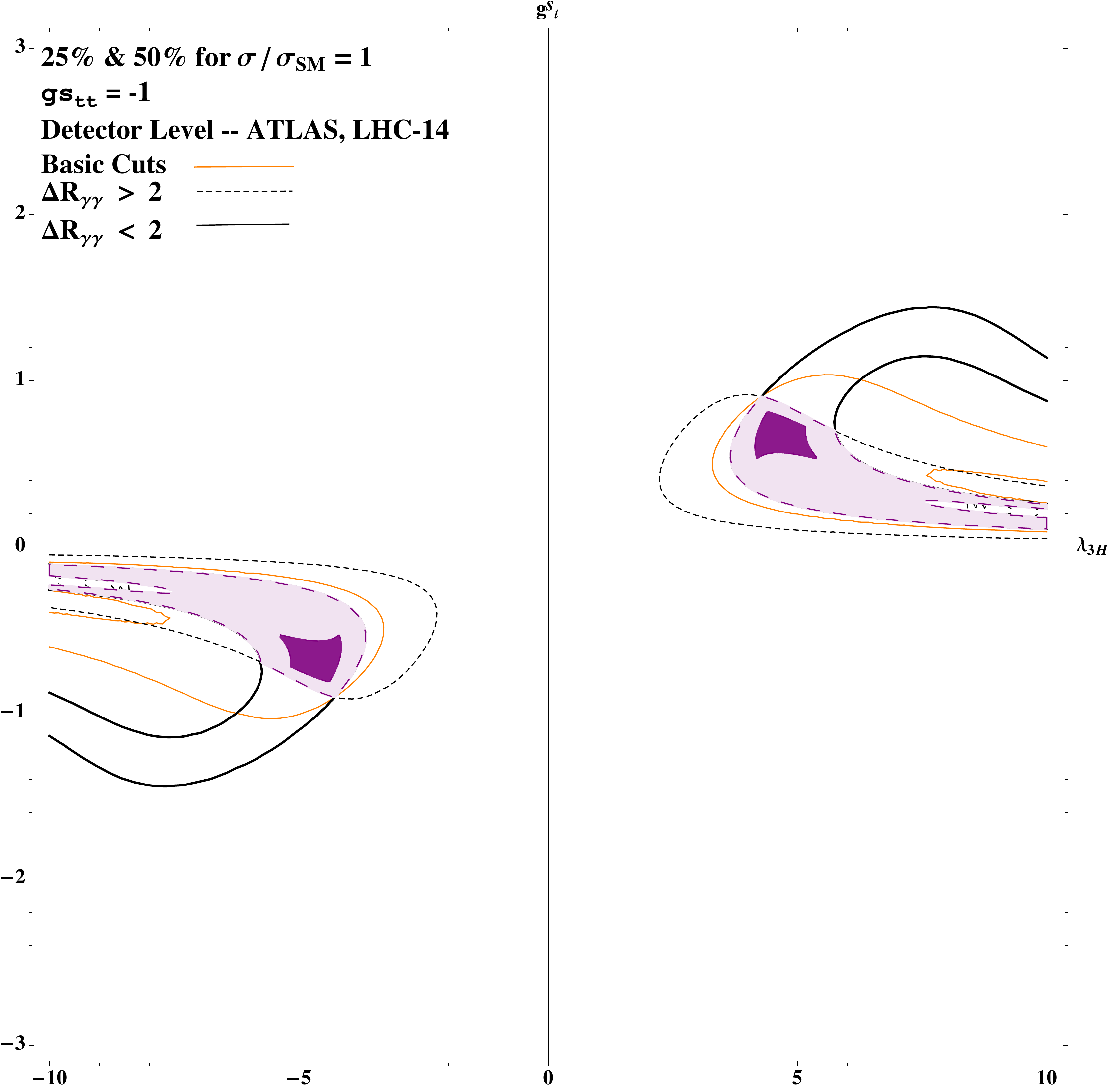}
\includegraphics[width=3in,height=2.7in]{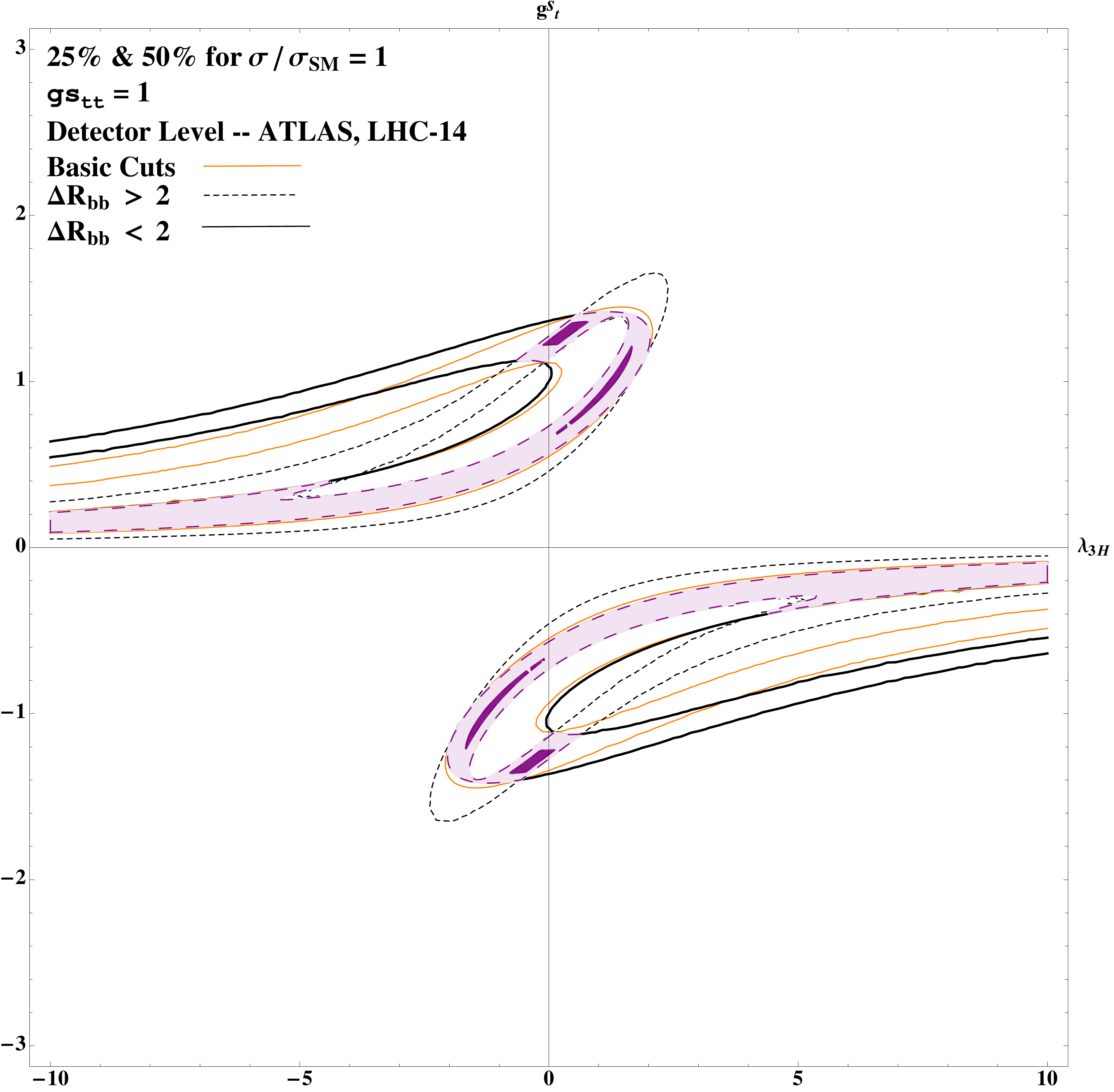}
\includegraphics[width=3in,height=2.7in]{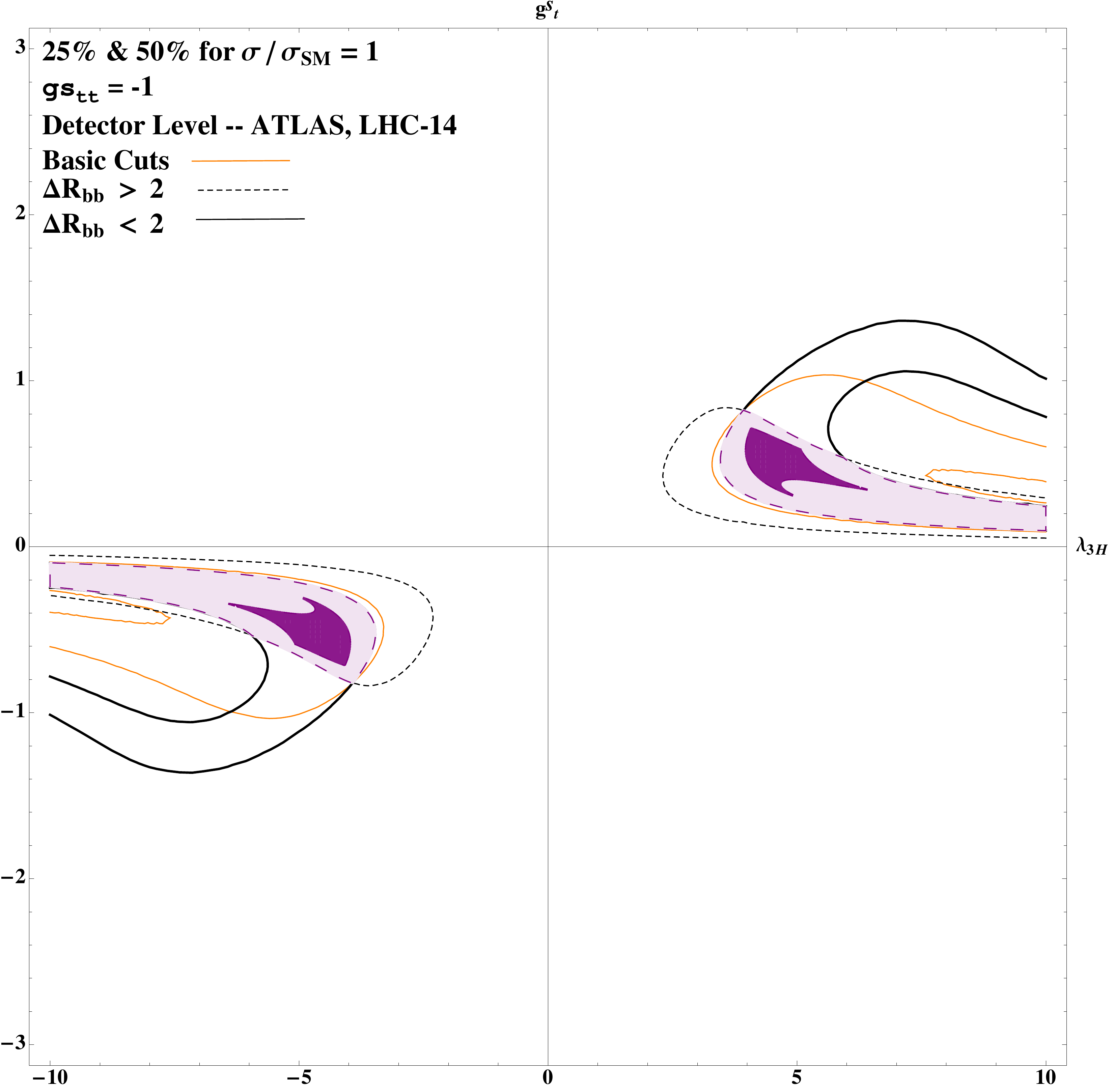}
\includegraphics[width=3in,height=2.7in]{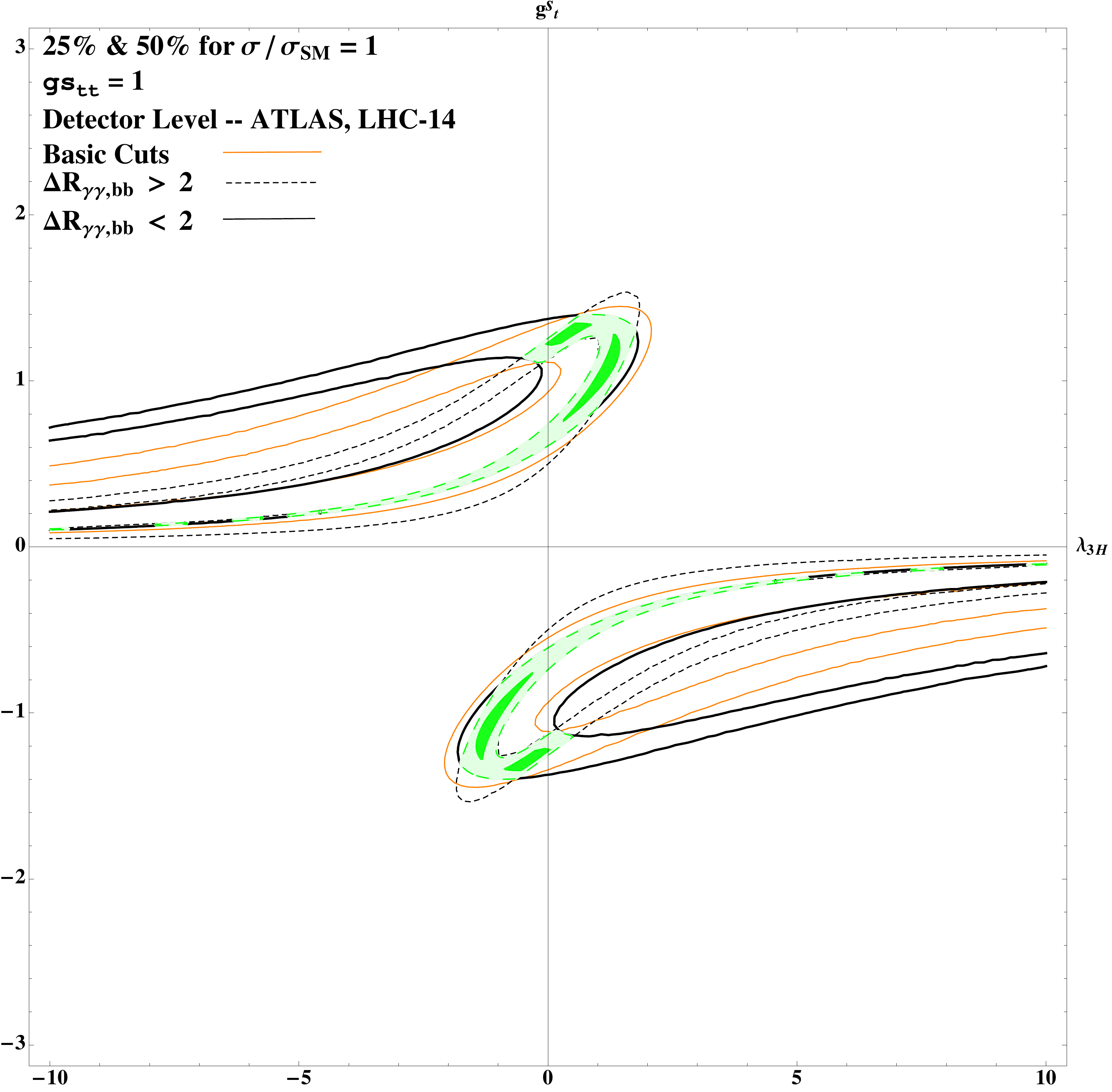}
\includegraphics[width=3in,height=2.7in]{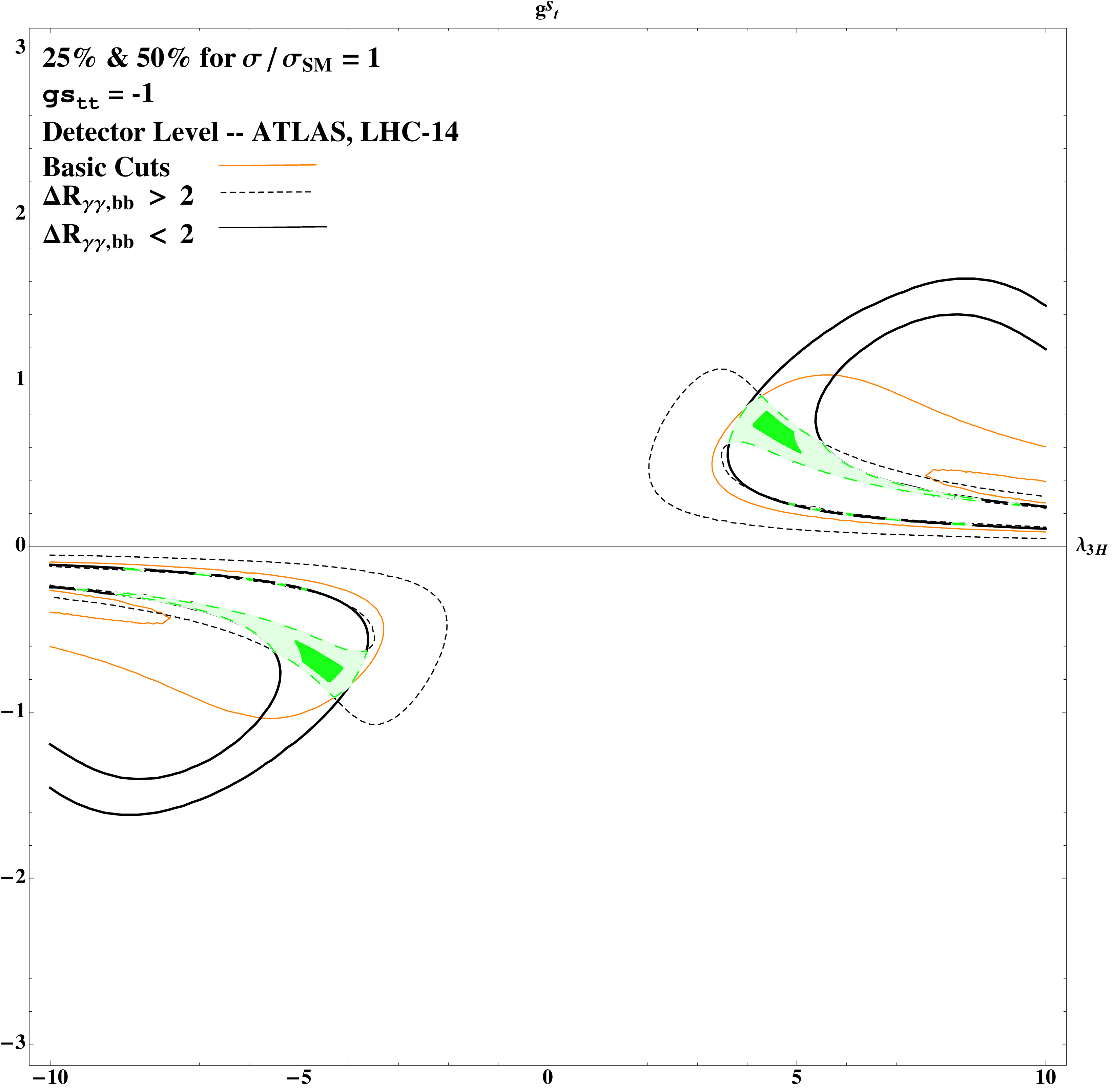}
\caption{\small \label{fig:cpc2-5}
{\bf CPC2:} 
The 25\% and 50\% sensitivity regions in $(\lambda_{3H}, g_t^S)$ bounded by three
measurements of cross sections with basic cuts, 
$\Delta R_{\gamma\gamma} > 2$, and 
$\Delta R_{\gamma\gamma} < 2$ (the upper panels);
with basic cuts, 
$\Delta R_{bb} > 2$, and 
$\Delta R_{bb } < 2$ (the middle panels); and
basic cuts,
$\Delta R_{\gamma\gamma},\,\Delta R_{b b} > 2$, and 
$\Delta R_{\gamma\gamma},\,\Delta R_{b b} < 2$ (the lower panels).
The left panels are for $g_{tt}^S=1$ while those on the right are for 
$g_{tt}^S=-1$.
We assume that the measurements agree with the SM values with 
uncertainties of 25\% and 50\%, respectively.
}
\end{figure}

\begin{figure}[th!]
\centering
\includegraphics[width=3in]{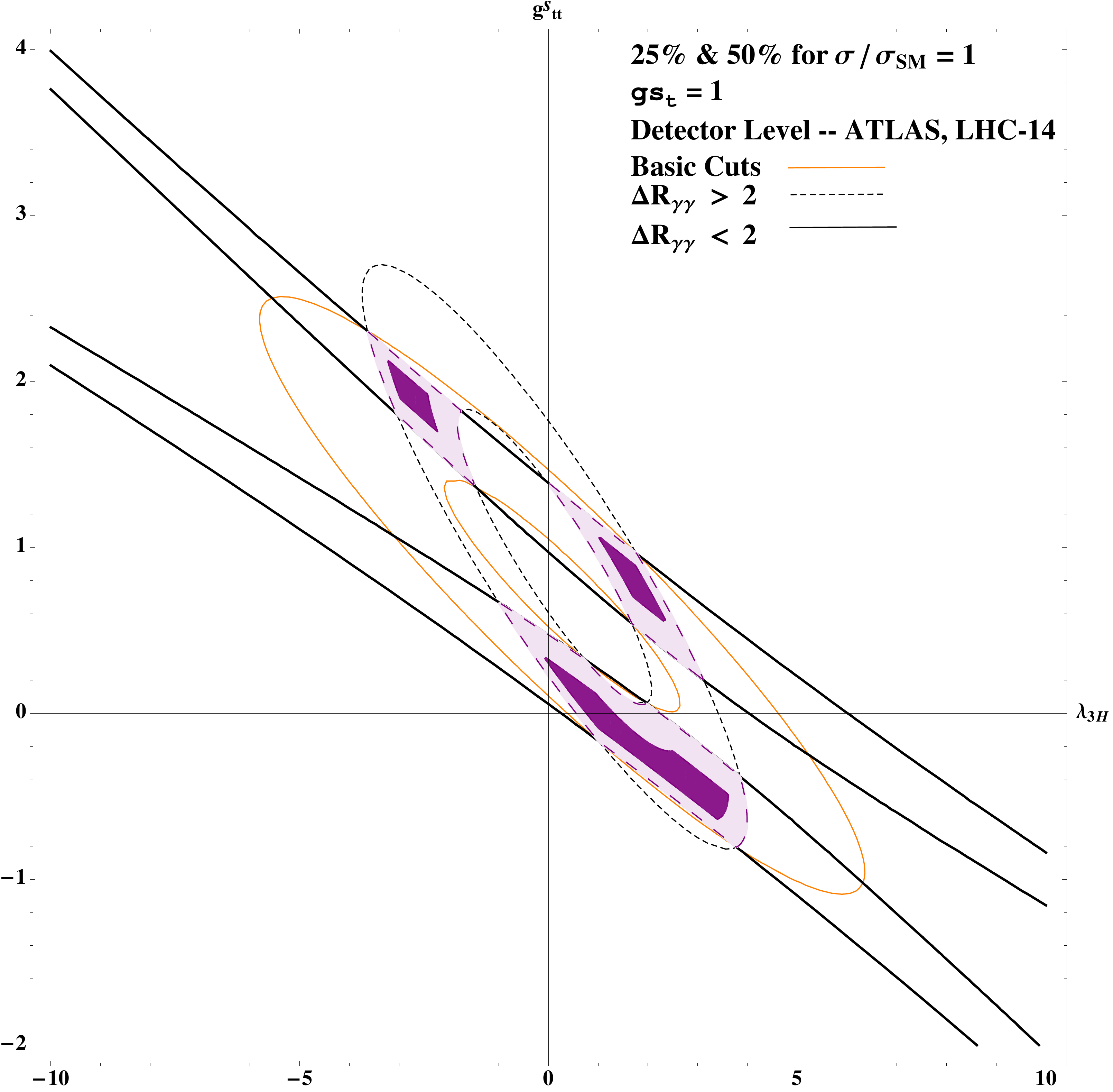}
\includegraphics[width=3in]{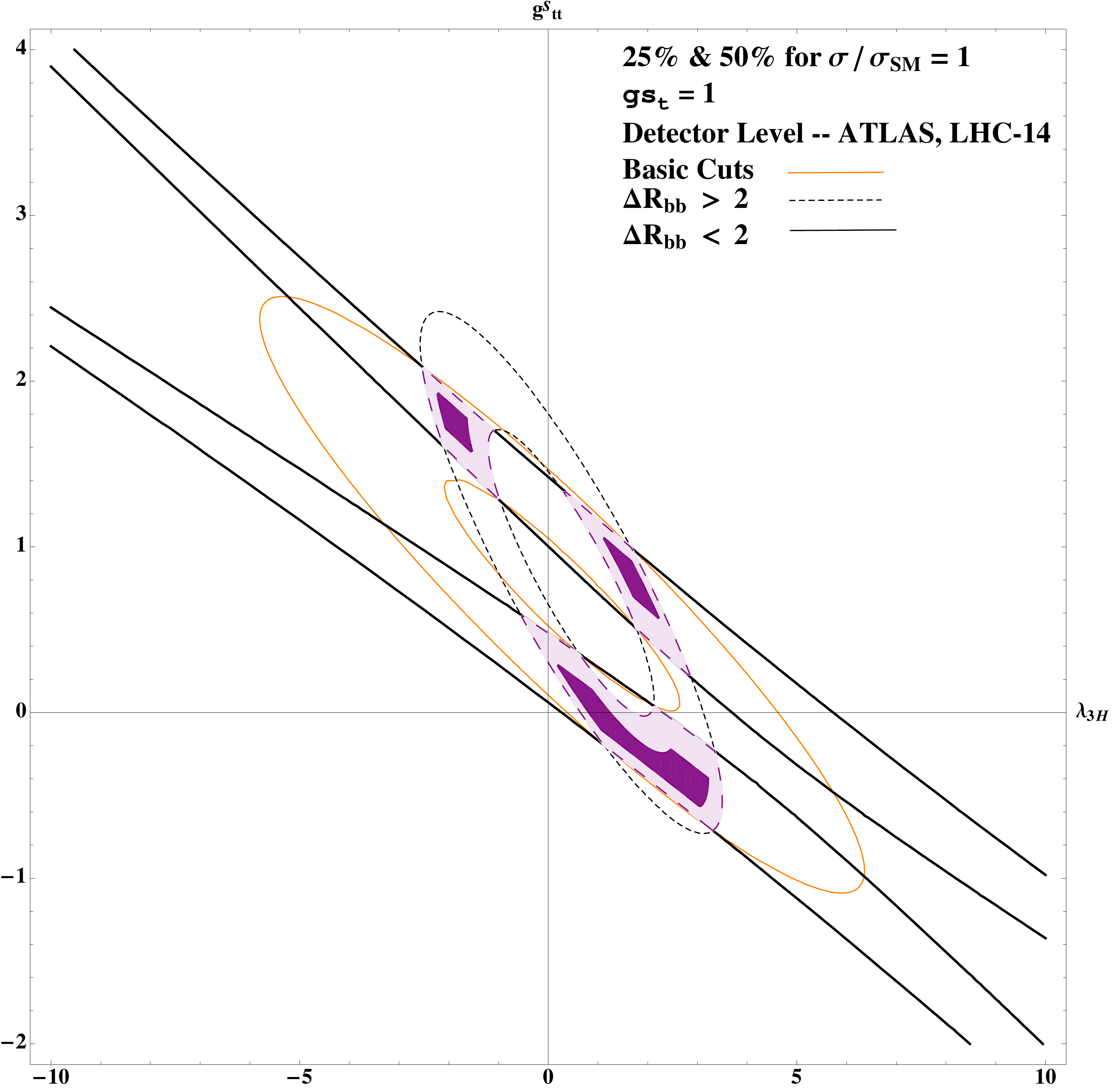}
\includegraphics[width=3in]{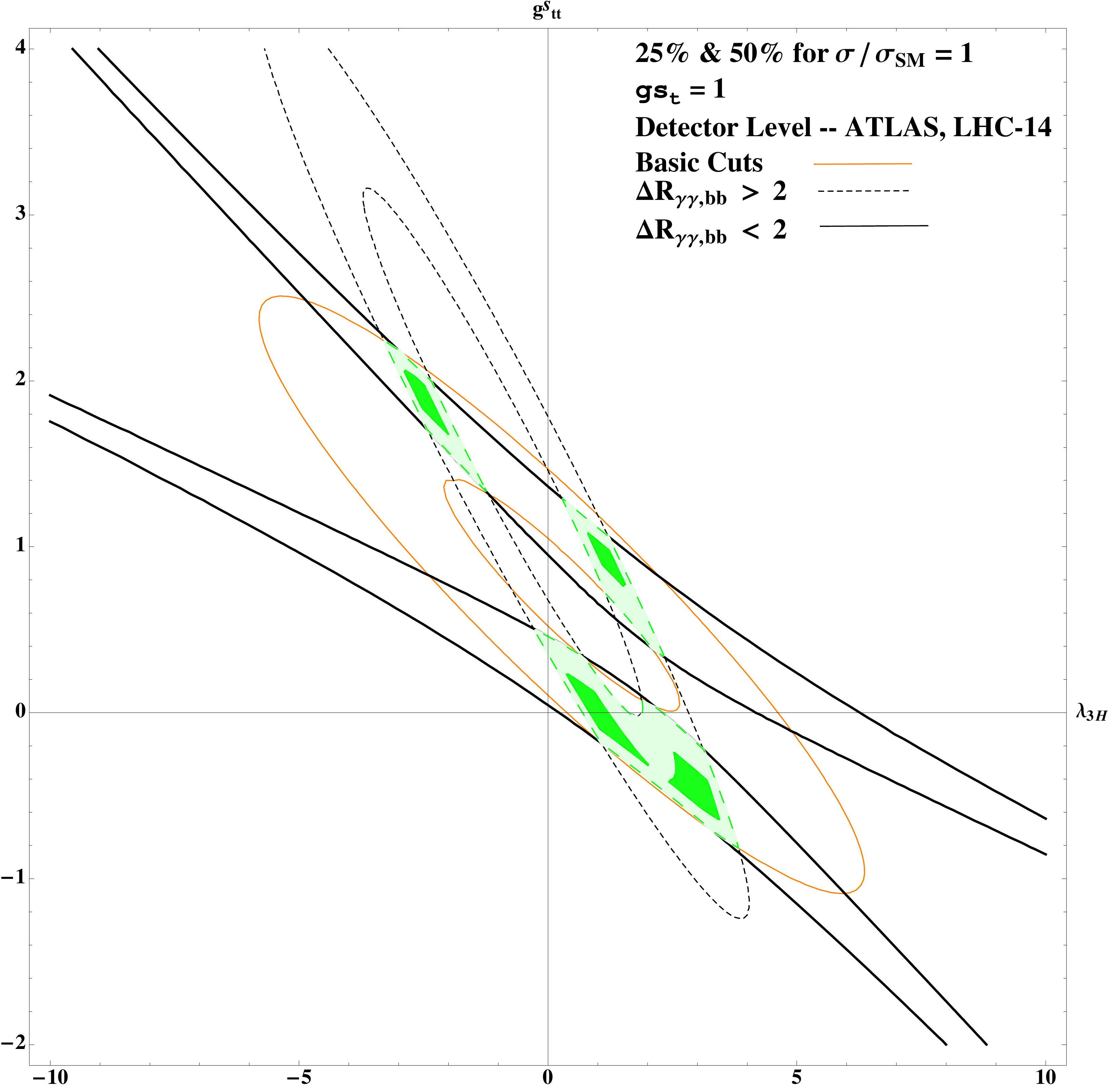}
\caption{\small \label{fig:cpc2-6}
{\bf CPC2:} similar to Fig.~\ref{fig:cpc2-5} but in the plane 
of $(\lambda_{3H},\, g_{tt}^S)$.
}
\end{figure}

\begin{figure}[th!]
\centering
\includegraphics[width=3in]{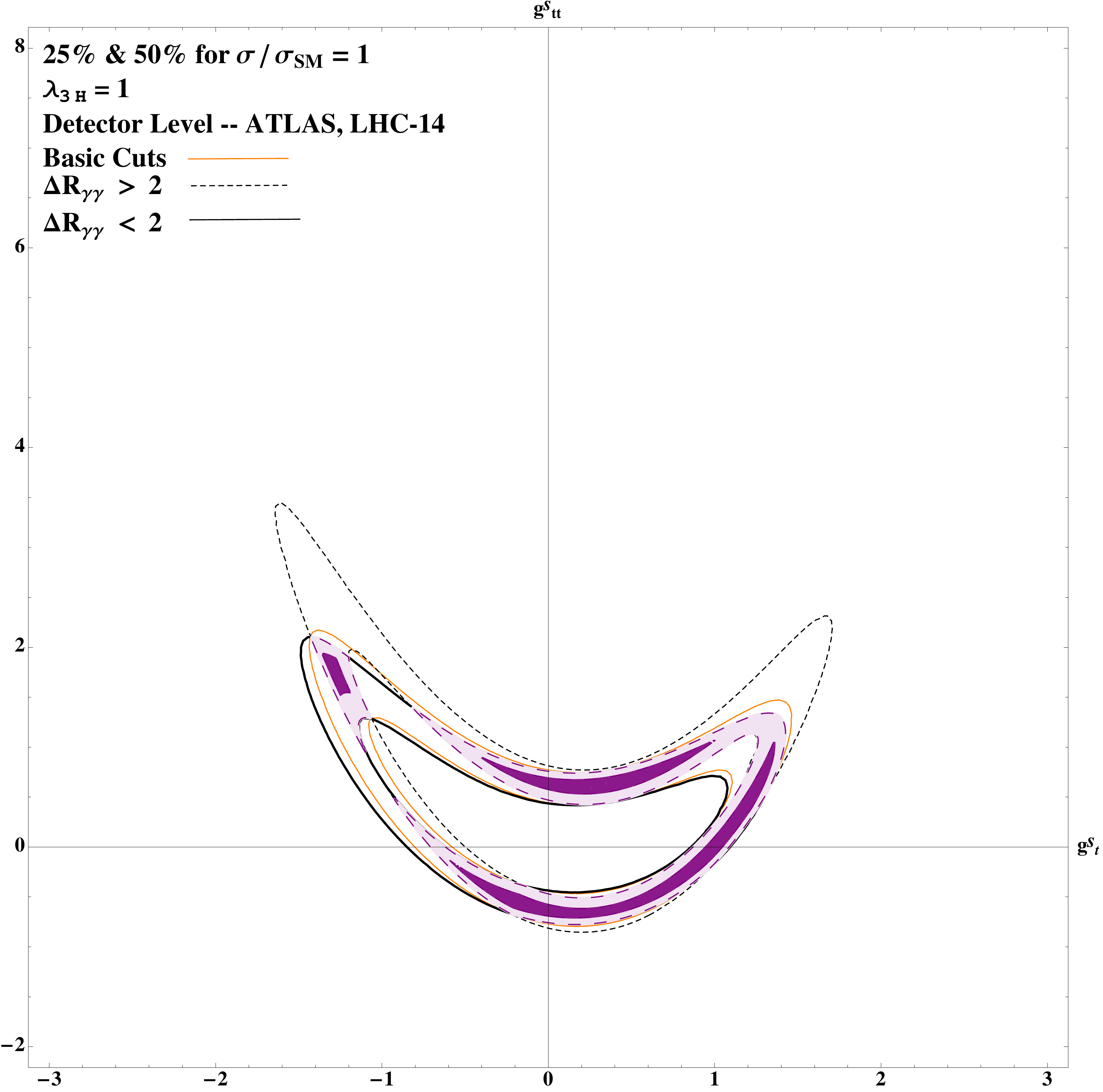}
\includegraphics[width=3in]{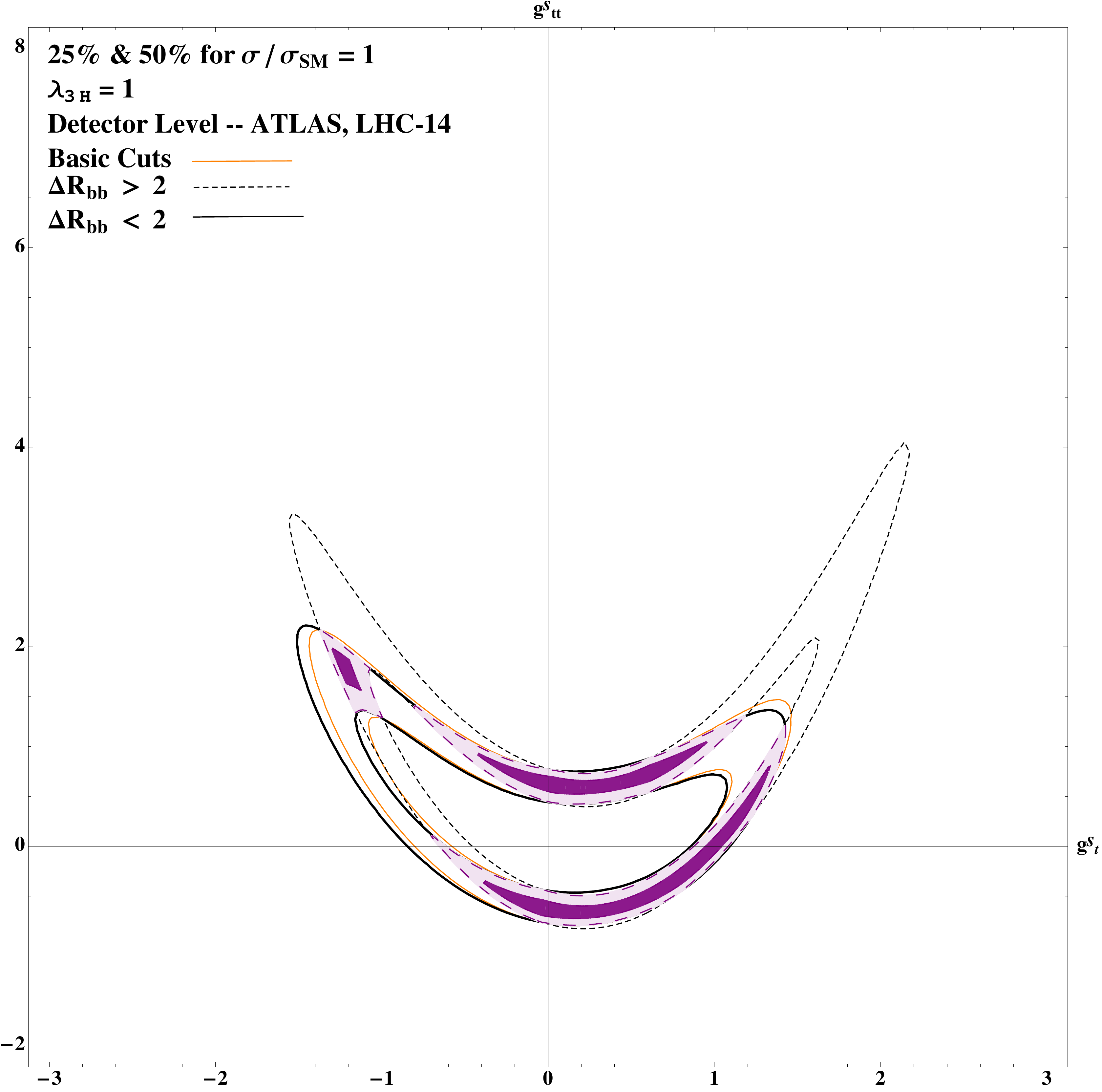}
\includegraphics[width=3in]{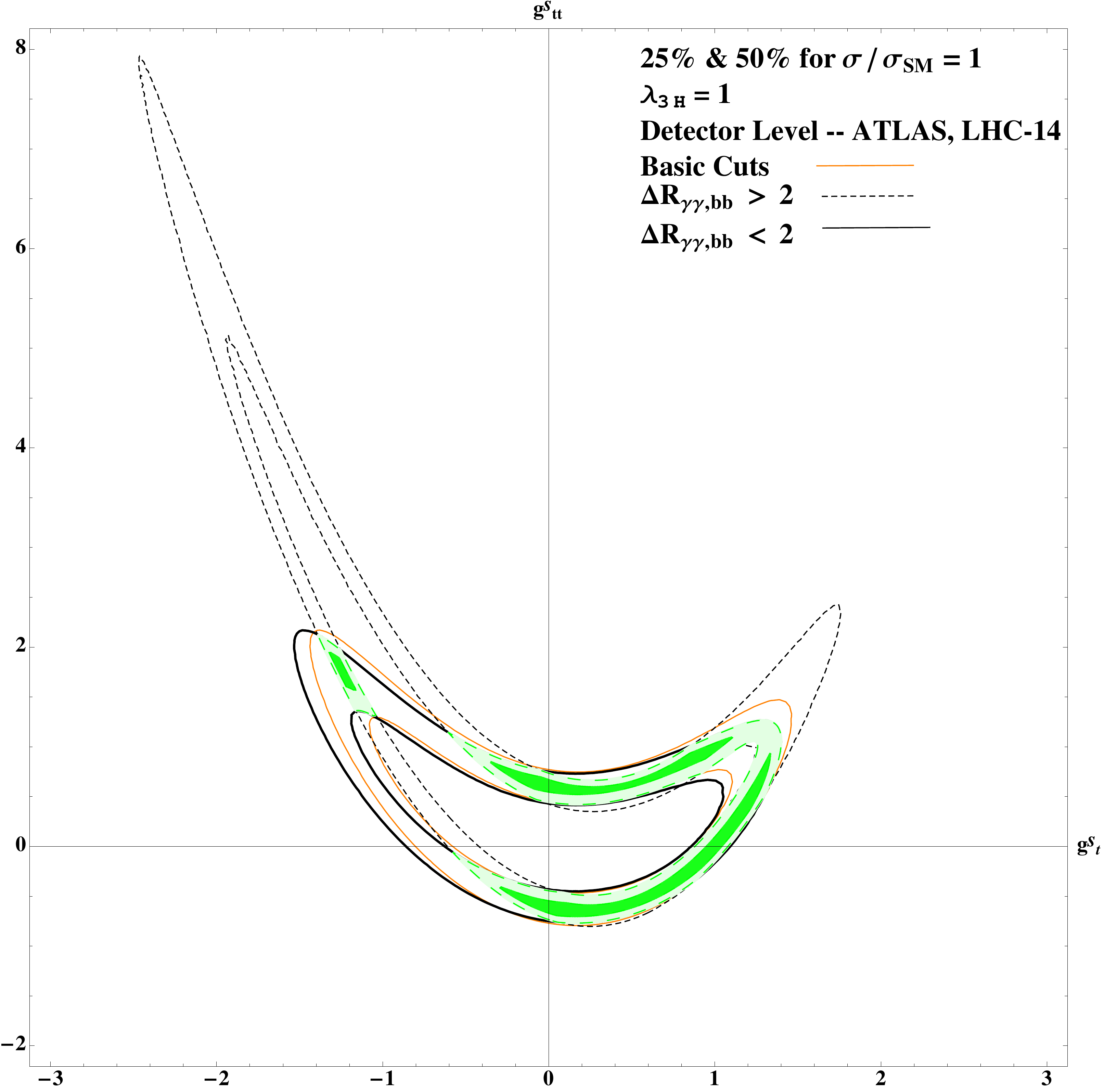}
\caption{\small \label{fig:cpc2-7}
{\bf CPC2:} similar to Fig.~\ref{fig:cpc2-5} but in the plane 
of $(g_t^S,\, g_{tt}^S)$.
}
\end{figure}

Similar to what we have done for {\bf CPC1}, we can make use of three 
simultaneous
measurements of cross sections with basic cuts,
 $\Delta R >2$, and $\Delta R < 2$. 
We show the region of parameter space that we can obtain using 
$\Delta R_{\gamma\gamma}$ (upper panels), $\Delta R_{b b}$ (middle panels), and
$\Delta R_{\gamma\gamma}$ and $\Delta R_{b b}$ (lower panels) 
in the plane of $(\lambda_{3H} ,\, g_t^S)$ in Fig.~\ref{fig:cpc2-5}.
Those on the left are for $g_{tt}^S=1$ while those on the right are for 
$g_{tt}^S=-1$.
Similarly, we show the parameter space in the plane of $(\lambda_{3H},\,g_t^S)$ 
in Fig.~\ref{fig:cpc2-6} and in the plane of $(g_t^S,\,g_{tt}^S)$ in
Fig.~\ref{fig:cpc2-7}.

\subsection{CPV1: $g_t^S$, $g_t^P$, and $\lambda_{3H}$}
In this scenario, we entertain the possibility that the top Yukawa
coupling allows an imaginary part. In most of the measurements of the
Higgs boson production cross sections, for example, Higgs boson
production cross section via gluon fusion and $t\bar tH$ production,
both the real and imaginary parts of the coupling come in the form
$|g_t^S|^2 + |g_t^P|^2$, therefore one cannot tell the phase in the
coupling
\footnote{
Single top plus Higgs production. on the other hand, has some chances to 
isolate the phase of the Yukawa coupling \cite{single}.  }.
The relevant coefficients for this {\bf CPV1} scenario are 
$c_1,c_2, c_3, d_1, d_2, d_3, d_4$. They are shown 
in Table~\ref{tab:all} at the
detector simulation level (ATLAS).

\begin{figure}[th!]
\centering
\includegraphics[width=5in]{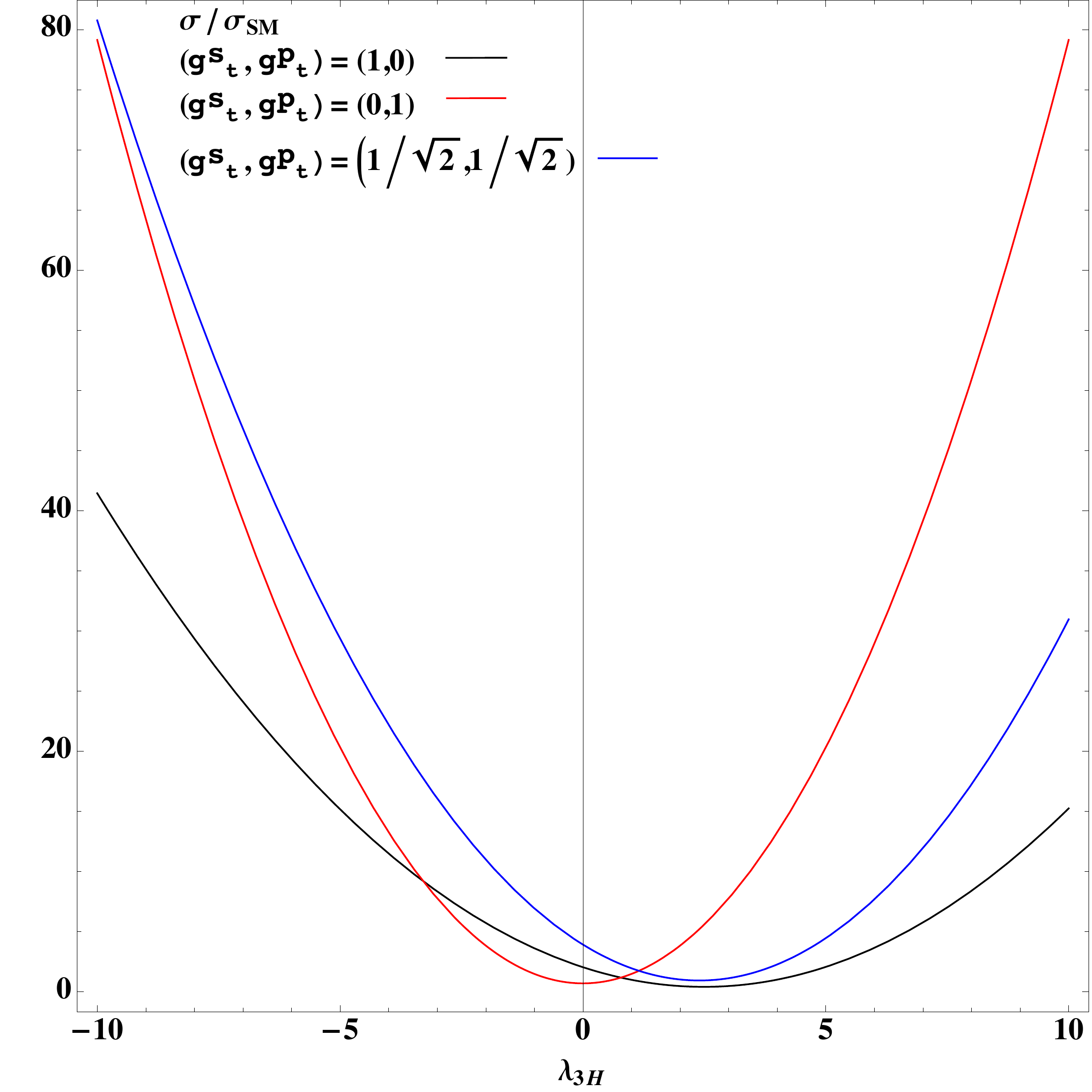}
\caption{\small \label{fig:cpv1-1}
{\bf CPV1:} The ratio $\sigma/\sigma_{\rm SM}$ versus 
$\lambda_{3H}$ for $(g_t^S,\,g_t^P)= (1,0),(0,1),(1/\sqrt{2},1/\sqrt{2})$
}
\end{figure}

\begin{figure}[th!]
\centering
\includegraphics[width=3.2in]{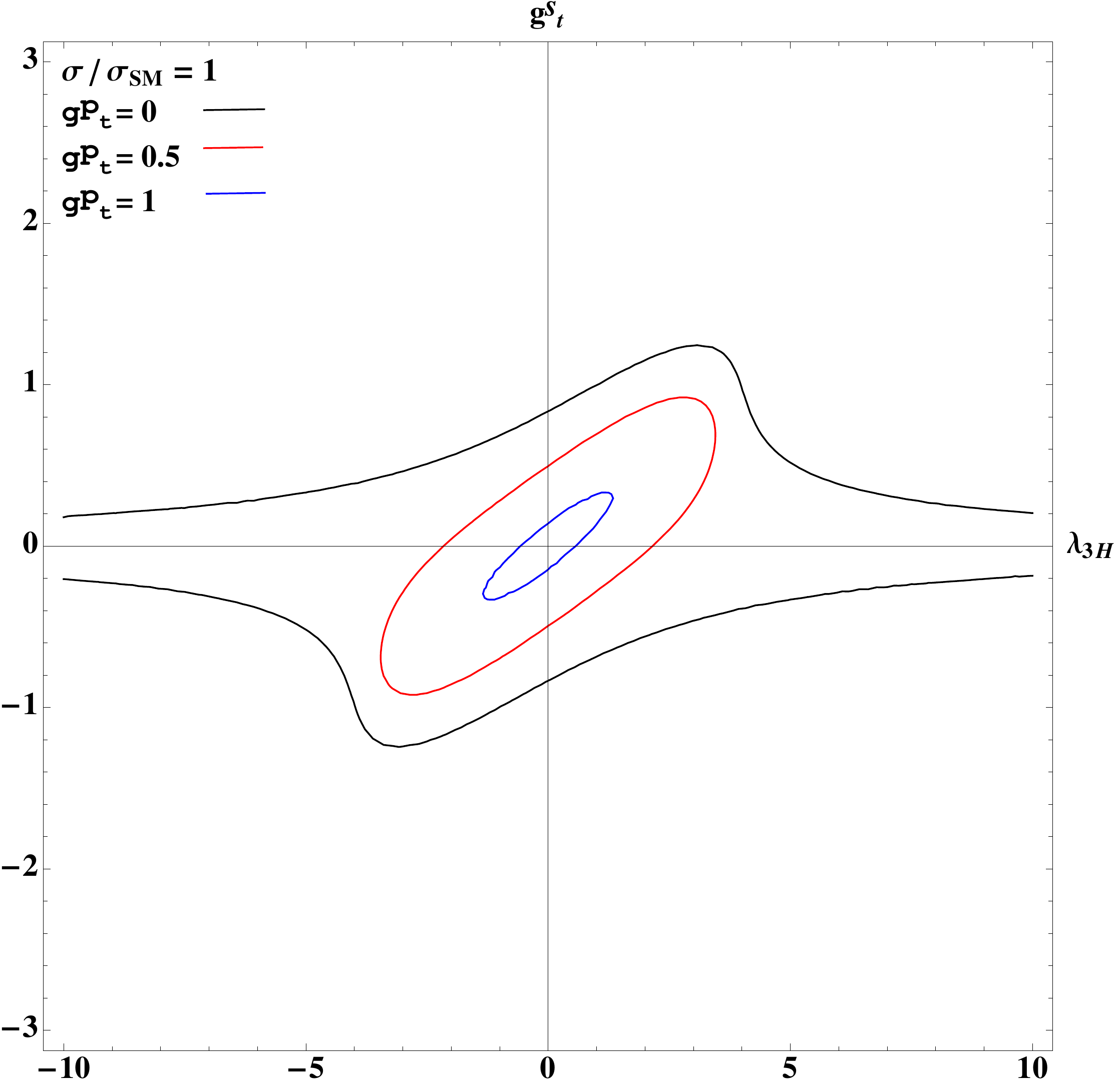}
\includegraphics[width=3.2in]{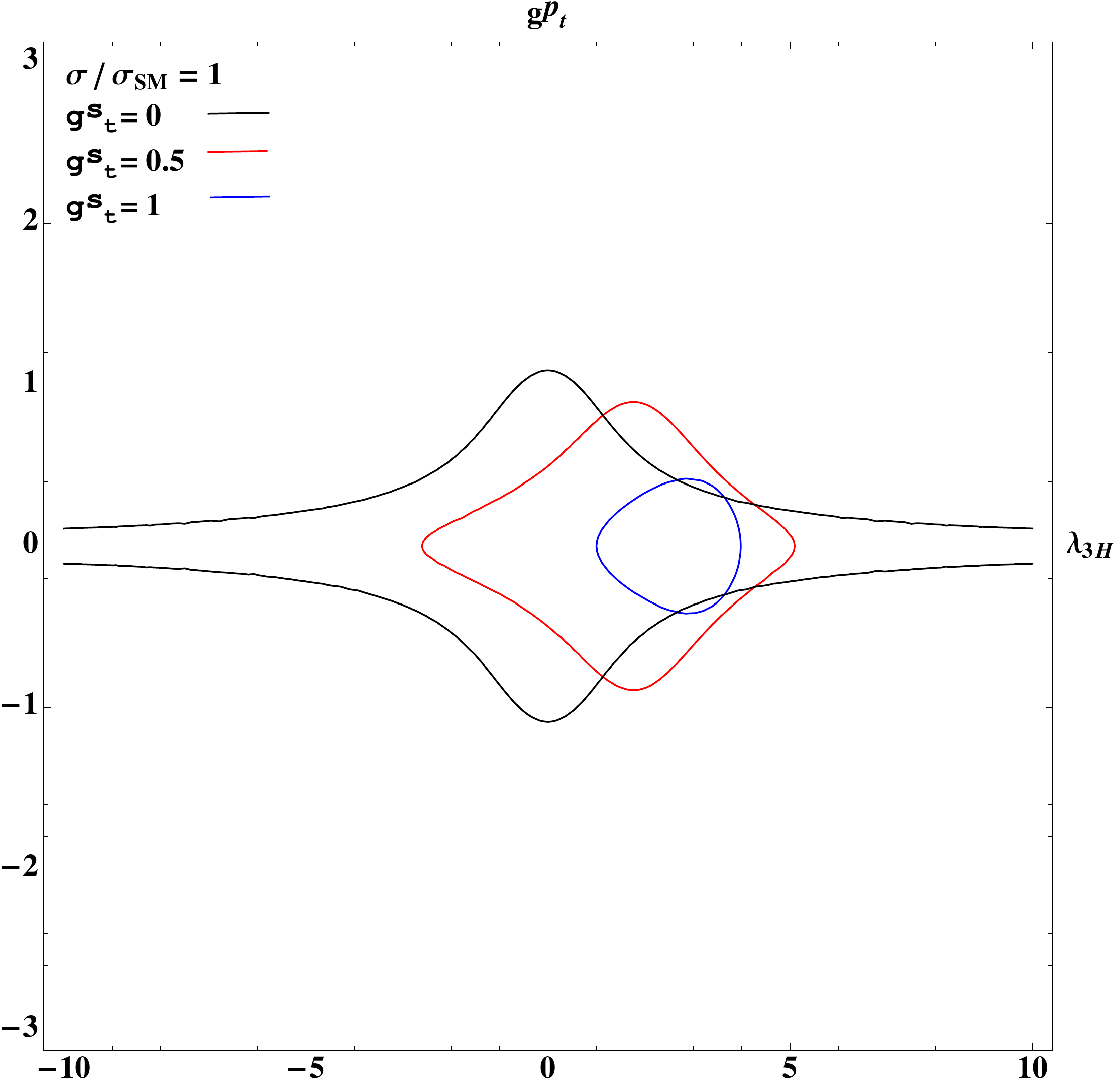}
\includegraphics[width=3.2in]{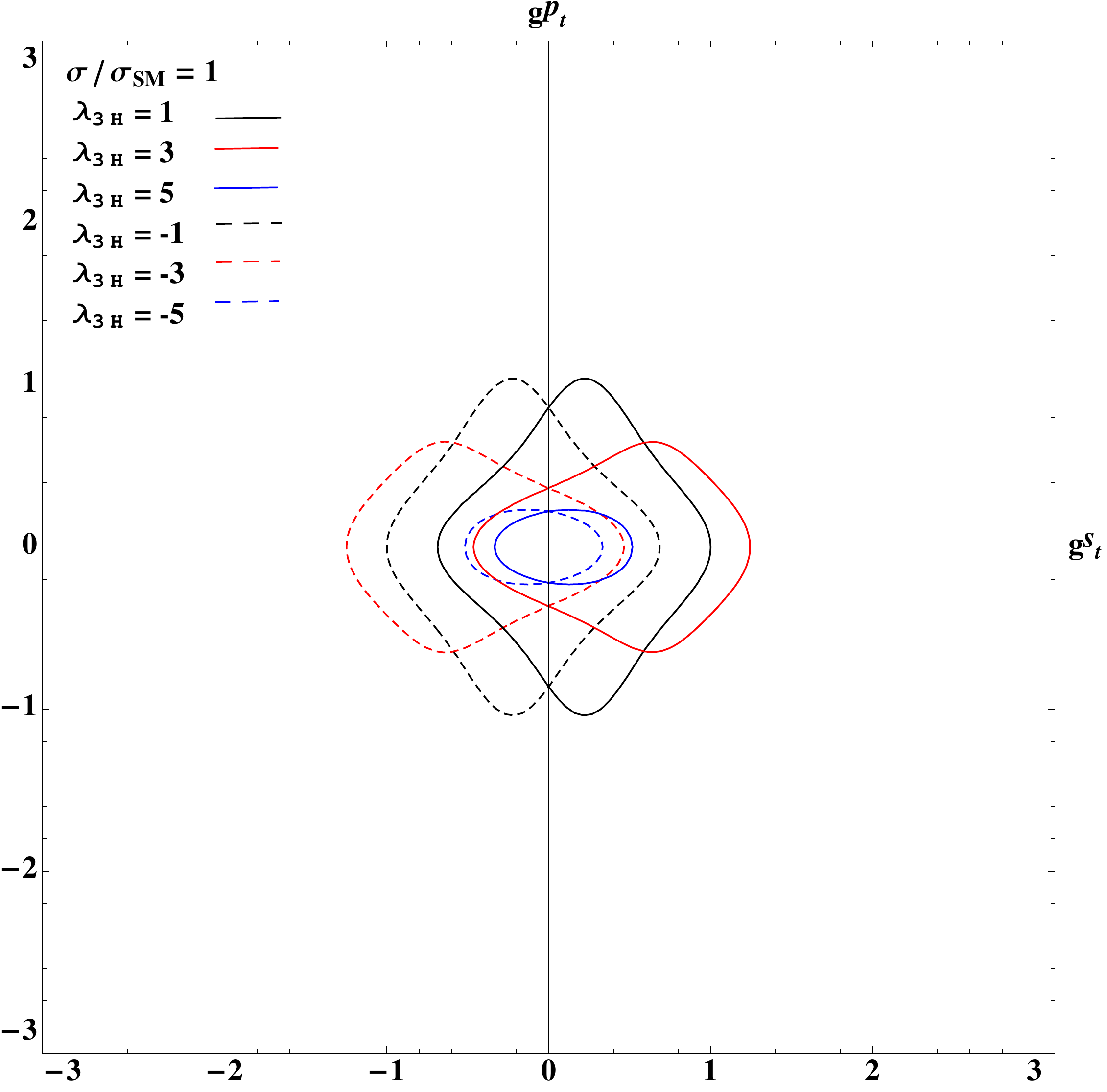}
\caption{\small \label{fig:cpv1-2}
{\bf CPV1:} Contours for the ratio $\sigma/\sigma_{\rm SM}=1$ in the plane of
$(\lambda_{3H},g_t^S)$ (upper-left), $(\lambda_{3H},\,g_t^P)$ (upper-right), and
$(g^S_t,g_t^P)$ (lower) for a few values of the third parameter.
}
\end{figure}

We first show the variations of cross sections versus $\lambda_{3H}$
with some fixed values of $g_t^S$ and $g_t^P$ in
Fig.~\ref{fig:cpv1-1}
\footnote{We note that the lines are the same for the 
negative values of $g^P_t$ since the cross section contains only the 
$(g^P_t)^2$ and $(g^P_t)^4$ terms.}.
Also, the contours for the ratio
$\sigma/\sigma_{\rm SM}=1$ in the plane of 
$(\lambda_{3H},\,g_t^S)$
(upper-left), $(\lambda_{3H},\,g_t^P)$ (upper-right), and
$(g_t^S,\,g_t^P)$ (lower) for a few values of the third
parameter are shown in Fig.~\ref{fig:cpv1-2}.

\begin{figure}[th!]
\centering
\includegraphics[width=5.2in]{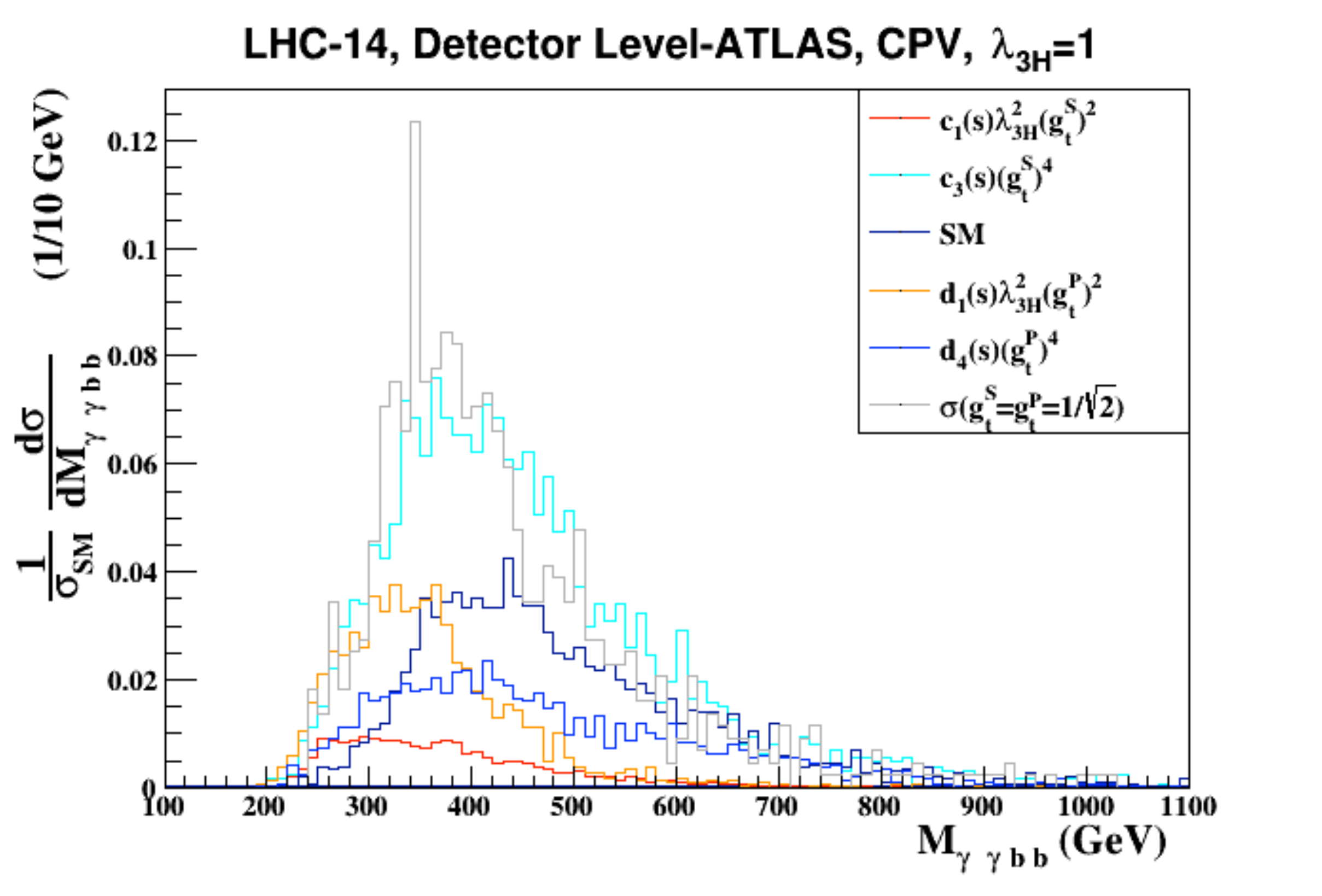}
\includegraphics[width=5.2in]{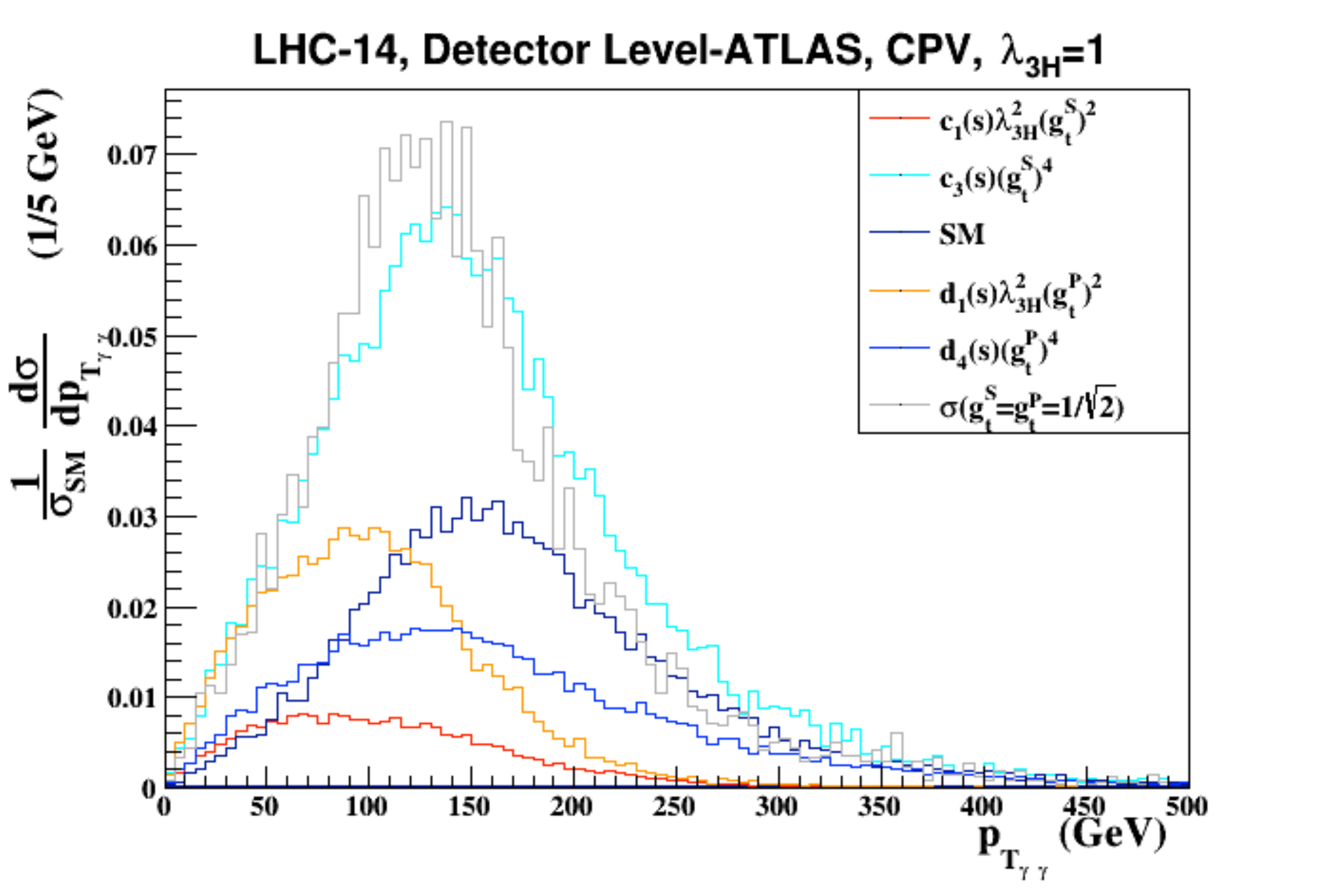}
\caption{\small \label{fig:cpv1-m}
{\bf CPV1:} 
Invariant mass distribution  $M_{\gamma\gamma b b}$ and 
$p_{T_{\gamma\gamma}}$ for the decay products of the Higgs-boson pair.
}
\end{figure}

\begin{figure}[th!]
\centering
\includegraphics[width=5.2in]{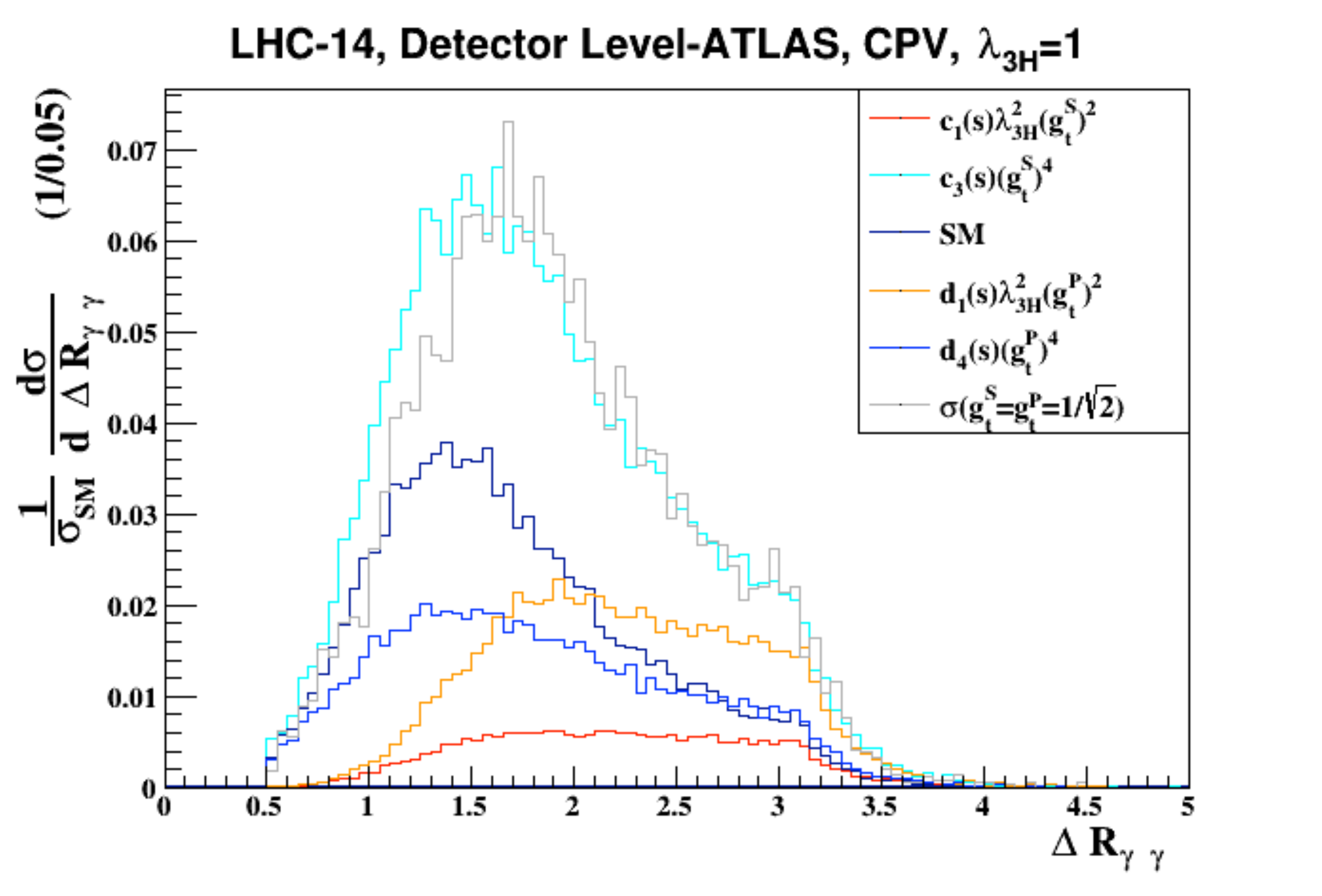}
\includegraphics[width=5.2in]{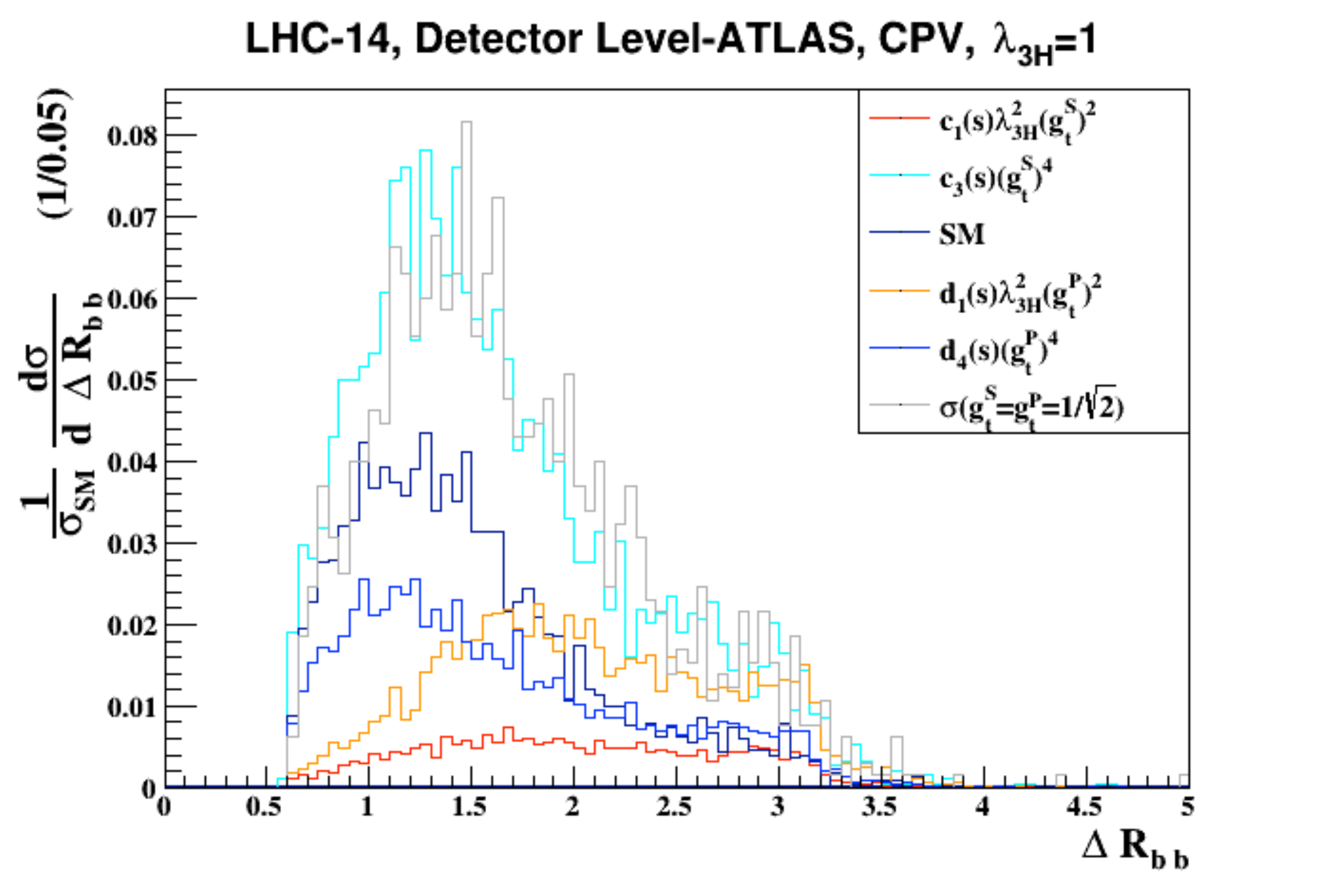}
\caption{\small \label{fig:cpv1-3}
{\bf CPV1:} 
Angular distributions of $\Delta R_{\gamma\gamma}$ and $\Delta R_{b b}$ 
between the two photons and between the two $b$ quarks 
}
\end{figure}

Similar to previous two scenarios, we use the same tools to analyze the 
decay channel $HH \to \gamma\gamma b \bar b$ with parton showering and 
detector simulations.
We show the invariant mass $M_{\gamma\gamma b \bar b}$ and $p_{T_{\gamma\gamma}}$
in Fig.~\ref{fig:cpv1-m}, and 
the angular distributions 
$\Delta R_{\gamma\gamma}$ and $\Delta R_{b b}$ 
between the two decay products of each Higgs boson in Fig.~\ref{fig:cpv1-3}.

The terms by the triangle diagram (proportional to $c_1$ and $d_1$
in red and orange lines, respectively)
give the widest separation among all the terms. 
The terms by the box diagram (proportional to $c_3$ and $d_4$
in skyblue and blue lines, respectively) 
give smaller angular separation. The full set of diagrams at the SM values
(darkblue lines)
and at $g_t^S = g_t^P = 1/\sqrt{2}$ (grey lines) 
give similar results as the box diagram.

\begin{figure}[th!]
\centering
\includegraphics[width=3in]{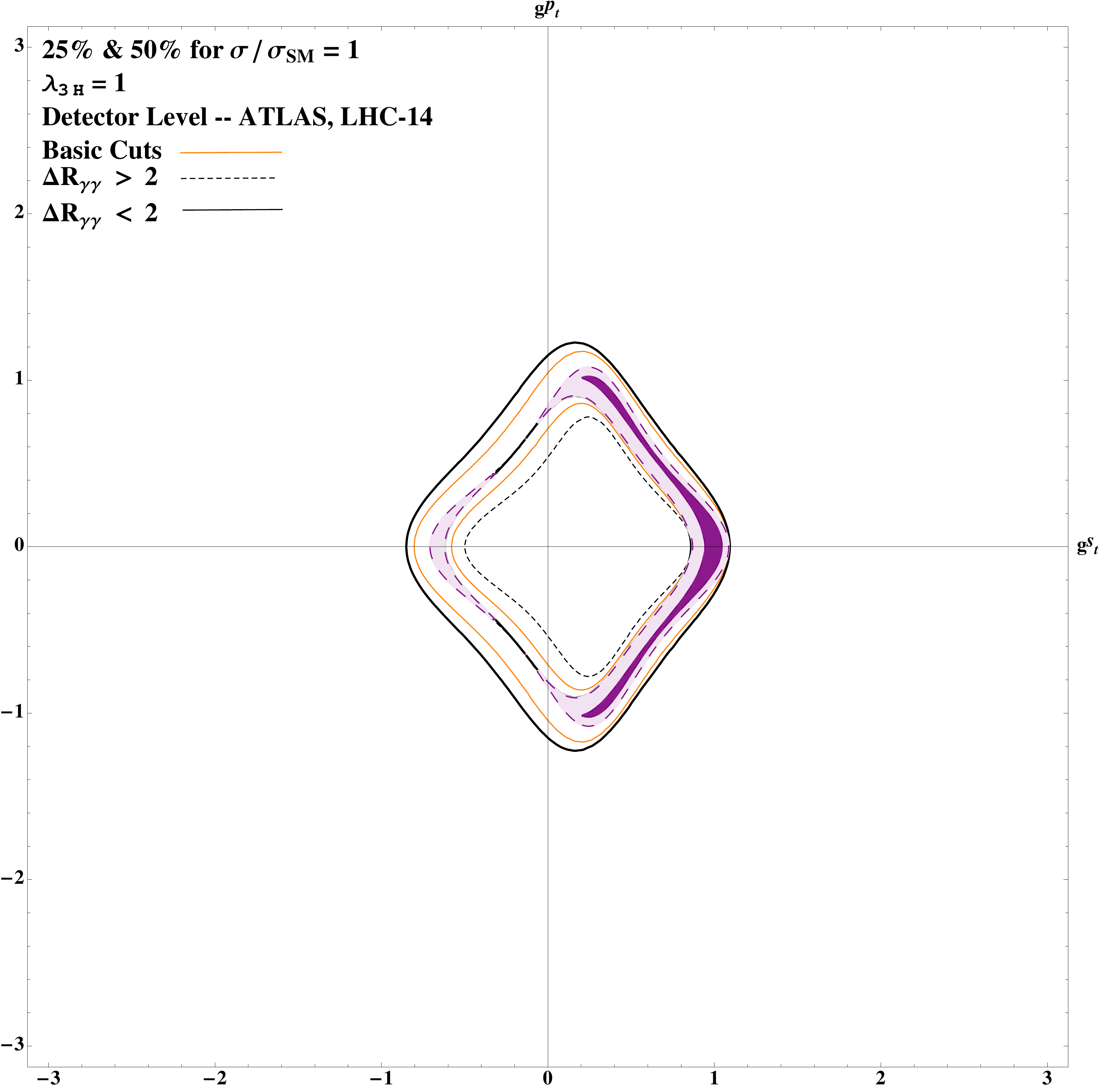}
\includegraphics[width=3in]{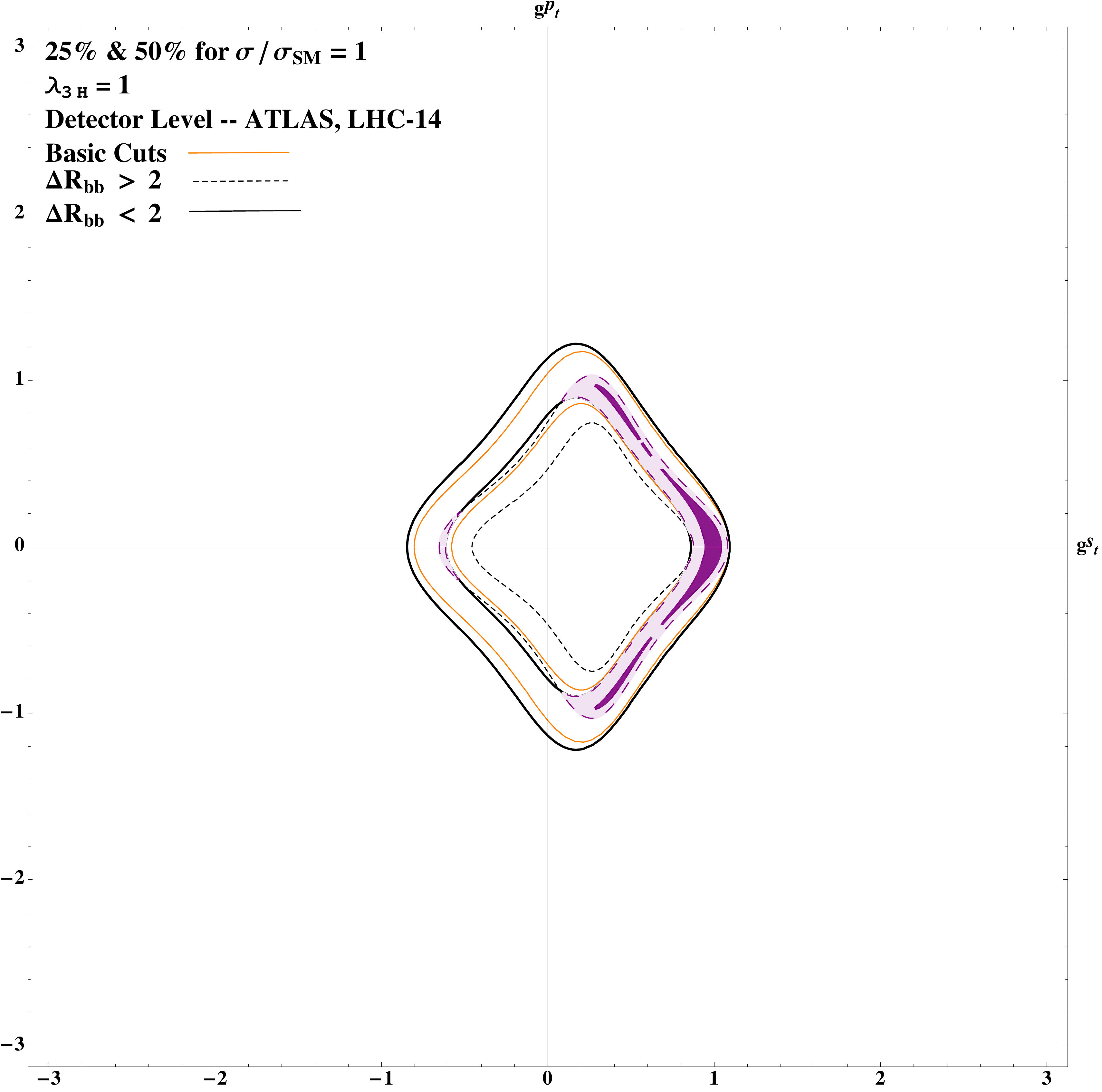}
\includegraphics[width=3in]{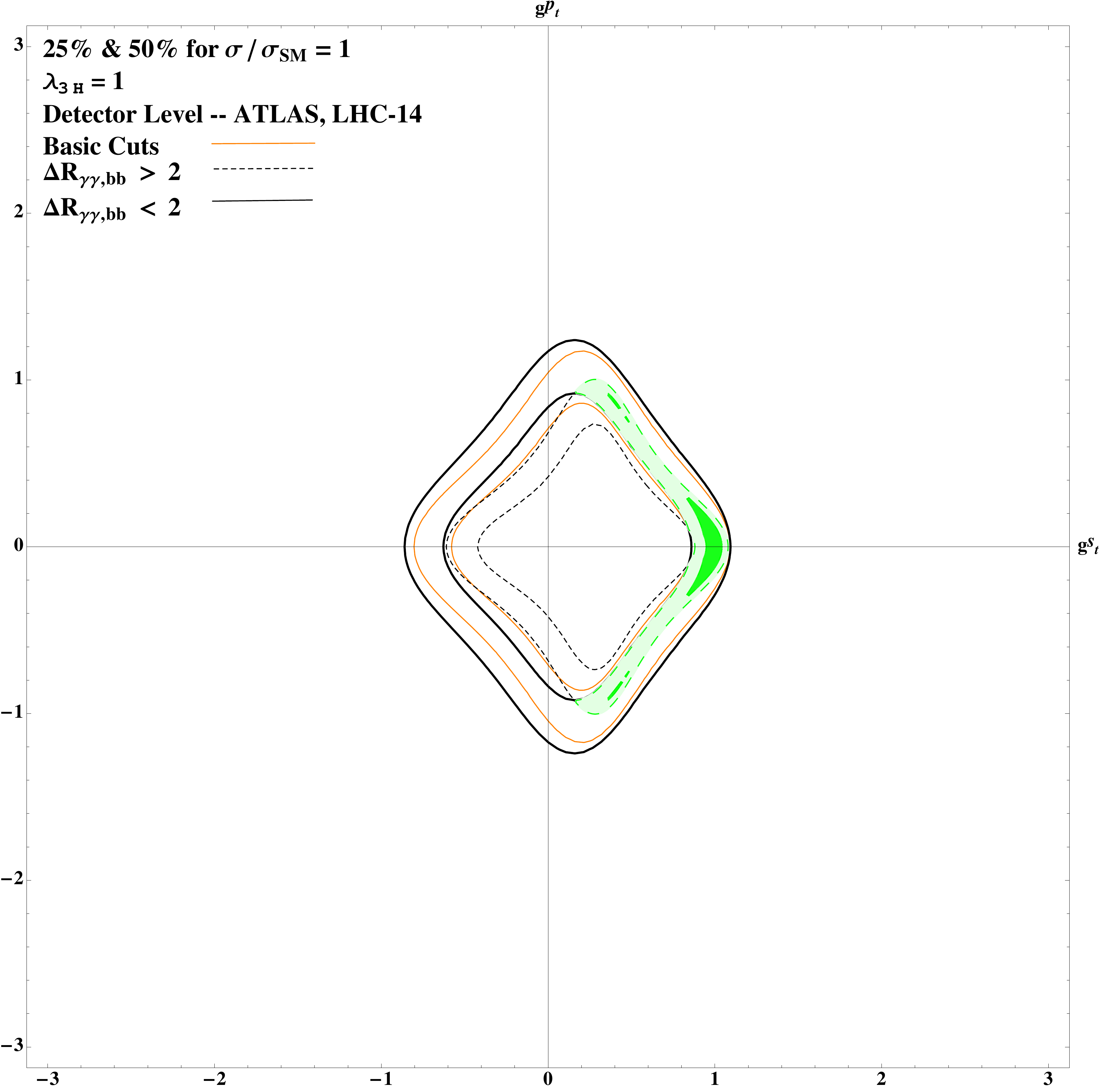}
\caption{\small \label{fig:cpv1-4}
{\bf CPV1:} The 25\% and 50\% sensitivity regions in the plane 
$(g_t^S,\, g_t^P)$ bounded by three
measurements of cross sections with basic cuts, 
$\Delta R_{\gamma\gamma} > 2$, and 
$\Delta R_{\gamma\gamma} < 2$ (the upper-left panel);
with basic cuts, 
$\Delta R_{b b} > 2$, and 
$\Delta R_{b b } < 2$ (the upper-right panel); and
basic cuts,
$\Delta R_{\gamma\gamma},\,\Delta R_{b b} > 2$, and 
$\Delta R_{\gamma\gamma},\,\Delta R_{b b} < 2$ (the lower panel).
We assume that the measurements agree with the SM values with 
uncertainties of 25\% and 50\%, respectively.
}
\end{figure}

\begin{figure}[th!]
\centering
\includegraphics[width=3in]{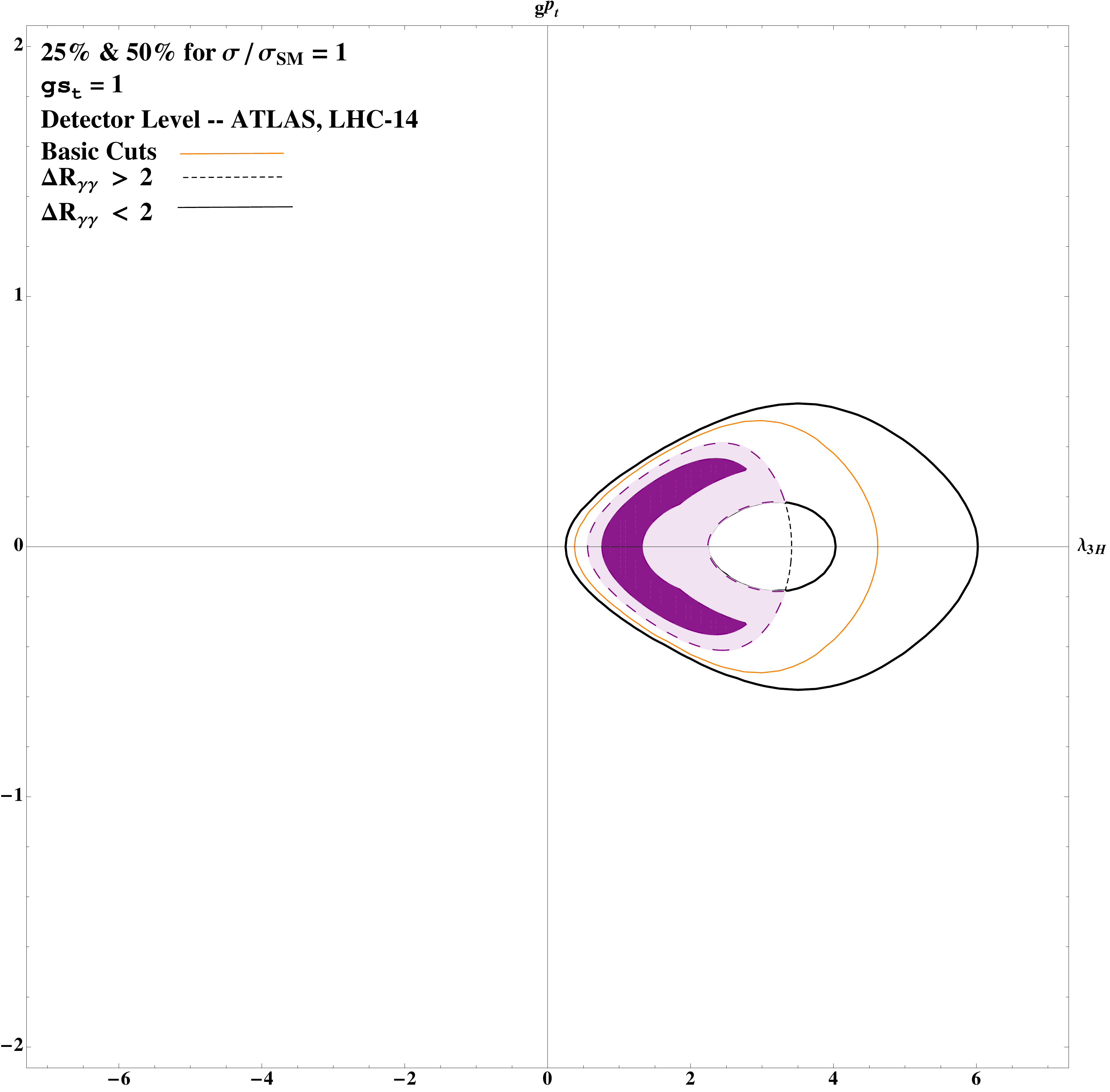}
\includegraphics[width=3in]{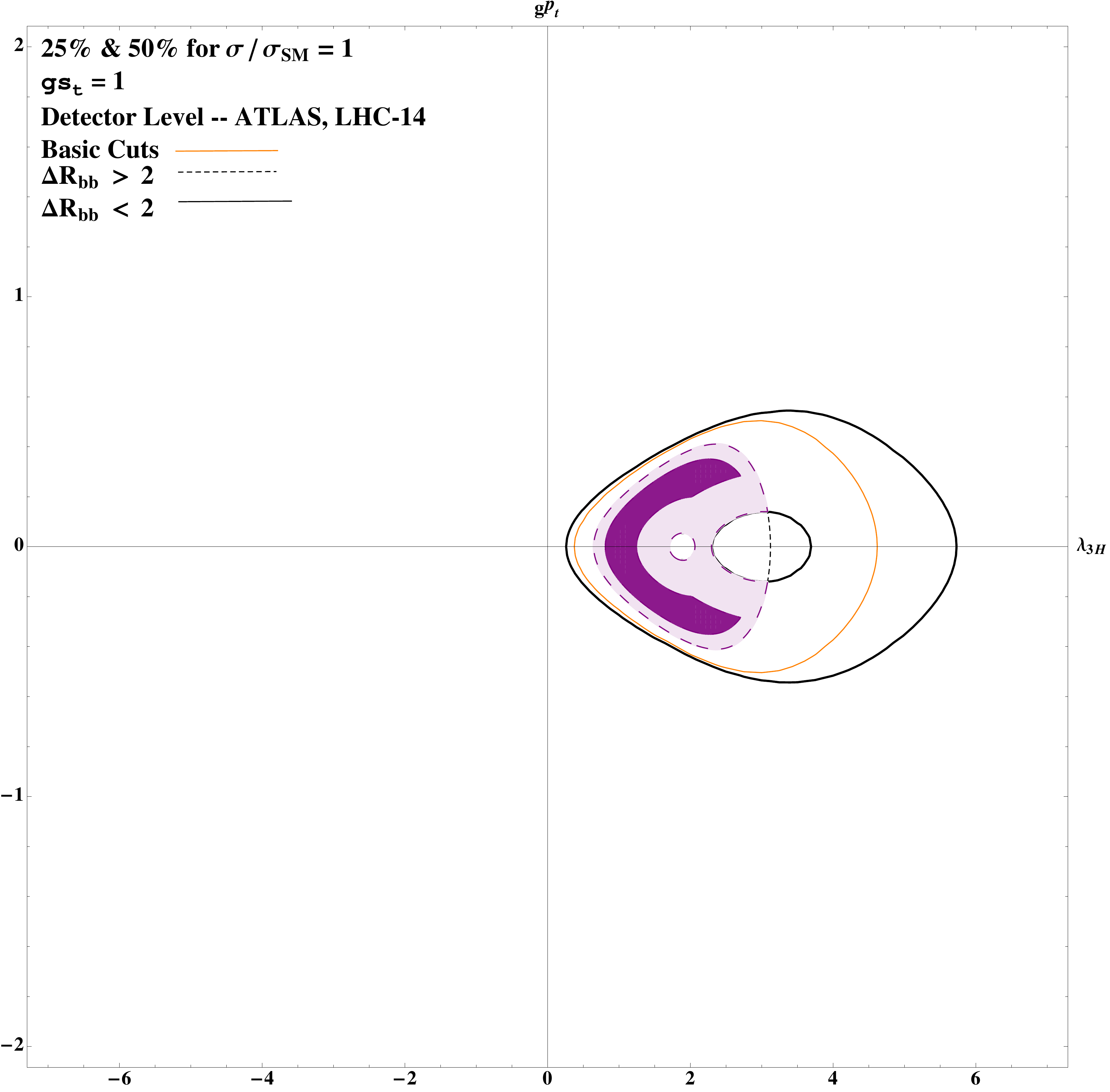}
\includegraphics[width=3in]{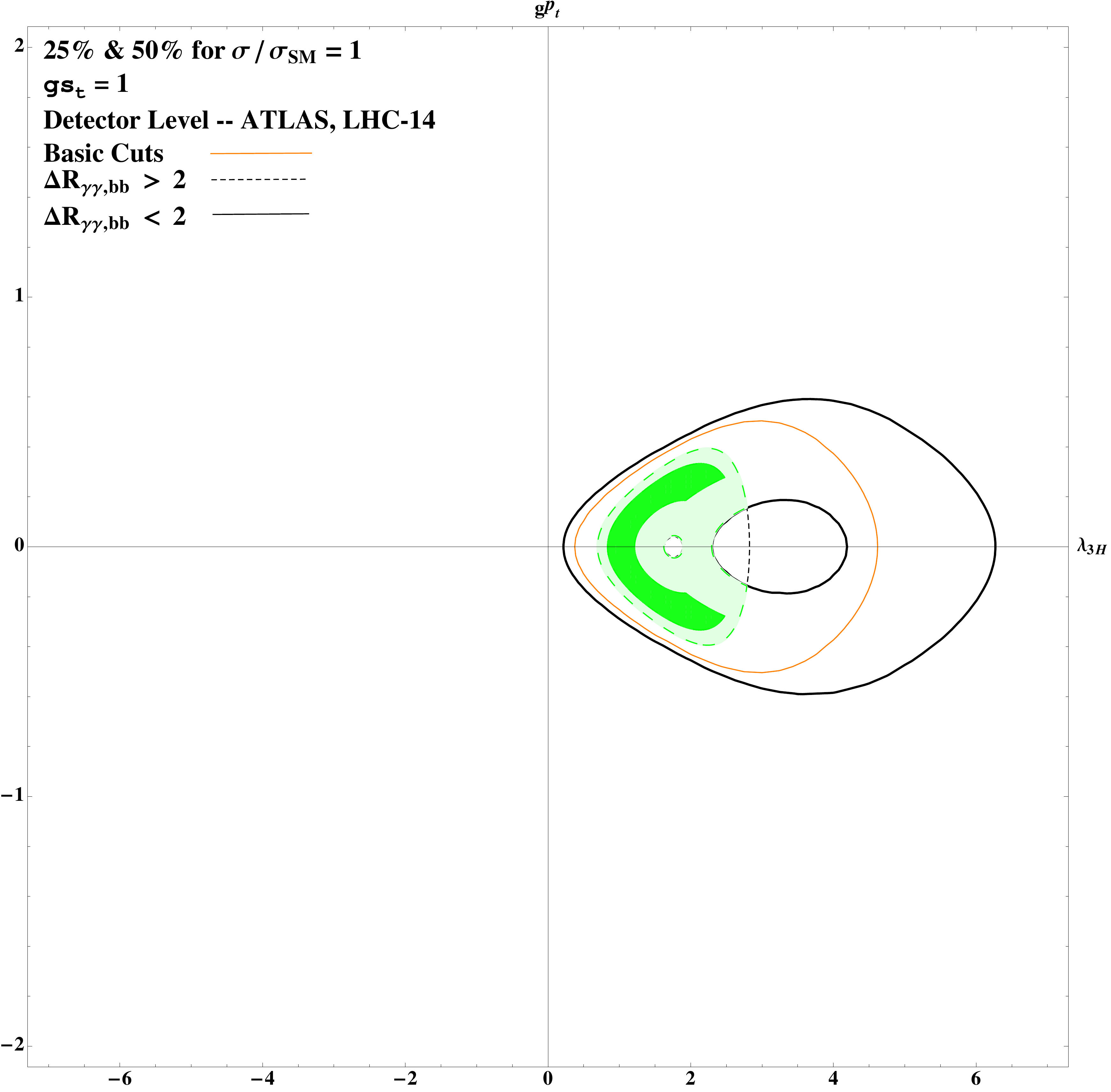}
\caption{\small \label{fig:cpv1-5}
{\bf CPV1:} similar to Fig.~\ref{fig:cpv1-4} but in the plane of
$(\lambda_{3H},\,g_t^P)$.
}
\end{figure}

\begin{figure}[th!]
\centering
\includegraphics[width=3in]{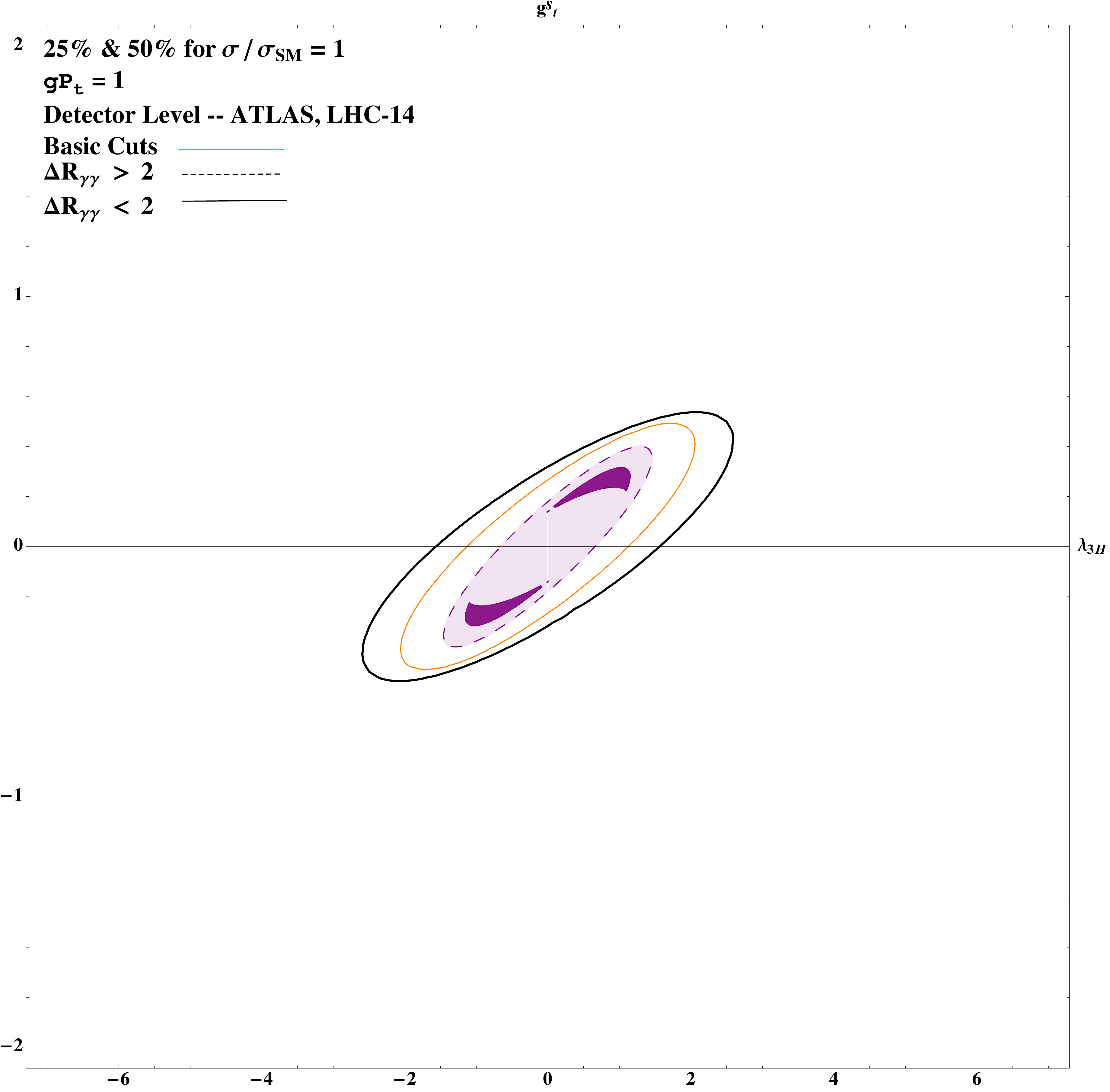}
\includegraphics[width=3in]{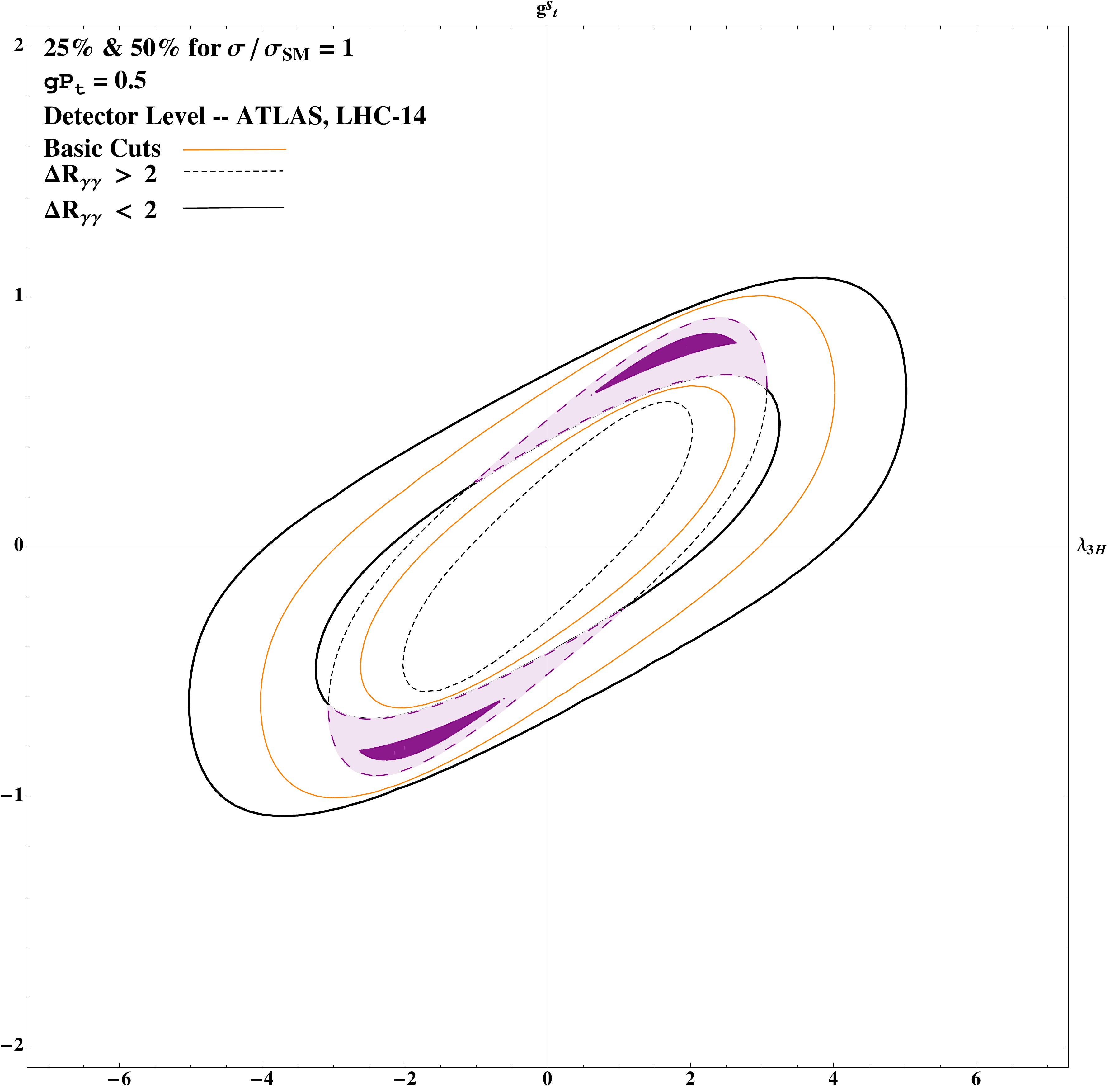}
\includegraphics[width=3in]{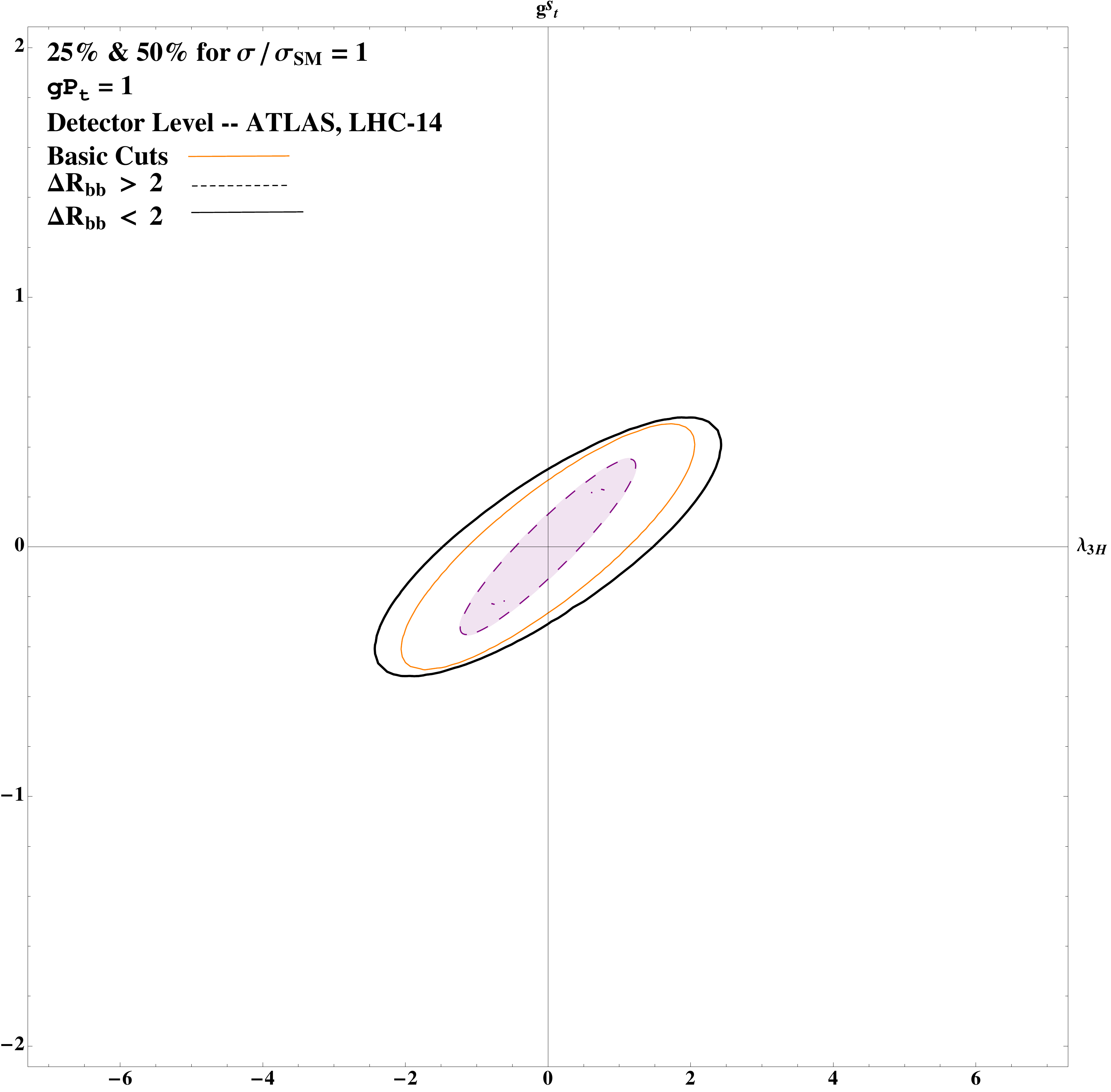}
\includegraphics[width=3in]{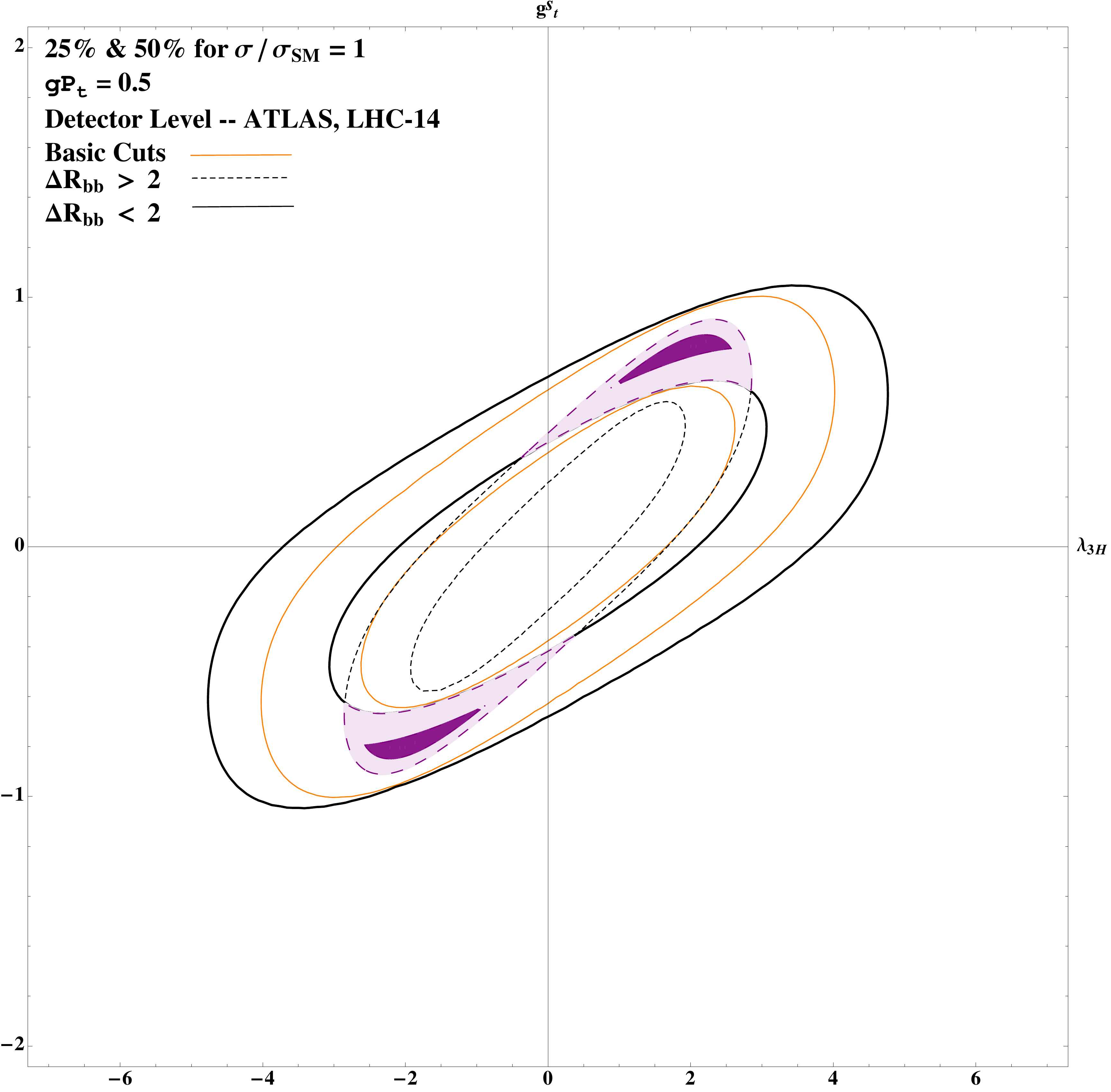}
\includegraphics[width=3in]{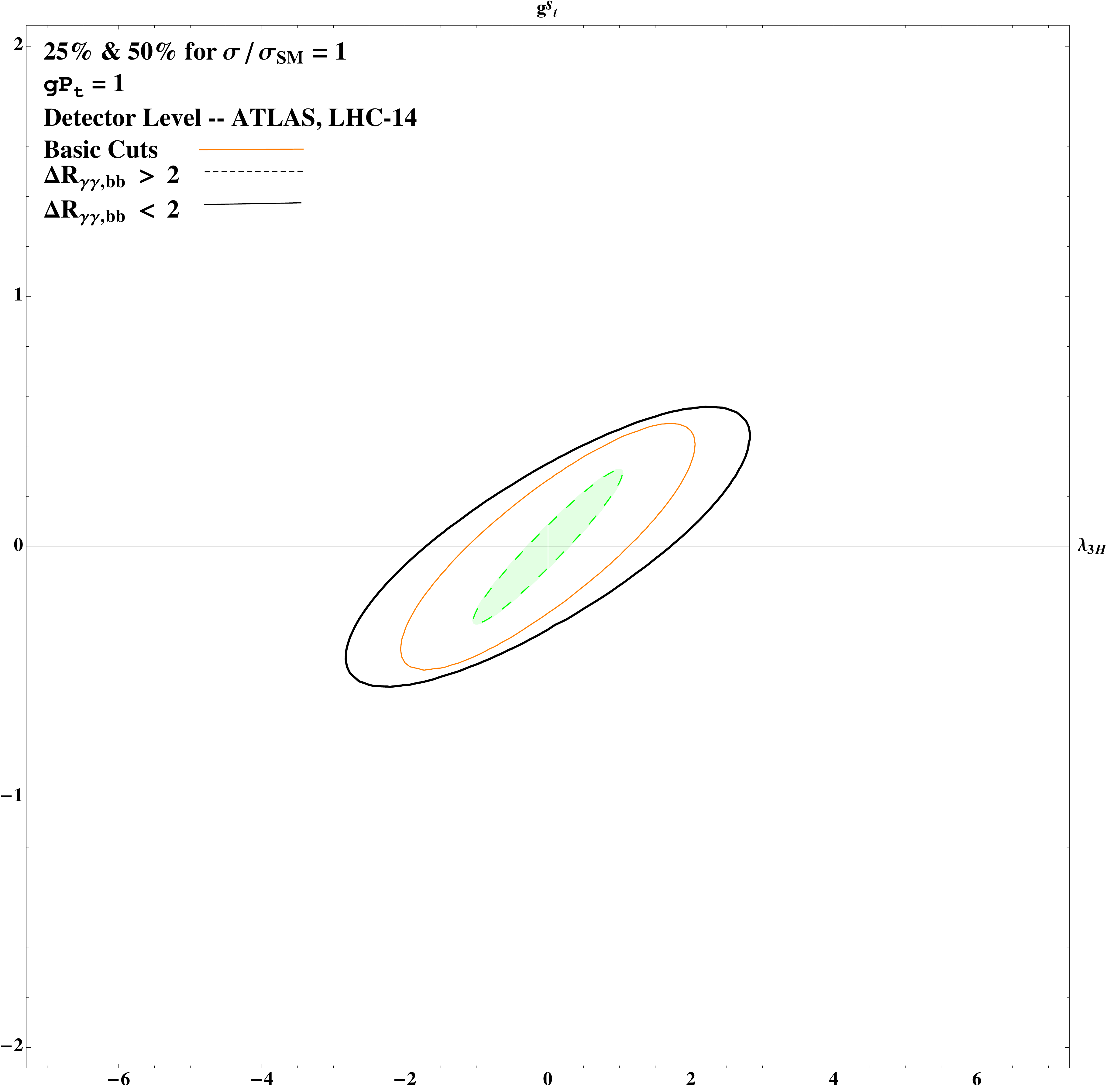}
\includegraphics[width=3in]{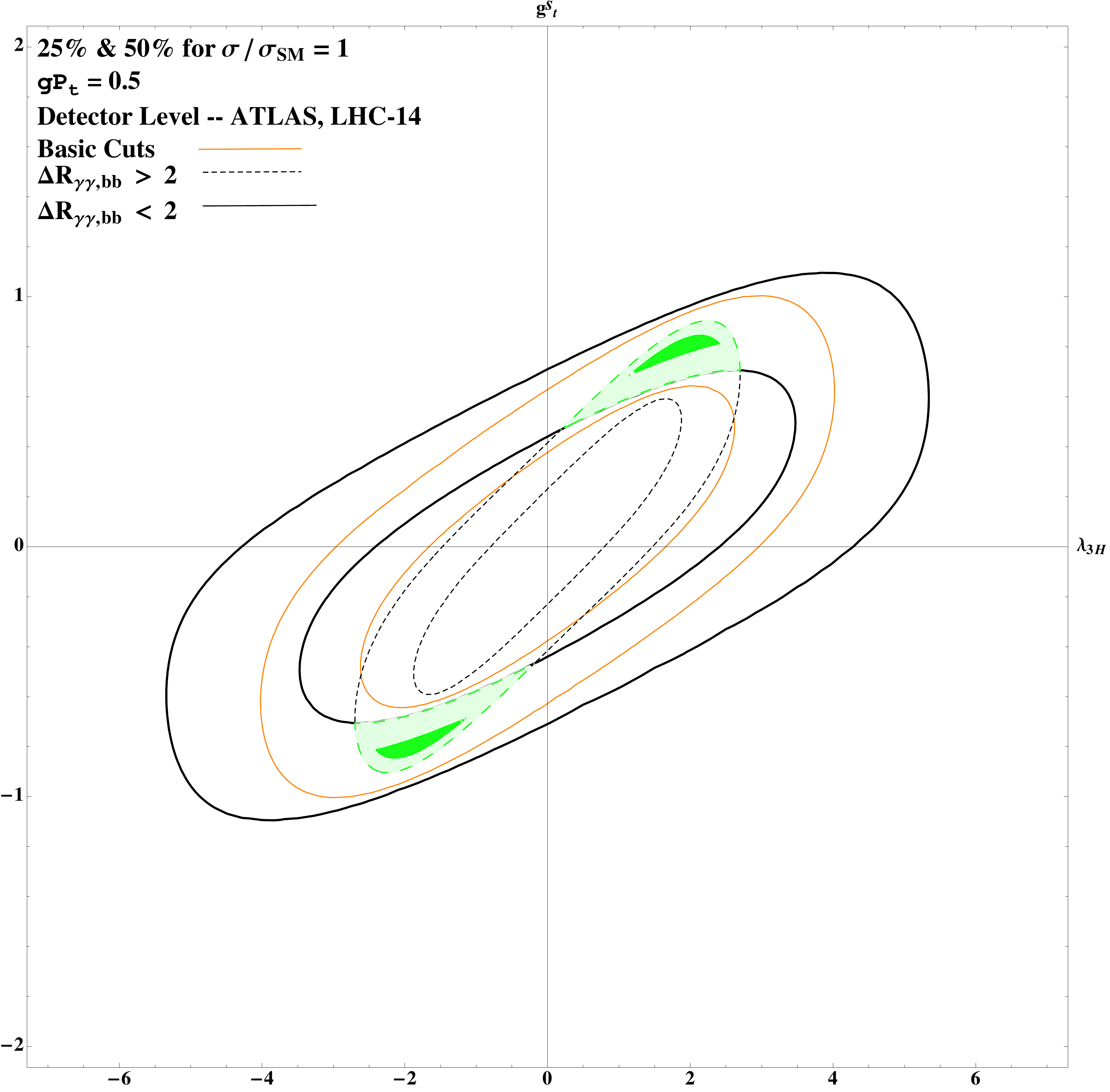}
\caption{\small \label{fig:cpv1-6}
{\bf CPV1:} similar to Fig.~\ref{fig:cpv1-4} but in the plane of
$(\lambda_{3H},\,g_t^S)$. The left panels are for $g_t^P=1$ while those
on the right are for $g_t^P=0.5$.
}
\end{figure}

Similar to the {\bf CPC1} and {\bf CPC2} cases, we use the cuts
$\Delta R$ larger or smaller than $2$ to discriminate the triangle and box
diagrams. We show in Table~\ref{tab:all} the coefficients 
$c_1,\, c_2,\, c_3,\, d_1,\, d_2,\, d_3, d_4$, which are relevant ones in the 
{\bf CPV1}
scenario, such that the ratio of cross sections to the
SM predictions can be obtained by Eq.~(\ref{cdef}).
We show the region of parameter space that we can obtain using 
$\Delta R_{\gamma\gamma}$, $\Delta R_{b b}$, and
$\Delta R_{\gamma\gamma}$ and $\Delta R_{b b}$
in the plane of $(g_t^S,\, g_t^P)$ in Fig.~\ref{fig:cpv1-4},
in the plane of $(\lambda_{3H},\, g_t^P)$ in Fig.~\ref{fig:cpv1-5}, and 
in the plane of $(\lambda_{3H},\, g_t^S)$ in Fig.~\ref{fig:cpv1-6}.

\subsection{CPV2: $g_{tt}^S$, $g_{tt}^P$, and $\lambda_{3H}$}
Here we study another CP-violating scenario with the CP-even 
and CP-odd components of the $ttHH$ coupling with the top-Yukawa
couplings $g_t^S$ and $g_t^P$  at fixed values.
Note that the CP-odd coupling $g_{tt}^P$ only appears in the product 
with $g_t^P$ or by itself squared. 
In this case, all the coefficients are relevant and they
are
shown in Table~\ref{tab:all} at the detector simulation level (ATLAS).

\begin{figure}[th!]
\centering
\includegraphics[width=3in]{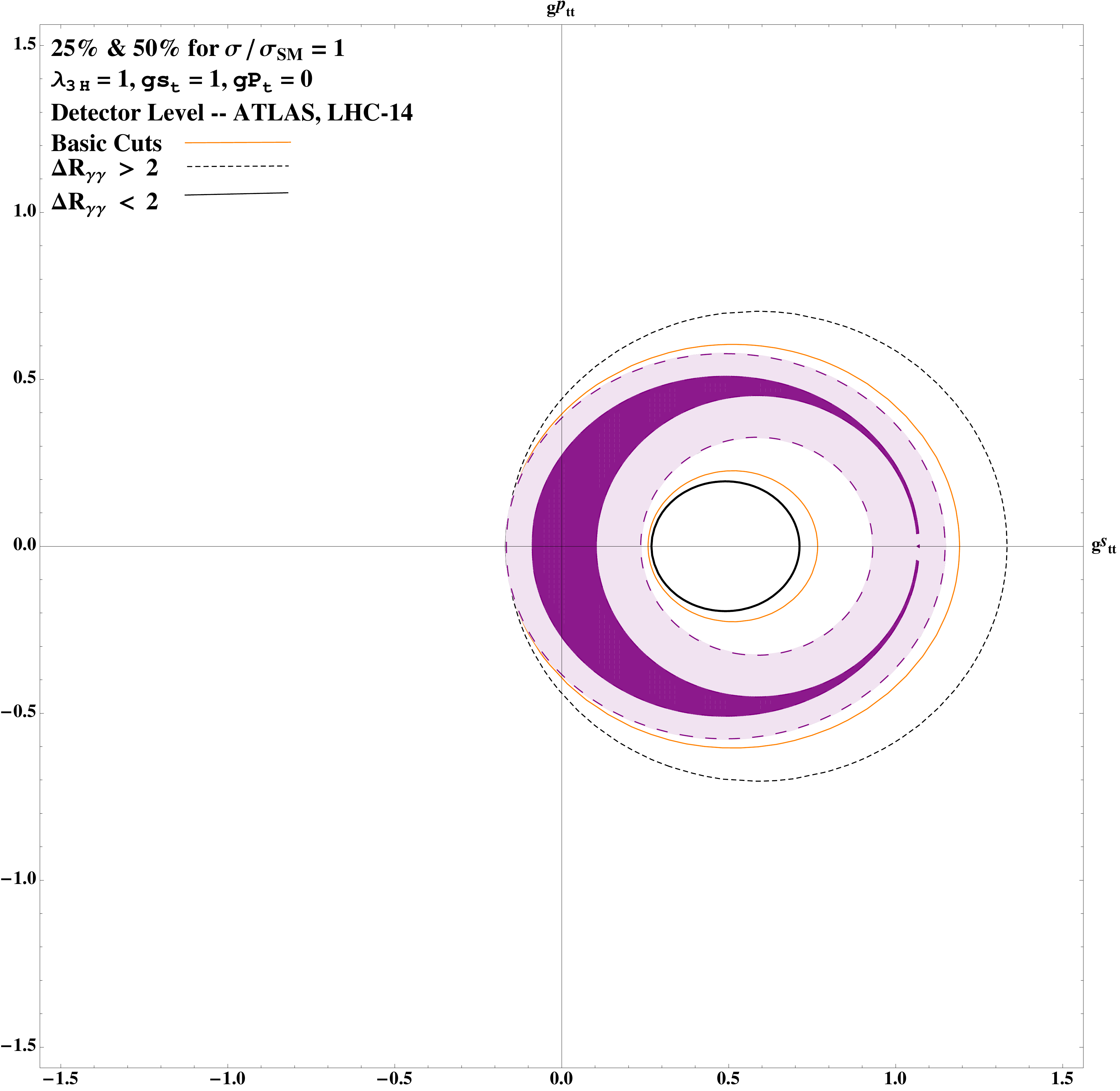}
\includegraphics[width=3in]{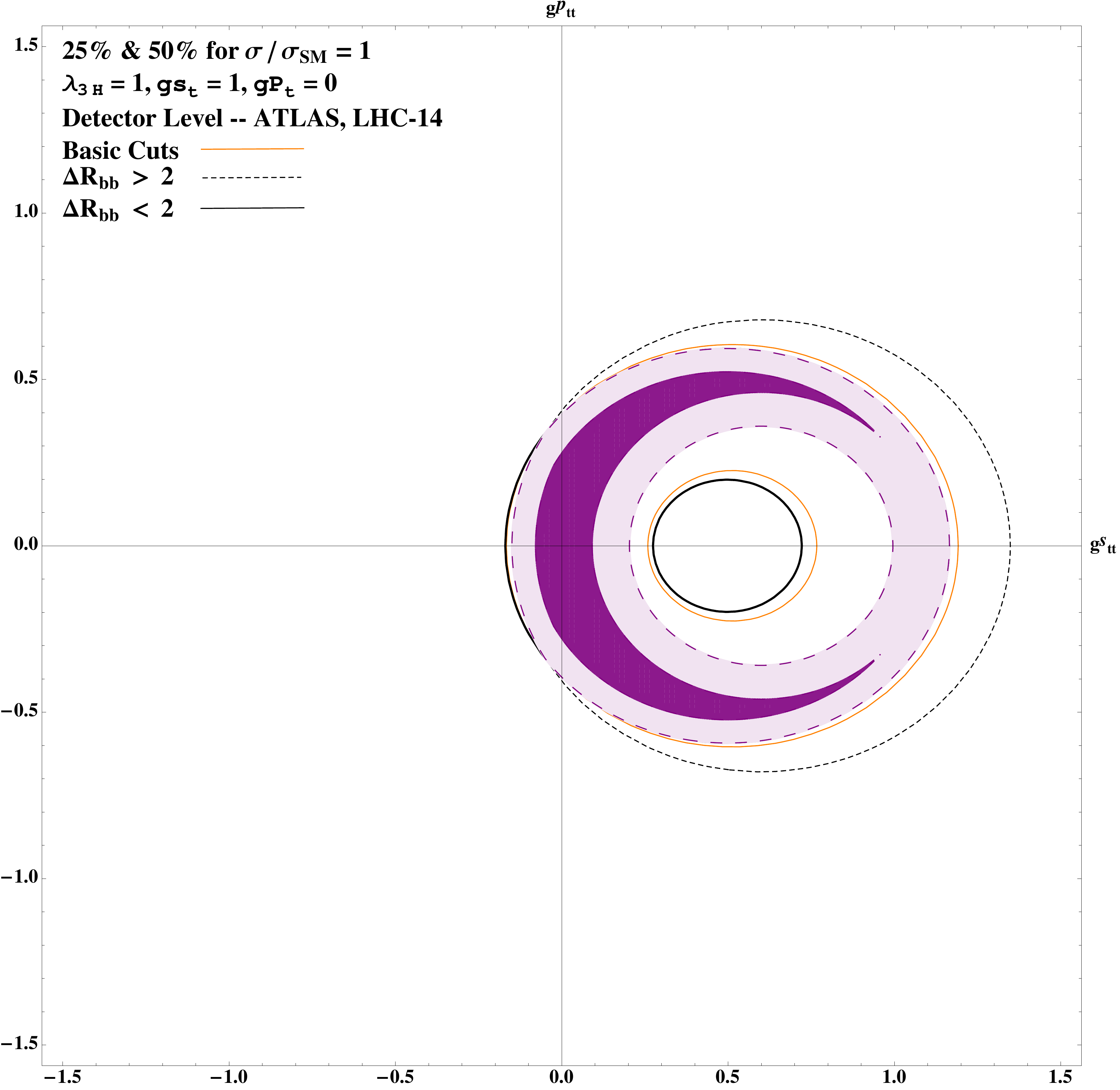}
\includegraphics[width=3in]{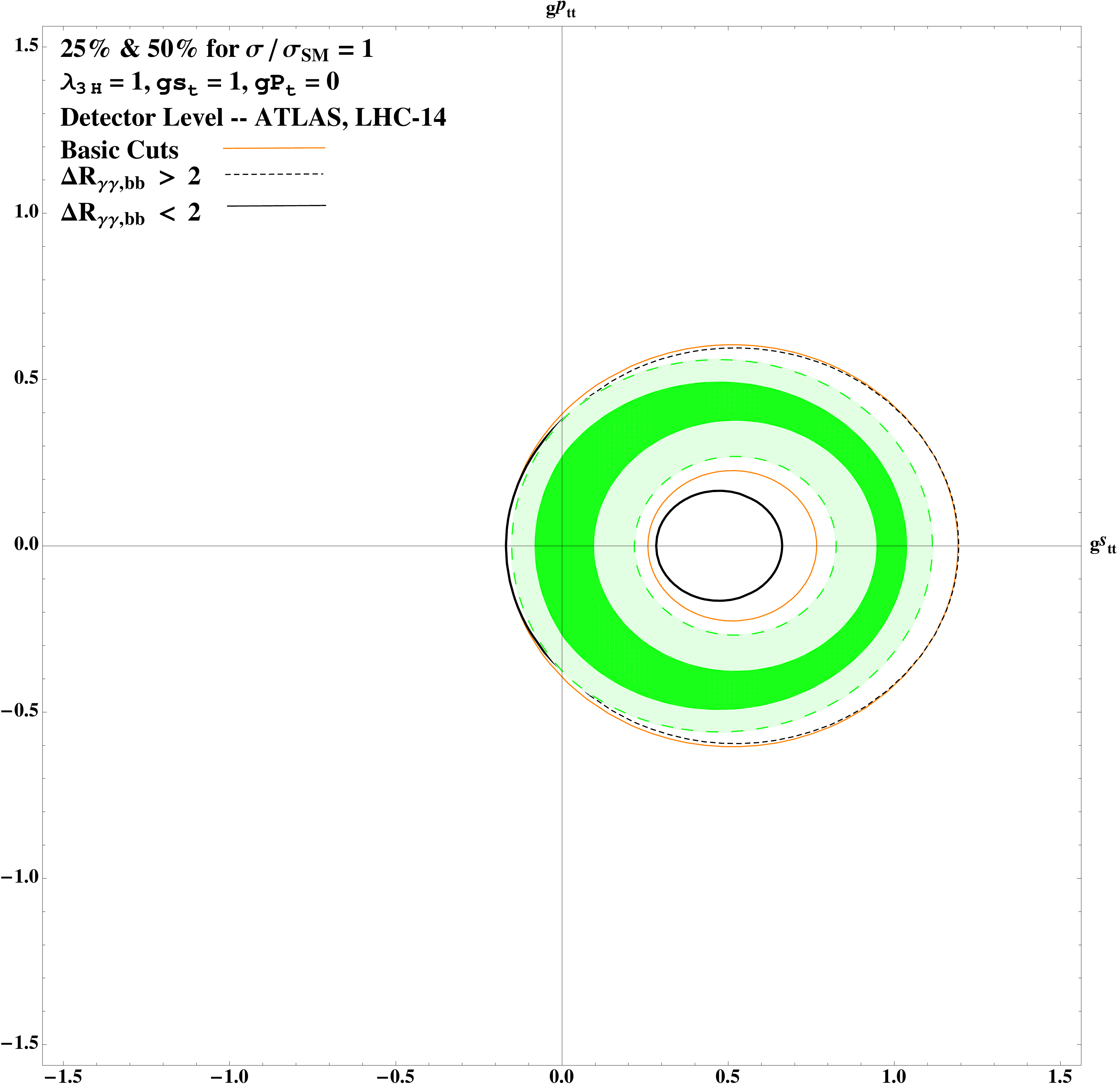}
\caption{\small \label{fig:cpv2-1}
{\bf CPV2:} 
The 25\% and 50\% sensitivity regions in the plane 
$(g_{tt}^S,\, g_{tt}^P)$ (with fixed $\lambda_{3H}=1$, $g_t^S=1$, and $g_t^P=0$)
bounded by three
measurements of cross sections with basic cuts, 
$\Delta R_{\gamma\gamma} > 2$, and 
$\Delta R_{\gamma\gamma} < 2$ (the upper-left panel);
with basic cuts, 
$\Delta R_{b b} > 2$, and 
$\Delta R_{b b } < 2$ (the upper-right panel); and
basic cuts,
$\Delta R_{\gamma\gamma},\,\Delta R_{b b} > 2$, and 
$\Delta R_{\gamma\gamma},\,\Delta R_{b b} < 2$ (the lower panel).
We assume that the measurements agree with the SM values with 
uncertainties of 25\% and 50\%, respectively.
}
\end{figure}

\begin{figure}[th!]
\centering
\includegraphics[width=3in]{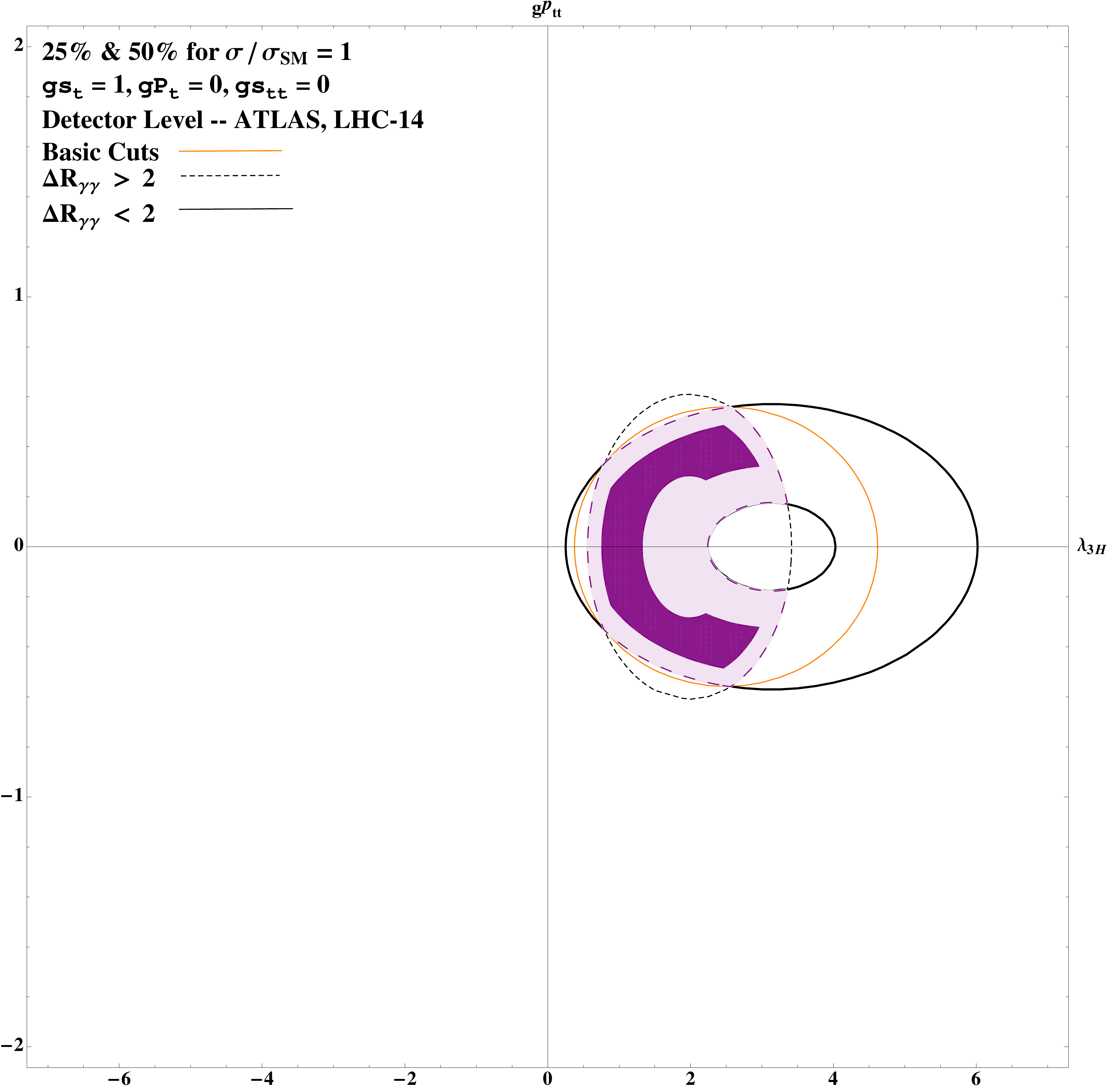}
\includegraphics[width=3in]{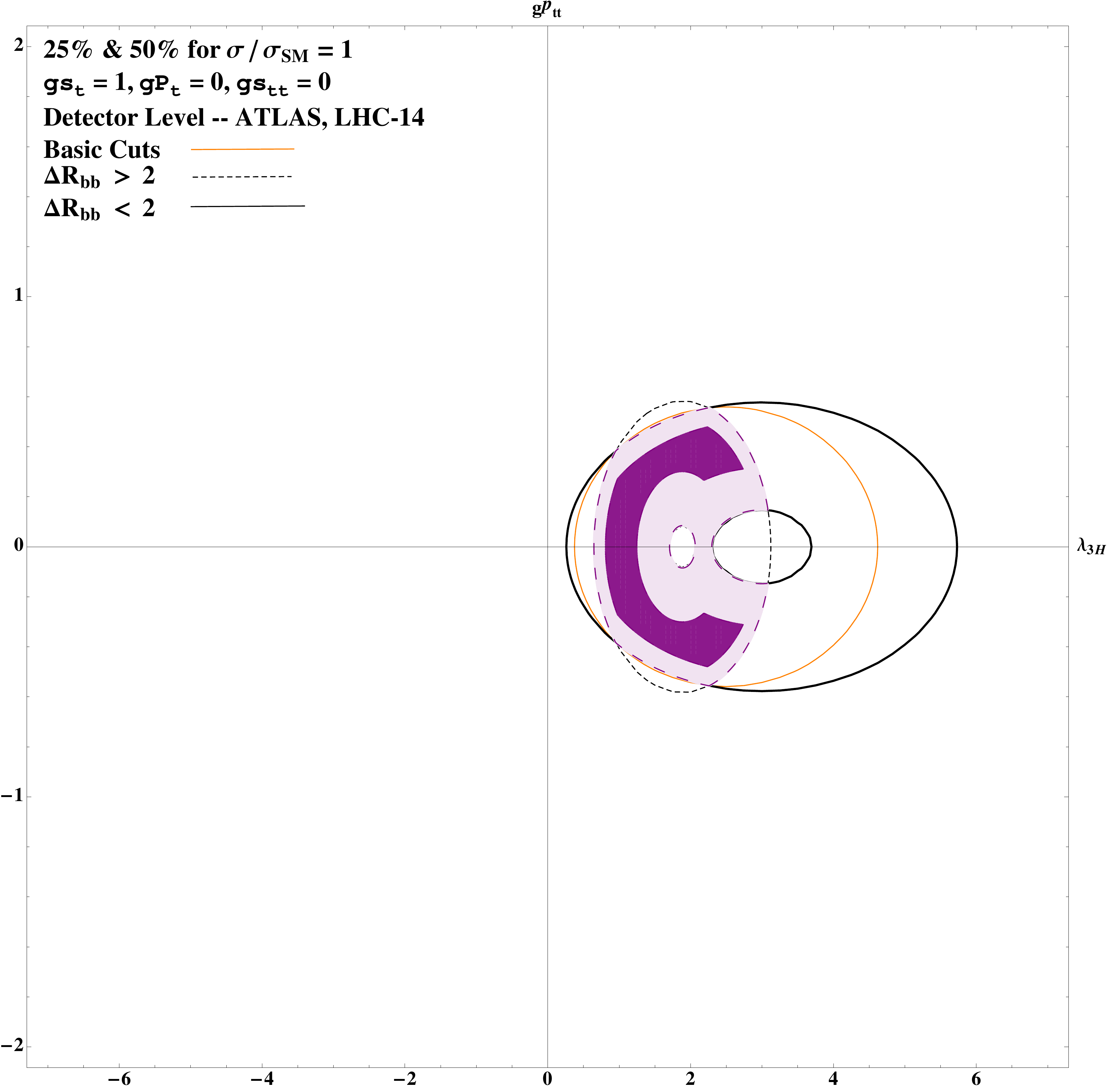}
\includegraphics[width=3in]{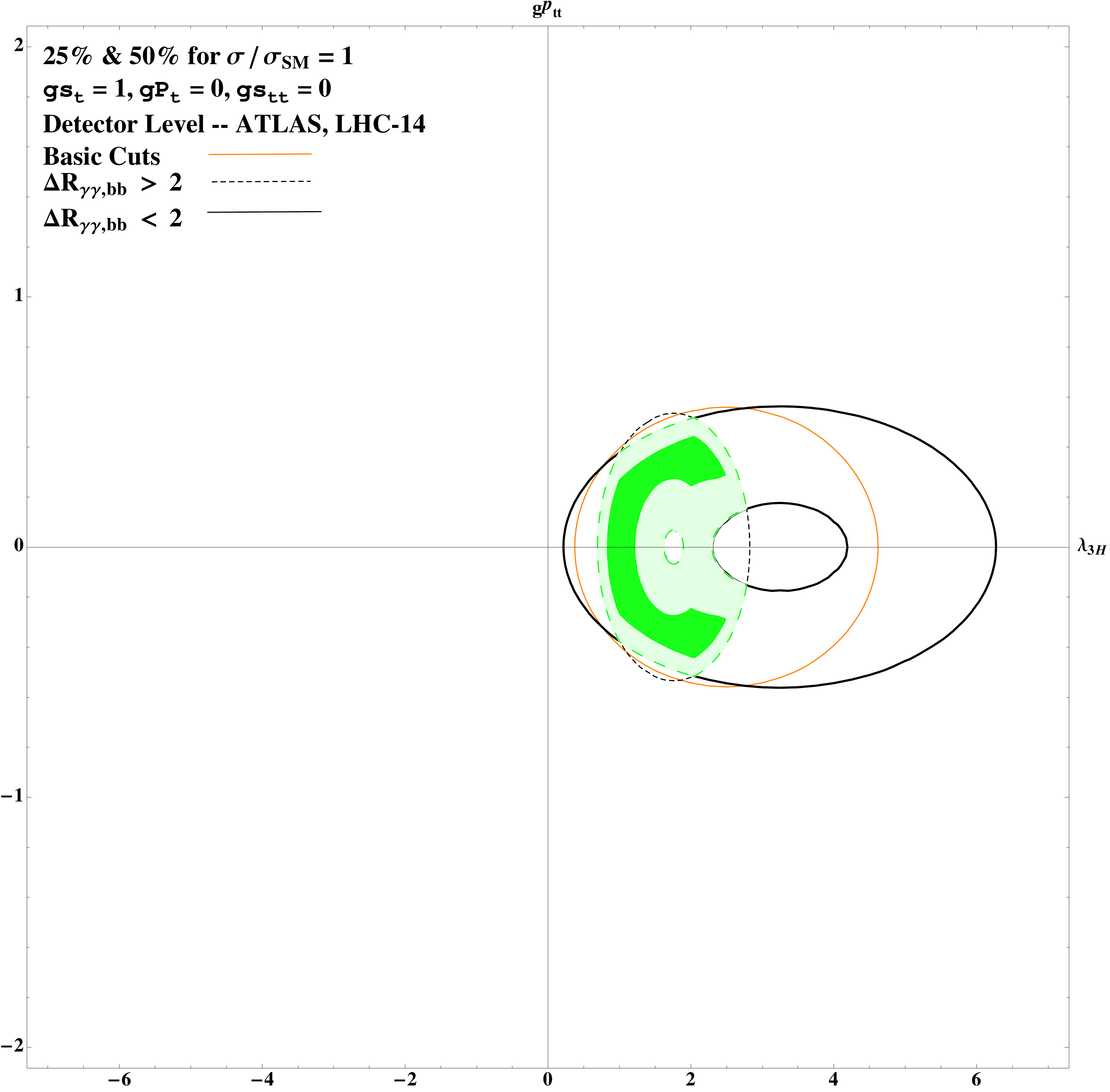}
\caption{\small \label{fig:cpv2-2}
{\bf CPV2:} similar to Fig.~\ref{fig:cpv2-1} but in the plane
of $(\lambda_{3H},\, g_{tt}^P)$ for $g_t^S=1$, $g_t^P= g_{tt}^S = 0$.
}
\end{figure}

\begin{figure}[th!]
\centering
\includegraphics[width=3in]{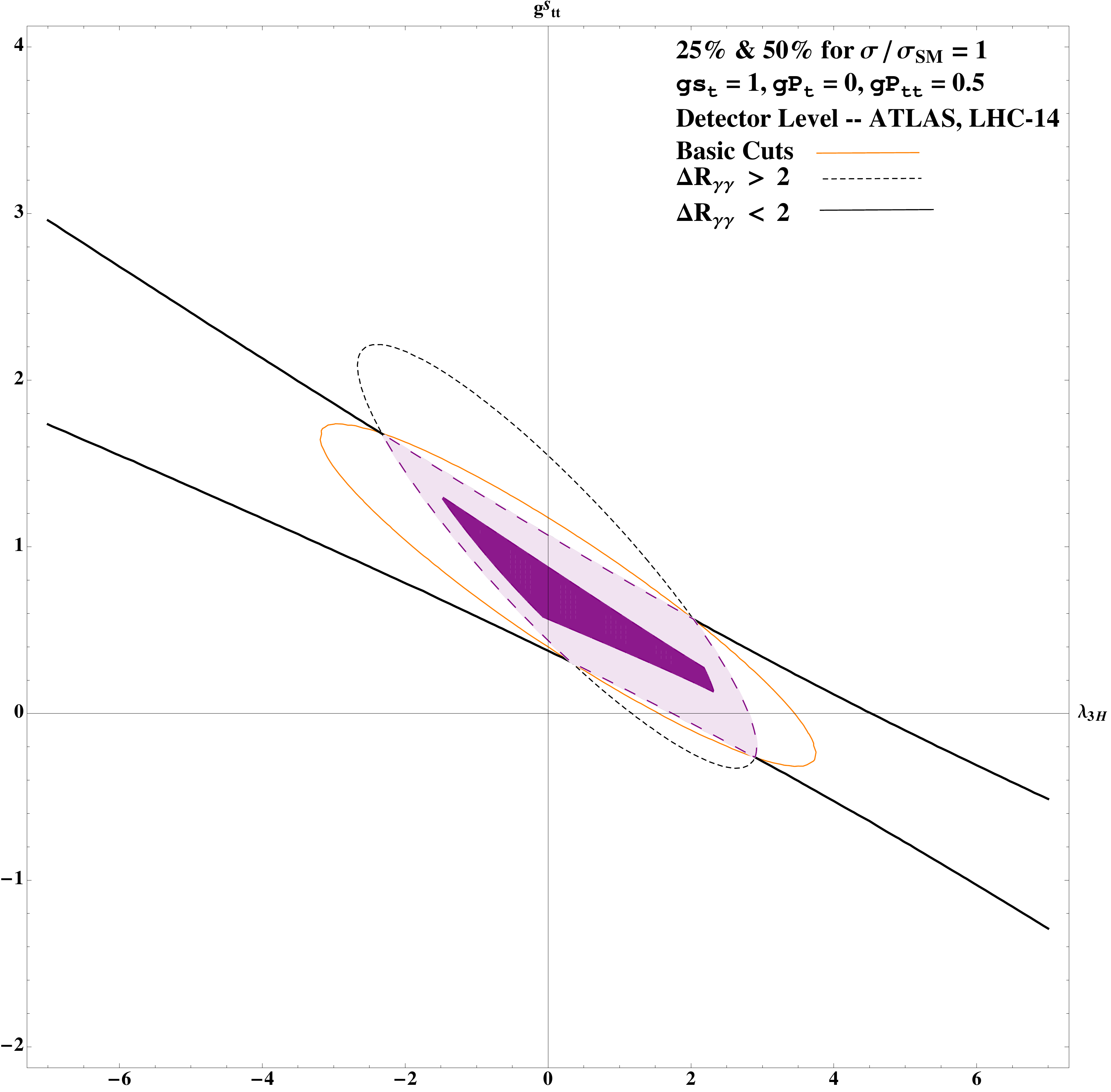}
\includegraphics[width=3in]{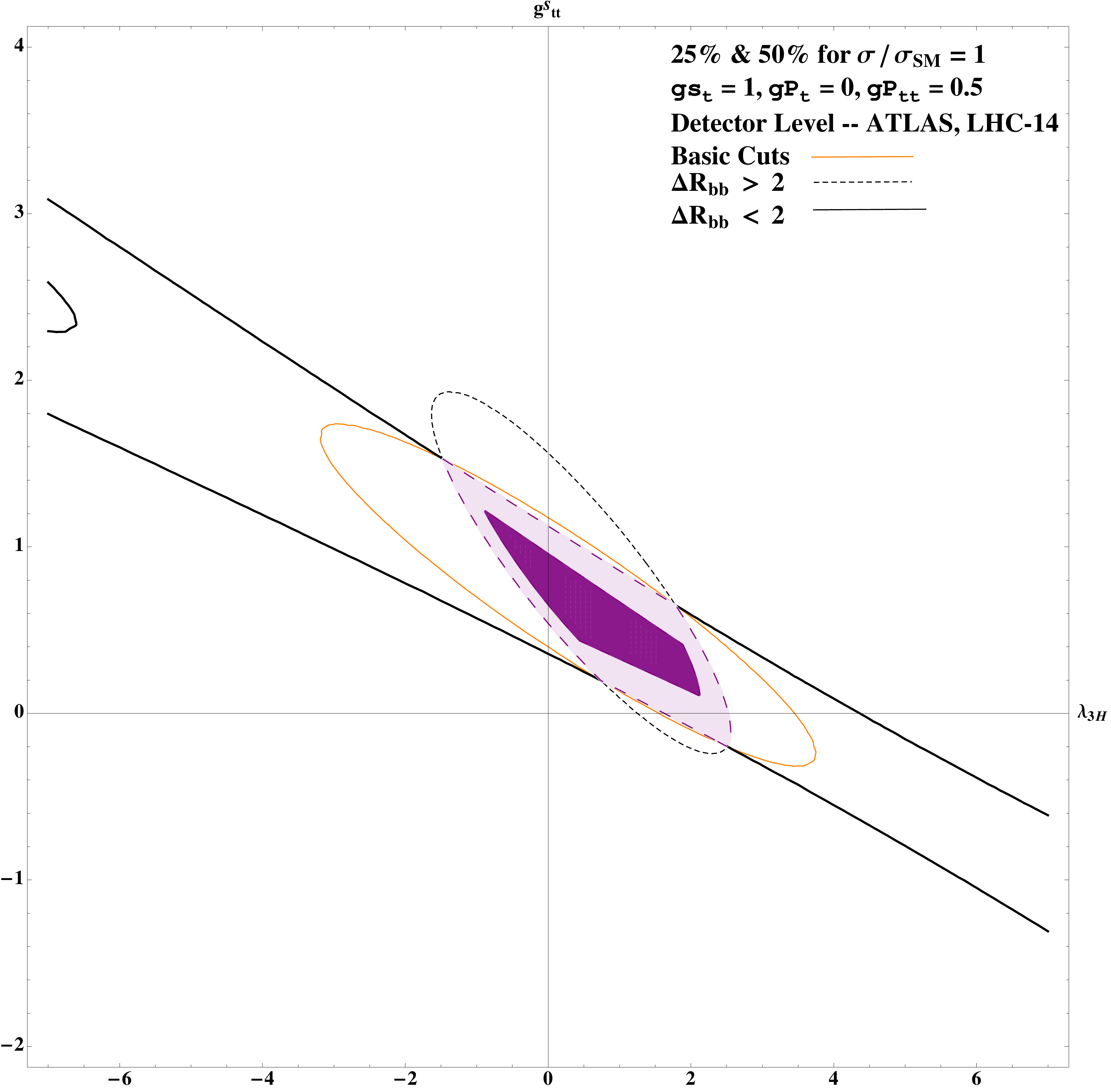}
\includegraphics[width=3in]{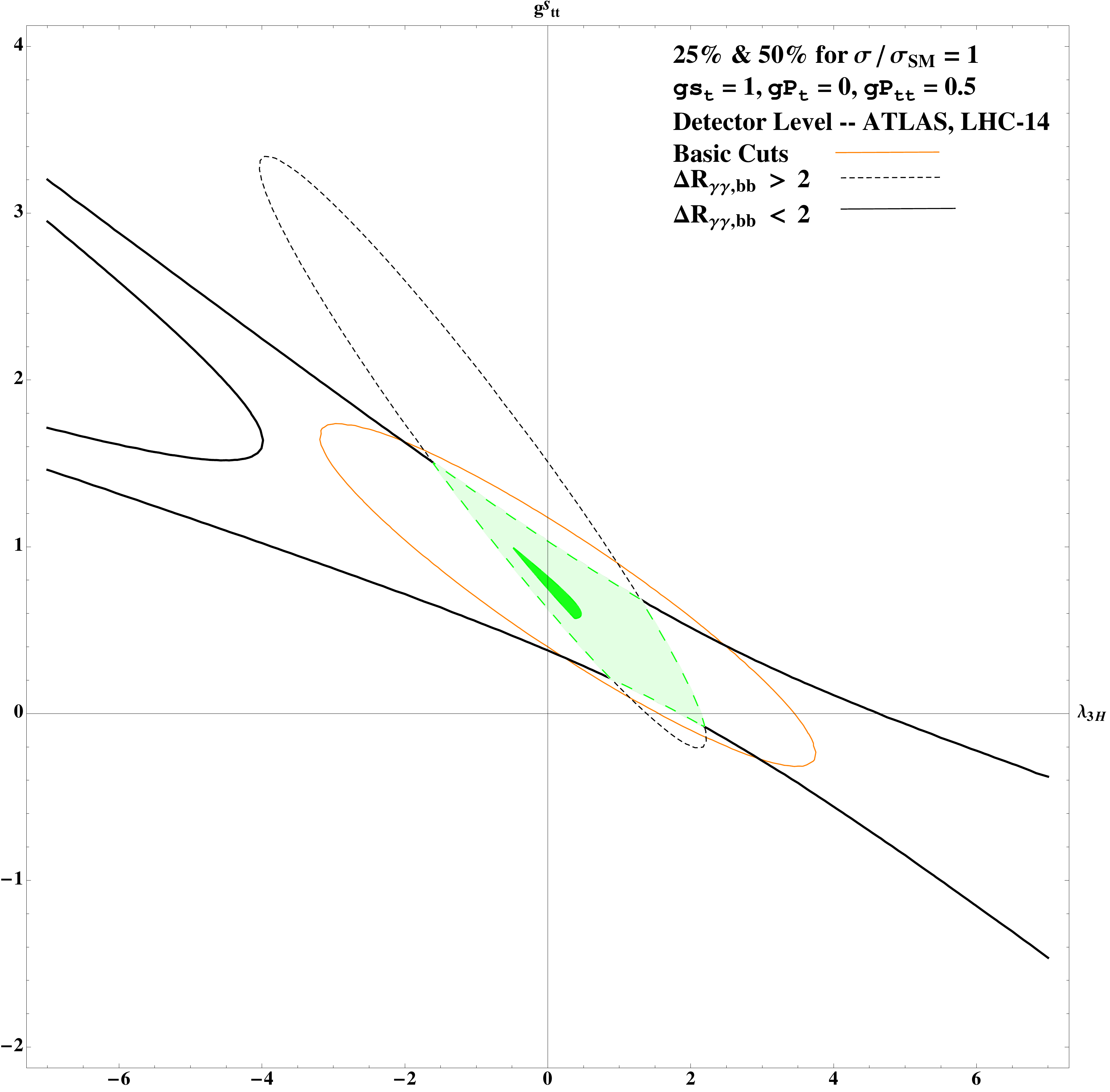}
\caption{\small \label{fig:cpv2-3}
{\bf CPV2:}  similar to Fig.~\ref{fig:cpv2-1} but in the plane
of $(\lambda_{3H},\, g_{tt}^S)$ for $g_t^S=1$, $g_t^P=0$, and $g_{tt}^P = 0.5$.
}
\end{figure}

Similar to previous scenarios we used $HH \to \gamma\gamma b \bar b$ with 
parton showering and detector simulations. We use the cuts $\Delta R$ larger
or smaller than 2 to discriminate the triangle and box diagrams.
Using the coefficients presented in Table~\ref{tab:all}, 
the ratio of cross sections to the SM predictions
can be obtained by Eq.~(\ref{cdef}).
We show the region of parameter space that we can obtain using 
$\Delta R_{\gamma\gamma}$, $\Delta R_{bb}$, and $\Delta R_{\gamma\gamma,bb}$
in the plane of $(g_{tt}^S,\, g_{tt}^P)$ for fixed $\lambda_{3H}=1$, $g_t^S=1$,
and $g_t^P=0$ in Fig.~\ref{fig:cpv2-1};
and similarly in the plane of 
$(\lambda_{3H},\, g_{tt}^P)$ for $g_t^S=1$, $g_t^P= g_{tt}^S = 0$ in
Fig.~\ref{fig:cpv2-2};
and finally in the plane of 
$(\lambda_{3H},\, g_{tt}^S)$ for $g_t^S=1$, $g_t^P=0$, and $g_{tt}^P = 0.5$ in
Fig.~\ref{fig:cpv2-3};

\subsection{100 TeV Prospect}
\begin{figure}[th!]
\centering
\includegraphics[width=5.2in]{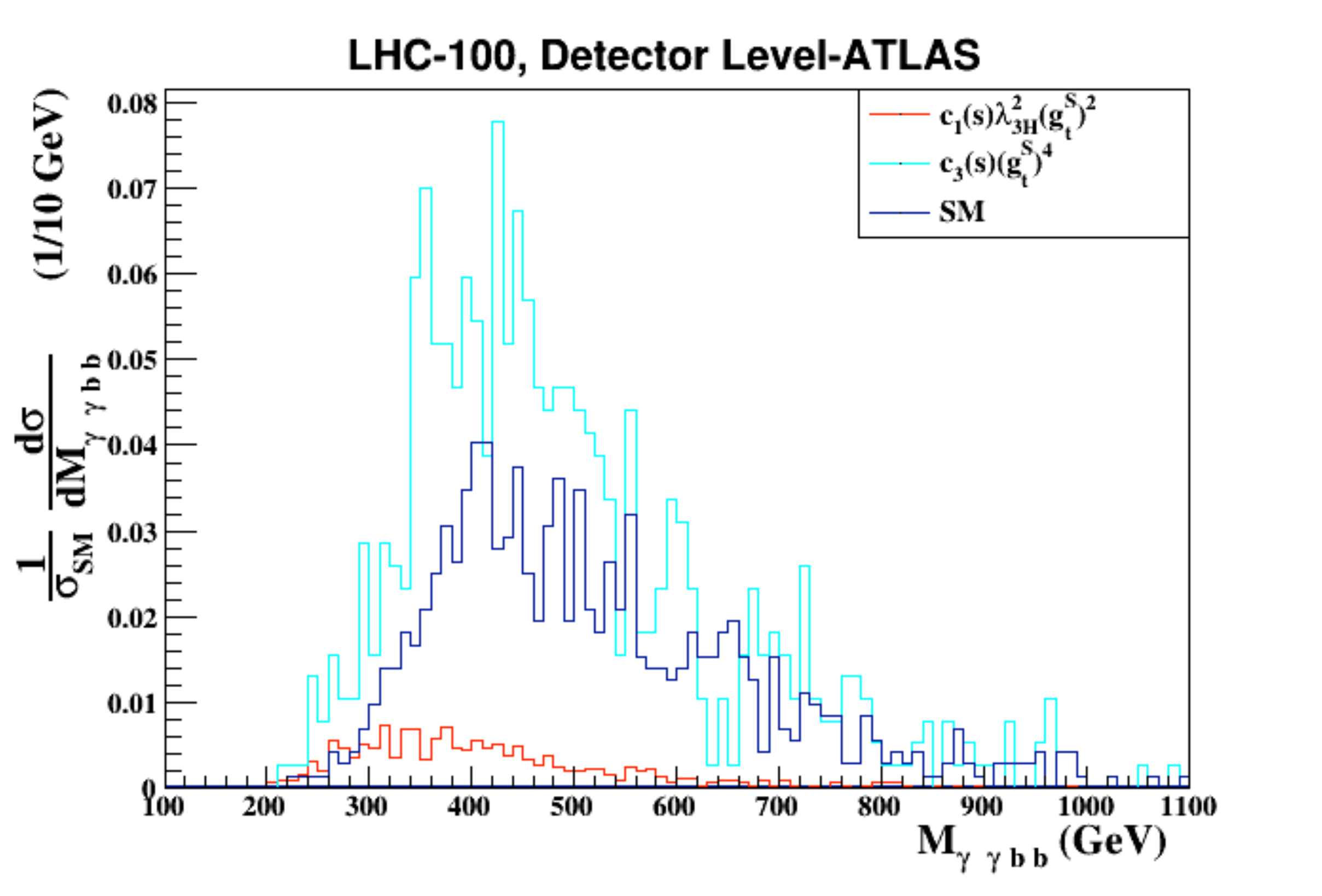}
\includegraphics[width=5.2in]{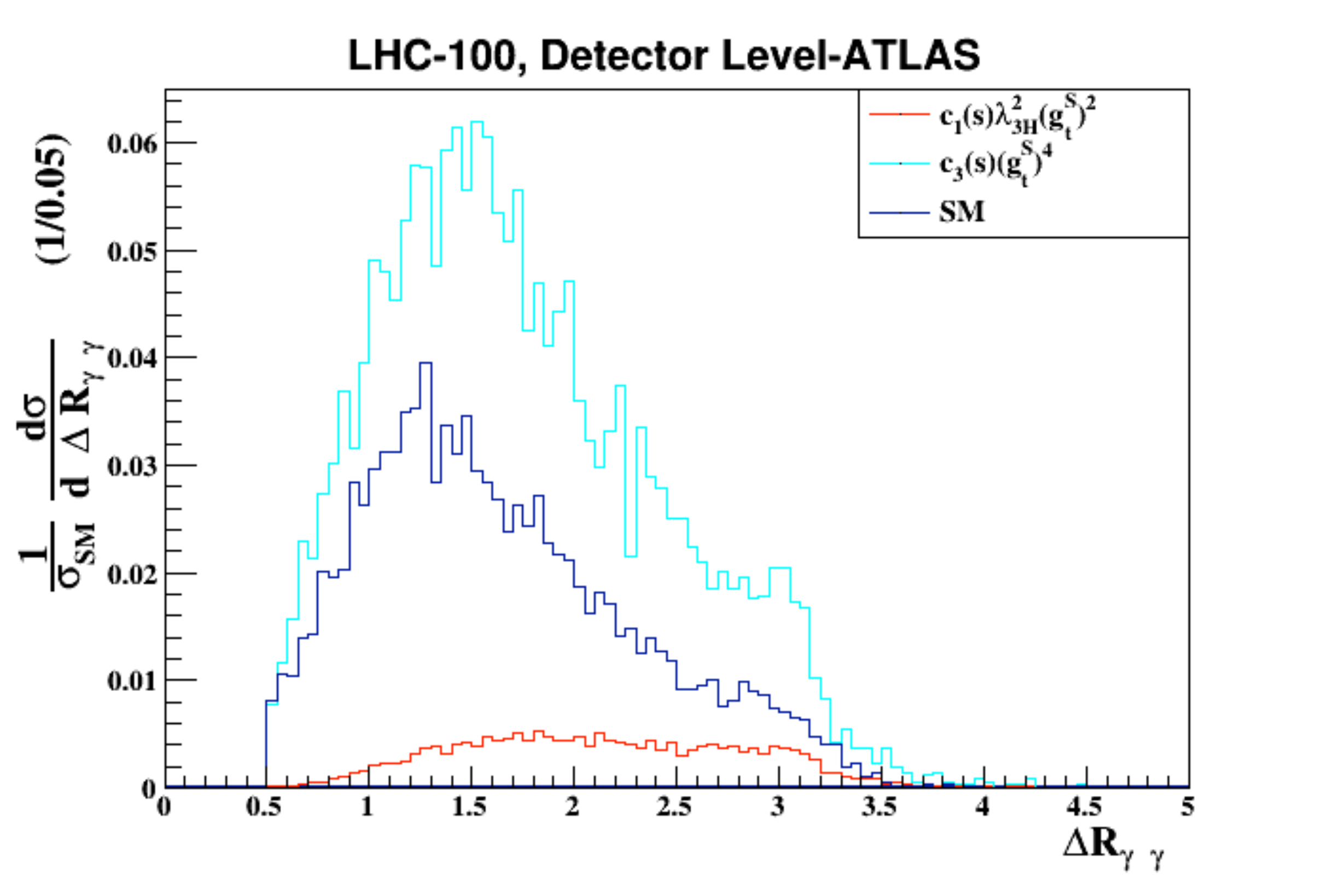}
\caption{\small \label{fig:100tev}
Distributions in the invariant mass distribution $M_{\gamma\gamma b b}$ and
the angular separation $\Delta R_{\gamma\gamma}$ 
in the decay products of the Higgs boson pair with detector simulation
at the 100 TeV $pp$ machine.
}
\end{figure}

\begin{figure}[th!]
\centering
\includegraphics[width=3.2in]{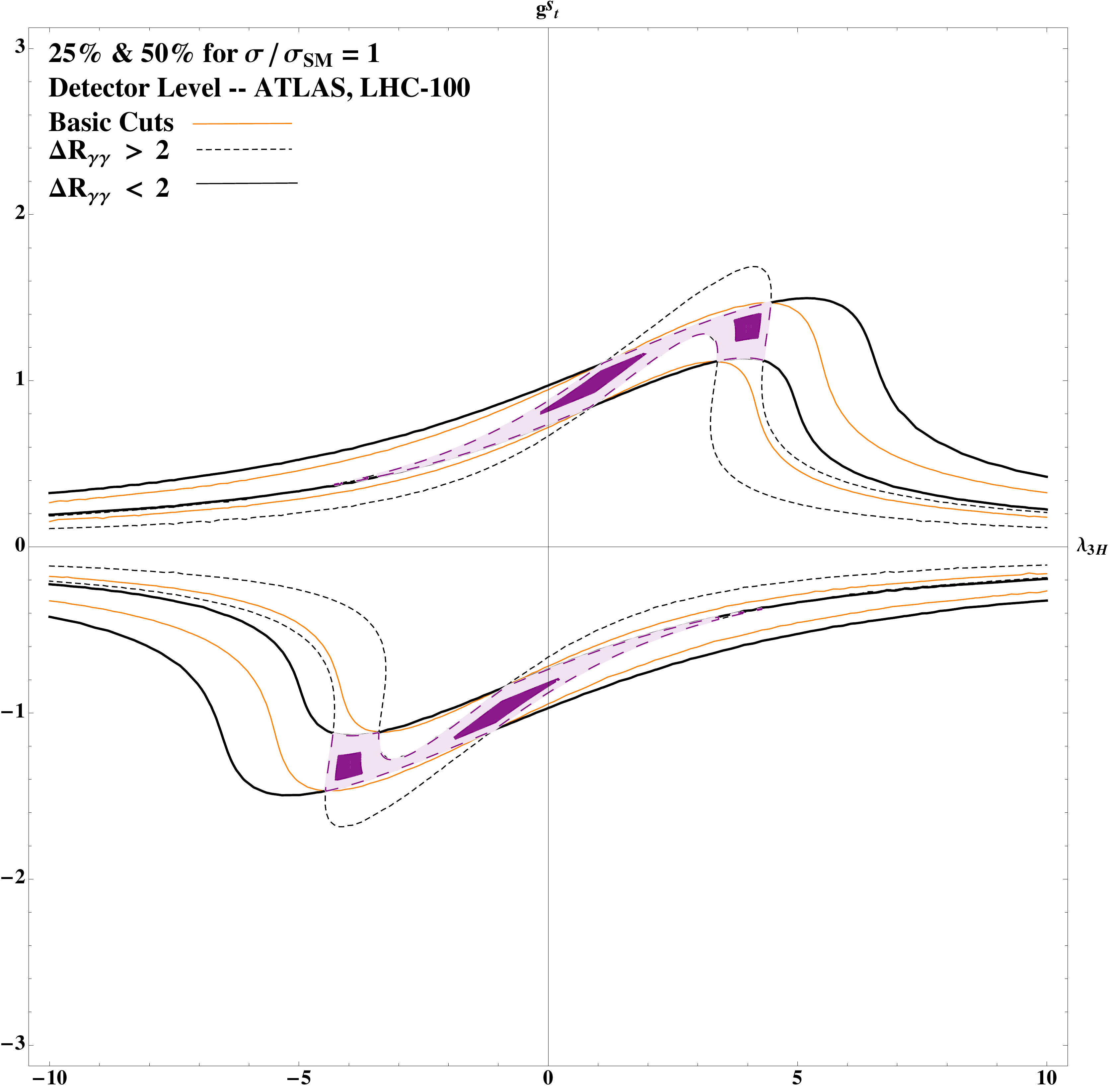}
\includegraphics[width=3.2in]{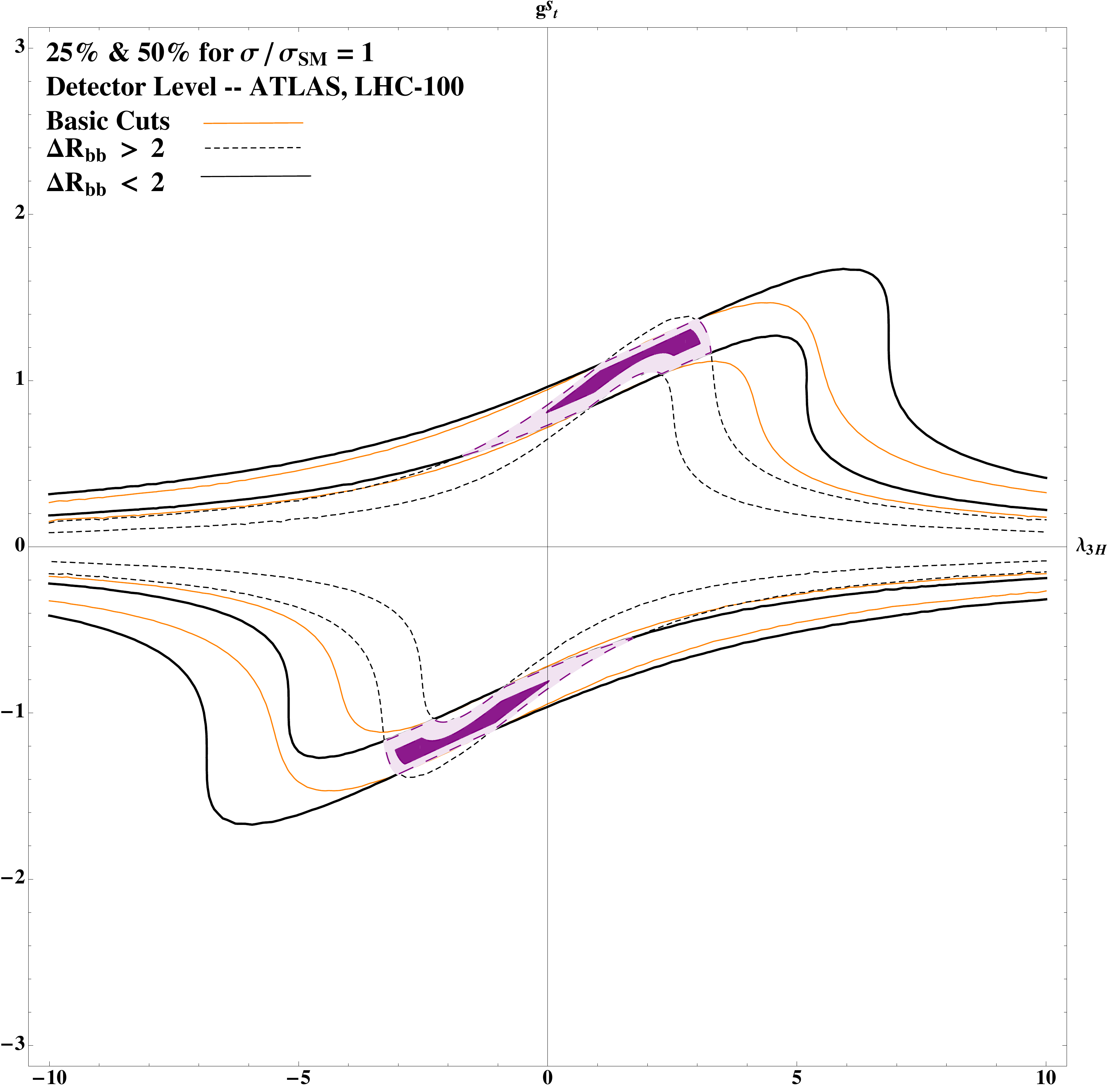}
\includegraphics[width=3.2in]{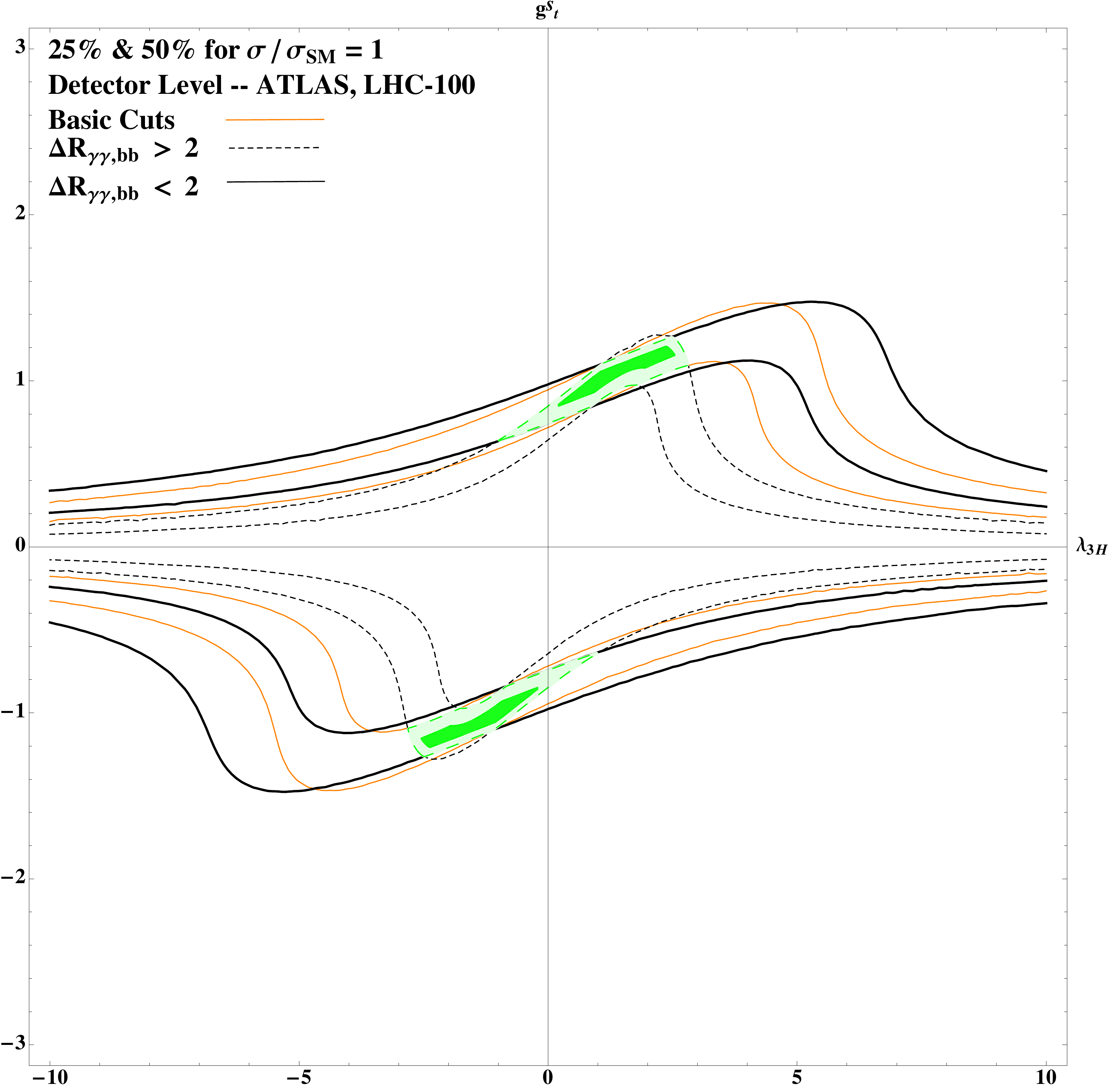}
\caption{\small \label{fig:cpc1_100}
{\bf CPC1 at 100 TeV:} 
Exactly the same as Fig.~\ref{fig:cpc1}, except that it is the case of 
the 100 TeV $pp$ machine. }
\end{figure}

All the results represented for the 14 TeV run were obtained by manipulating
the coefficients represented in Table~\ref{tab:all}. 
We represent the coefficients 
$c_{1,2,3},\, d_{1,2,3,4},\, e_{1,2,3},\, f_{1,2,3,4}$ 
for the 100 TeV $pp$ machine in Table~\ref{tab:100}.
Just for illustrations, we show the distributions of the 
invariant mass $M_{\gamma\gamma b\bar b}$ and angular separation
$\Delta R_{\gamma\gamma}$ for the {\bf CPC1} case at the 100 TeV machine 
in Fig.~\ref{fig:100tev}.
We found that the behavior of the distributions at 100 TeV 
is very similar to those at 14 TeV.  Therefore, the kinematic regions 
of interests separated by $\Delta R$ can be taken to be the same as 
14 TeV.  We can make simultaneous measurements of cross sections at 
100 TeV $pp$ machine to isolate the Higgs trilinear coupling. 
We show the sensitivity regions of parameter space in the {\bf CPC1} case 
at the 100 TeV $pp$ machine in Fig.~\ref{fig:cpc1_100}. The regions
are very similar to those in 14 TeV, though not exactly the same.
Sensitivity reach for each coupling in other cases 
can be obtained by similar methods with
the assumed luminosity. 

\section{Conclusions}

In this work, we have studied the behavior of Higgs-boson pair production
via gluon fusion at the 14 TeV LHC and the 100 TeV $pp$ machine. We have 
performed an exploratory study with heavy degrees of freedom being
integrated out and resulting in possible modifications of the top-Yukawa
coupling, Higgs trilinear coupling, and a new contact $ttHH$ coupling,
as well as the potential CP-odd component in the Yukawa and contact 
couplings.  We have identified useful variables -- the angular separation
between the decay products of the Higgs boson -- to discriminate among
the contributions from the triangle, box, and contact diagrams.
We have successfully demonstrated that with three simultaneous measurements
of the Higgs-pair production cross sections, defined by the kinematic
cuts, one can statistically show a nonzero value for the Higgs trilinear
coupling $\lambda_{3H}$ if we can measure the cross sections with
less than 25\% uncertainty.  This is the key result of this work.

We also offer the following comments with regards to our findings.
\begin{enumerate}
\item
The triangle diagram, which contains an $s$-channel
Higgs propagator, does not increase as much as the box diagram or 
the contact diagram with the center-of-mass energy 
$\sqrt{\hat s}$. This explains why the opening angle 
($\Delta R_{\gamma\gamma}$ or $\Delta R_{bb}$) in the decay products
of each Higgs boson is a useful variable to separate between
the triangle and the box diagram.  Thus, it helps to isolate the
Higgs trilinear coupling $\lambda_{3H}$.

\item
The contact diagram contains a dim-5 operator $ttHH$, which actually
breaks the unitarity at about $\sqrt{\hat s} \sim 17.6/g_{tt}^S$ TeV.
This implies that it could become dominant at high invariant mass.

\item 
Suppose we take a measurement of cross sections, we can map out the
possible region of parameter space. Since in different kinematic regions
the regions of parameter space are mapped out differently, such that
simultaneous measurements can map out the intersected regions.
With measurement uncertainties less than 25\% one can statistically 
show a nonzero value for the Higgs trilinear coupling,
and also obtain the sensitivity regions of $\lambda_{3H}$:
 $0.3 \alt |\lambda_{3H}| \alt 2.6$.
for $\sigma/\sigma_{\rm SM} =1$.

\item
We found that the behavior of the distributions of the 
invariant mass $M_{\gamma\gamma b\bar b}$ and angular separation
$\Delta R_{\gamma\gamma}$ or $\Delta R_{b b}$ at 14 TeV are very similar to
those at 100 TeV.  We can then use the same method as in 14 TeV to
isolate the Higgs trilinear coupling. 

\item
It is difficult, if not impossible, to determine the 
Higgs trilinear coupling uniquely at the LHC and 100 TeV $pp$ machine
even in the simplest case assuming very high luminosity and 
precise independent input for the top-Yukawa coupling.
We suggest to combine the LHC results with information which can be obtained at 
a future $e^+e^-$ linear collider.

\item If the couplings deviate from their SM values,
the Higgs-boson 
pair production cross section can easily increase by an order of
magnitude. For example, in the {\bf CPC2} case, $\sigma/\sigma_{\rm SM}>10$ for 
$\lambda_{3H}>9\,{\rm or}\,<-4$ when $g^S_t=1$ and $g^S_{tt}=0$,
$g^S_t>1.7\,{\rm or}\,<-1.3$ when $\lambda_{3H}=1$ and $g^S_{tt}=0$, and
$g^S_{tt}>2.6\,{\rm or}\,<-1.4$ when $\lambda_{3H}=g^S_t=1$:
see Fig.~\ref{fig:cpc2-3}. 
The cross section larger  than the SM prediction
may reveal the new physics hidden behind the SM and we can have better
prospect to measure the Higgs self coupling at the LHC.

\end{enumerate}

\section*{Acknowledgment}  
We thank Olivier Mattelaer for useful discussion about the simulation 
of Higgs-boson pair production in MADGRAPH v.5, and special thanks to 
Eleni Vryonidou for sending us the original code of Higgs pair 
production and  enlightening discussion.  
This work was supported by the National Science
Council of Taiwan under Grants No. 102-2112-M-007-015-MY3,
and by
the National Research Foundation of Korea (NRF) grant
(No. 2013R1A2A2A01015406).

\appendix
\section{}

In Table~\ref{tab:cx}, 
we show the SM cross sections
for $pp \to HH \to \gamma\gamma b\bar b$ at the 14 TeV LHC
with and without angular-separation cuts.
Note that the cross section
before applying any cuts is about $0.09$ fb 
and it becomes $0.005$ fb after applying the basic cuts.
In the region of $\Delta R_{\gamma\gamma}>2 (<2)$, the cross
section is  0.0013 fb (0.0038 fb)
where it is dominated by the triangle (box)
diagram. The ratio is about $1:2.8$. 
We also show the cross sections
for the 100
TeV $pp$ machine, and the corresponding ratio is about 
$1:3.7$. 
It shows the fact that the triangle diagram is more
suppressed because of the $s$-channel Higgs propagator at higher energy.
In the regions of $\Delta R_{b  b}$ larger and smaller than 2, the
ratios are $1:5.7$ and $1:8.2$ at the  14 TeV LHC and the 100
TeV $pp$ machine, respectively.

As we have promised, we are going to comment on the $HH \to \tau^+ \tau^- b\bar b$ 
decay mode. This mode  has the obvious advantage of a larger branching ratio
than the $\gamma\gamma b\bar b$ mode, but the identification efficiency and 
momentum measurements of $\tau$ leptons are much weaker than photons. 
In Table~\ref{tab:cx}, we show the 
SM cross sections
for $pp \to HH \to \tau^+\tau^- b\bar b$ at the 14 TeV LHC
with and without angular-separation cuts.
%
%
Taking into account the branching ratios, and the identification and selection
efficiencies, 
the event rates of $\tau^+ \tau^- b\bar b$
are similar to those of the $\gamma\gamma b\bar b$ mode.
%

\begin{table}[h!]
   \centering
   \caption{
{\bf 14 TeV LHC:} The coeffcients for the ratio 
of the cross sections $\sigma(gg \to HH)/\sigma_{\rm SM}(gg \to
HH)$ as in Eq.~(\ref{cdef})
with and without the angular-separation cuts of
$\Delta R_{\gamma\gamma} > \;{\rm or} < 2$;
$\Delta R_{b b} > \;{\rm or} < 2$; and 
$\Delta R_{\gamma\gamma}$ and $\Delta R_{b b}$ both $> 2$ or $<2$.
The relevant coefficients for the {\bf CPC1} scenario are 
$c_1,\, c_2,\, c_3$; those for the {\bf CPC2} scenario are
$c_1,\, c_2,\, c_3, e_1, e_2, e_3$;
and those for the {\bf CPV1} scenario are
$c_1,\, c_2,\, c_3, d_1, d_2, d_3, d_4$.
All the coefficients are involved in the {\bf CPV2} case.
Results are at the detector level using the ATLAS template in Delphes v.3.
   \label{tab:all} 
}
   \begin{tabular}{ c c c c c c c c }
      \multicolumn{8}{}{}\\
\hline $\sqrt{s} :14$ TeV& $c_1(s)$ & $c_2(s)$ & $c_3(s)$ &
$d_1(s)$ & $d_2(s)$ & $d_3(s)$ & $d_4(s)$  \\ 
$$ Cuts & $\left[\lambda_{3H}^2(g_t^S)^2\right]$ & 
$\left[\lambda_{3H}(g_t^S)^3\right]$ & 
$\left[(g_t^S)^4\right]$ &$\left[\lambda_{3H}^2(g_t^P)^2\right]$ & 
$\left[\lambda_{3H}g_t^S(g_t^P)^2\right]$ & 
$\left[(g_t^S)^2(g_t^P)^2\right]$ & 
$\left[(g_t^P)^4\right]$ \\

 \hline 
No cuts & $0.263$ & $-1.310$ & $2.047$ &
$0.820$ & $-5.961$ & $13.348$ & $0.707$ \\
Basic Cuts &0.221	&	-1.104	&	1.883	&	0.665	&	-4.738	&	11.757	&	0.650\\ 
    $\Delta R_{\gamma\gamma} > 2$ &0.470	&	-1.868	&	2.398	&	1.481	&	-9.754	&	19.859	&	0.858\\

    $\Delta R_{\gamma\gamma}  < 2$ &0.133	&	-0.834	&	1.701	&	0.376	&	-2.959	&	8.884	&	0.576\\
 
  $\Delta R_{bb} > 2$ &0.666	&	-2.512	&	2.847	&	2.040	&	-13.425	&	25.316	&	1.074\\ 
$\Delta R_{bb} < 2$ & 0.143	&	-0.857	&	1.714	&	0.424	&	-3.214	&	9.378	&	0.575\\
  $\Delta R_{bb} > 2 \ \& \ \Delta R_{\gamma\gamma}  > 2$ &0.895	&	-3.150	&	3.255	&	2.613	&	-17.210	&	30.456	&	1.278\\
  $\Delta R_{bb} < 2 \ \& \ \Delta R_{\gamma\gamma}  < 2$  &0.121	&	-0.785	&	1.664	&	0.319	&	-2.630	&	8.257	&	0.563\\
  \hline   
  \hline  $\sqrt{s} :14 $TeV&$e_1(s)$ & $e_2(s)$ & $e_3(s)$&$f_1(s)$ & $f_2(s)$ & $f_3(s)$ & $f_4(s)$ \\ 
$$ Cuts &  $\left[\lambda_{3H}g_t^S g_{tt}^S\right]$ &
$\left[g_{tt}^S (g_t^S)^2\right]$ &
$\left[(g_{tt}^S)^2\right]$ &
$\left[\lambda_{3H}g_t^P g_{tt}^P\right]$ &
$\left[g_{tt}^S (g_t^P)^2\right]$ &
$\left[g_t^S g_t^P g_{tt}^P\right]$ &
$\left[(g_{tt}^P)^2\right]$  
\\

 \hline 
No cuts & $1.364$ & $-4.224$ & $2.617$ &
$1.848$ & $2.269$ & $-6.886$ & $3.769$\\
Basic Cuts &1.381	&	-3.966	&	2.521	&	1.939	&	2.328	&	-5.239	&	3.178\\ 

    $\Delta R_{\gamma\gamma}  > 2$ &1.857	&	-4.506	&	2.267	&	4.014	&	2.555	&	-11.188	&	2.569\\

    $\Delta R_{\gamma\gamma}  < 2$ &1.212	&	-3.774	&	2.611	&	1.203	&	2.247	&	-3.130	&	3.394\\
 
  $\Delta R_{bb} > 2$ & 2.248	&	-5.214	&	2.474	&	5.517	&	3.367	&	-16.349	&	3.003		\\ 
$\Delta R_{bb} < 2$ & 1.229	&	-3.747	&	2.529	&	1.311	&	2.146	&	-3.290	&	3.208\\
  $\Delta R_{bb} > 2 \ \& \ \Delta R_{\gamma\gamma} > 2$ &3.047	&	-5.947	&	2.780	&	7.274	&	3.759	&	-21.142	&	3.547\\
  $\Delta R_{bb} < 2 \ \& \ \Delta R_{\gamma\gamma} < 2$  &1.238	&	-3.758	&	2.664	&	1.095	&	2.211	&	-2.716	&	3.500\\
  \hline   

   \end{tabular}
\end{table}


\begin{table}[htbp!]
\centering
\caption{\label{tab:cx}
The SM cross sections for the process
$pp \to HH \to \gamma\gamma b\bar b$  with various 
angular-separation cuts on $\Delta R_{bb\,,\gamma\gamma}$
at the 14 TeV LHC (second column) and at the 100 TeV $pp$ machine (third
column).
The last column shows them for the process
$pp \to HH \to \tau^+\tau^- b\bar b$   at the 14 TeV LHC.
We have taken account of the SM NLO cross section
$\sigma_{\rm SM}(pp\to HH) \simeq 34$ fb, the Higgs branching fractions,
and both the photon and $b$-quark reconstruction efficiencies.
The $p_T$ dependence of $b$-tagging efficiency is considered and
$0.5$ is taken for the $\tau$-tagging efficiency.
Also considered is the mis-tagging
probability of $P_{j\rightarrow \tau}= 0.01$.
Results are at the detector level using the ATLAS template in Delphes v.3.
}
\begin{tabular}{ c l l l}
\multicolumn{4}{}{}\\
\hline
 Cuts
&SM-14$~(\gamma \gamma b\bar{b})~~~~$
&SM-100$~(\gamma \gamma b\bar{b})~~~~$
&SM-14$~(\tau^+ \tau^- b\bar{b})~~~~$ \\
 & \multicolumn{3}{c}{Cross Section (fb)}  \\
\hline \\
No cuts & $8.92 \times 10^{-2}$ & 3.73 & 2.41 \\
Basic Cuts  & $5.1\times 10^{-3}$ &  $2.05\times 10^{-1}$ &  $3.53\times 10^{-3}$  \\
$\Delta R_{\gamma\gamma / \tau^+\tau^-} > 2$
&$1.34\times 10^{-3}$ & $4.34\times 10^{-2}$ & $6.43\times 10^{-4}$ \\
$\Delta R_{\gamma\gamma / \tau^+\tau^-} < 2$
& $3.76\times 10^{-3}$ & $1.61\times 10^{-1}$ & $2.89\times 10^{-3} $ \\
$\Delta R_{bb} > 2$
& $7.61\times 10^{-4}$ & $2.23\times 10^{-2}$ & $4.82\times 10^{-4}$ \\
$\Delta R_{bb} < 2$
& $4.34\times 10^{-3}$ & $1.83\times 10^{-1}$ & $3.05\times 10^{-3}$ \\
$\Delta R_{bb} > 2 \ \& \ \Delta R_{\gamma\gamma / \tau^+\tau^-} > 2~~~~$
& $4.79\times 10^{-4}$ & $1.40\times 10^{-2}$ & $3.21\times 10^{-4}$ \\
$\Delta R_{bb} < 2 \ \& \ \Delta R_{\gamma\gamma / \tau^+\tau^-} < 2~~~~$
& $3.48\times 10^{-3}$ & $1.53\times 10^{-1}$ & $2.73\times 10^{-3}$ \\ \\
\hline   \\
\end{tabular}
\end{table}


\begin{table}[h!]
   \centering
   \caption{{\bf 100 TeV $pp$ machine:} 
The coeffcients for the ratio 
of the cross sections $\sigma(gg \to HH)/\sigma_{\rm SM}(gg \to
HH)$ as in Eq.~(\ref{cdef})
with and without the angular-separation cuts of
$\Delta R_{\gamma\gamma} > \;{\rm or} < 2$;
$\Delta R_{b b} > \;{\rm or} < 2$; and 
$\Delta R_{\gamma\gamma}$ and $\Delta R_{b b}$ both $> 2$ or $<2$.
The relevant coefficients for the {\bf CPC1} scenario are 
$c_1,\, c_2,\, c_3$; those for the {\bf CPC2} scenario are
$c_1,\, c_2,\, c_3, e_1, e_2, e_3$;
and those for the {\bf CPV1} scenario are
$c_1,\, c_2,\, c_3, d_1, d_2, d_3, d_4$.
All the coefficients are involved in the {\bf CPV2} case.
Results are at the detector level using the ATLAS template in Delphes v.3.
   \label{tab:100} 
}
   \begin{tabular}{ c c c c c c c c }
      \multicolumn{8}{}{}\\
\hline $\sqrt{s} :100$ TeV& $c_1(s)$ & $c_2(s)$ & $c_3(s)$ &
$d_1(s)$ & $d_2(s)$ & $d_3(s)$ & $d_4(s)$  \\ 
$$ Cuts & $\left[\lambda_{3H}^2(g_t^S)^2\right]$ & 
$\left[\lambda_{3H}(g_t^S)^3\right]$ & 
$\left[(g_t^S)^4\right]$ &$\left[\lambda_{3H}^2(g_t^P)^2\right]$ & 
$\left[\lambda_{3H}g_t^S(g_t^P)^2\right]$ & 
$\left[(g_t^S)^2(g_t^P)^2\right]$ & 
$\left[(g_t^P)^4\right]$\\

 \hline 
No cuts & $0.208$ & $-1.108$ & $1.900$ &
$0.635$ & $-4.789$ & $11.225$ & $0.683$\\
Basic Cuts &	0.173	&	-1.032	&	1.860	&	0.503	&	-4.045	&	10.019	&	0.633									\\ 

    $\Delta R_{\gamma\gamma} > 2$ &	0.389	&	-1.904	&	2.515	&	1.275	&	-6.972	&	13.375	&	0.853									\\

    $\Delta R_{\gamma\gamma} < 2$ &	0.115	&	-0.798	&	1.683	&	0.295	&	-3.258	&	9.116	&	0.574									\\
 
  $\Delta R_{bb} > 2$ &	0.607	&	-2.419	&	2.813	&	1.845	&	-9.336	&	17.393	&	1.057									\\ 
$\Delta R_{bb} < 2$ &	0.120	&	-0.863	&	1.743	&	0.340	&	-3.400	&	9.119	&	0.581									\\
  $\Delta R_{bb} > 2 \ \& \ \Delta R_{\gamma\gamma} > 2$ & 	0.753	&	-2.662	&	2.909	&	2.248	&	-10.518	&	17.691	&	1.245									  \\
  $\Delta R_{bb} < 2 \ \& \ \Delta R_{\gamma\gamma} < 2$  &  	0.102	&	-0.733	&	1.632	&	0.249	&	-3.041	&	8.700	&	0.565									\\
 \hline   
  \hline  $\sqrt{s} :100 $TeV&$e_1(s)$ & $e_2(s)$ & $e_3(s)$&$f_1(s)$ & $f_2(s)$ & $f_3(s)$ & $f_4(s)$ \\ 
$$ Cuts &  $\left[\lambda_{3H}g_t^S g_{tt}^S\right]$ &
$\left[g_{tt}^S (g_t^S)^2\right]$ &
$\left[(g_{tt}^S)^2\right]$ &
$\left[\lambda_{3H}g_t^P g_{tt}^P\right]$ &
$\left[g_{tt}^S (g_t^P)^2\right]$ &
$\left[g_t^S g_t^P g_{tt}^P\right]$ &
$\left[(g_{tt}^P)^2\right]$
\\

 \hline 
No cuts & $1.214$ & $-4.137$ & $2.974$ &
$1.474$ & $2.154$ & $-5.342$ & $10.568$\\ 
Basic Cuts &	1.170	&	-4.081	&	2.848	&	1.300	&	1.935	&	-3.379	&	7.802									\\ 

    $\Delta R_{\gamma\gamma} > 2$ &	1.782	&	-4.886	&	2.591	&	3.675	&	2.151	&	-2.696	&	5.511									\\

    $\Delta R_{\gamma\gamma} < 2$ &	1.006	&	-3.865	&	2.917	&	0.662	&	1.878	&	-3.563	&	8.419									\\
 
  $\Delta R_{bb} > 2$ &	2.011	&	-5.585	&	2.957	&	6.947	&	2.576	&	-4.961	&	5.373									\\ 
$\Delta R_{bb} < 2$ &	1.068	&	-3.898	&	2.834	&	0.612	&	1.857	&	-3.186	&	8.099									\\
  $\Delta R_{bb} > 2 \ \& \ \Delta R_{\gamma\gamma} > 2$ &	2.483	&	-5.858	&	3.106	&	8.165	&	2.694	&	-4.722	&	6.079									\\
  $\Delta R_{bb} < 2 \ \& \ \Delta R_{\gamma\gamma} < 2$  &	0.995	&	-3.798	&	2.928	&	0.437	&	1.851	&	-3.466	&	8.647									\\
  \hline   

   \end{tabular}
\end{table}

\clearpage

\clearpage

\end{document}